\documentclass[preprint,12pt,authoryear]{elsarticle}

\usepackage[margin=2.5cm]{geometry}
\usepackage{amssymb}
\usepackage{amsthm}
\usepackage{lineno}
\usepackage{xcolor}
\usepackage{graphicx}
\graphicspath{{plot/}}
\usepackage{subcaption}
\usepackage{tabularx}
\usepackage{array}
\usepackage{amsmath,bm}
\usepackage{multirow}
\usepackage{hyperref}
\usepackage{upgreek}
\usepackage{booktabs}
\usepackage{placeins}
\usepackage{algorithm}
\usepackage{algpseudocode}
\usepackage{float}
\usepackage{setspace}
\usepackage{microtype}
\usepackage{fancyhdr}
\usepackage{titlesec}

\begin{document}


\begin{frontmatter}

\title{Basin-Scale Modeling of Multiple Storage Projects in the Broom Creek Formation, North Dakota, USA: Unified Model and Uncertainty Quantification}

\author[1]{Keisuke Yamamura}
\ead{keiyama@stanford.edu}

\author[1]{Louis J.~Durlofsky}
\ead{lou@stanford.edu}

\address[1]{Department of Energy Science and Engineering, Stanford University, Stanford, CA 94305, USA}

\begin{abstract}

Quantifying the interactions between neighboring projects will be essential as carbon storage operations expand in scale. However, many existing basin-scale studies are deterministic or rely on hypothetical scenarios. In addition, simulations that do not fully account for neighboring projects may suffice for initial permitting. In this study, we develop a unified basin-scale model for the Broom Creek Formation in North Dakota. The model includes a total of five (existing and planned) projects, involving 18~injection wells with a maximum potential injection rate of 30~MTPA. The unified model, constructed by integrating publicly available geological and project data, contains $44\times10^6$ cells and requires over two days to run. To enable faster simulations, we present a multilevel treatment that combines four-level nested local grid refinement, constructed through the solution of a constraint-satisfaction problem, with optimized power averaging for permeability in coarse regions. After demonstrating the accuracy of the coarse model, we use it to assess uncertainty in key quantities of interest (QoIs), including total CO$_2$ injected and plume area after 50~years, across a wide range of geological realizations and model parameters. In addition, we develop a deep learning surrogate model, which is used in global sensitivity analyses to identify dominant contributors to QoI variance. Comparisons between the unified model and standalone project simulations are presented to quantify inter-project interference. This can be substantial in some cases, e.g., for one project, the median total CO$_2$ injected decreases from 271~MT (standalone) to 192~MT (unified model).

\end{abstract}

\begin{keyword}
Geological carbon storage; Basin-scale modeling; Pressure interference; Multilevel upscaling; Local grid refinement; Uncertainty quantification; Global sensitivity analysis
\end{keyword}

\end{frontmatter}


\section{Introduction}\label{sec:Introduction}
The large-scale deployment of carbon capture and storage (CCS) is expected to be an important component of global decarbonization efforts. As CCS operations expand, it is increasingly likely that multiple projects will target the same basin (geological formation). For example, in the US, several CCS projects have been approved to inject into the Broom Creek Formation in North Dakota~\citep{BURTONKELLY2025104456}. Similar situations have been reported in other regions, including Alberta (Basal Cambrian Sandstone), offshore UK (Bunter Formation), offshore Norway (Hordar Platform), and Denmark (Gassum Formation)~\citep{Gibson-Poole2025}. In such settings, storage resources, particularly ``pressure space,'' are shared among multiple projects~\citep{BUMP2023103967,BUMP2024104174,OUGIERSIMONIN2026104610}. Pressure interference between projects can impact project economics, injection performance, and operational planning. Despite these inevitable challenges, existing regulatory frameworks often follow a “first come, first served” approach that does not specify formal rules for treating inter-project interactions. This may lead to conflicts and the inefficient utilization of storage resources~\citep{Thibeau2025}. An essential first step in addressing these issues is the construction of reliable basin-scale models capable of quantifying inter-project interactions and associated uncertainties.

In this study, we develop open-source, unified, basin-scale numerical simulation models for eight interacting storage projects in the Broom Creek Formation in North Dakota. The reference (finest scale) model contains grid cells of approximate dimensions 100~m~$\times$ 100~m~$\times$ 2~m and covers a storage formation region spanning 110~km~$\times$ 80~km~$\times$ 81~m. This model includes a total of $44\times10^6$~grid cells and is thus expensive to run. For this reason, we develop a specialized coarsening procedure that includes nested local grid refinement (LGR) regions to provide higher resolution in key portions of the domain, along with power averaging of permeability in coarsened regions.
Upscaling errors for the main quantities of interest (QoIs) demonstrate the high level of accuracy that can be achieved with this upscaling strategy.
The resulting model setup is used to assess inter-project pressure interference and to perform multifidelity uncertainty quantification. The framework is also applicable for optimization, which will be considered in a subsequent study.

There have been a number of simulation-based studies that have assessed inter-project pressure interference. While some investigators have relied on hypothetical geological settings and injection scenarios to investigate generalized pressure-interference behavior~\citep{GHADERI20093113, WIJAYA2024205422}, others have employed basin-scale geological models based on real formations and storage complexes. For example,~\citet{BIRKHOLZER2009745} examined regional-scale pressurization in the Illinois Basin. Similar studies have been conducted for the Troll aquifer in the North Sea \citep{PetterssonKrumscheidGasda2024_1000174144,Tveit2024,Mykkeltvedt202522154} and the Bjarmeland Formation located in the Barents Sea \citep{Sandve2025MultiScaleSS}. These studies have demonstrated the importance of understanding basin-scale pressure propagation and its implications for regional operations. Although many studies have considered simplified or hypothetical injection strategies, a few have incorporated more realistic setups. \citet{Rowbotham202510447}, for example, combined actual licensed storage locations with hypothetical additional sites in regional-scale simulations of the UK North Sea. They showed that neighboring storage operations can reduce injectivity. Similarly, \citet{Gibson-Poole2025} demonstrated that cumulative injection from multiple projects can increase regional pressure, thereby reducing available pressure buildup and potentially requiring injection-rate curtailment. Although these studies have improved the realism of basin-scale CCS models, most existing work is still largely deterministic. More systematic uncertainty quantification and sensitivity analyses are required.

Eight Storage Facility Permits have already been granted for the Broom Creek Formation~\citep{BURTONKELLY2025104456}. Although not all these projects are currently operating, a substantial amount of modeling and simulation work has been conducted. However, individual developers have often performed their modeling independently, without accounting for interactions between projects~\citep{NDIC2021EastTundra, NDIC2022DGC, NDIC2023BF, NDIC2024SCS1}. \citet{osti_1606011}, by contrast, incorporated two potential storage sites into a single simulation model, though pressure interference between projects was not discussed. This is probably because the lateral boundaries were open, which corresponds to an infinite-acting aquifer. However, as noted by~\citet{OUGIERSIMONIN2026104610}, the infinite-acting approximation may break down with increasing numbers of projects and scale of injection. In such cases, the identification of formation extent and boundaries is necessary to quantify interactions. More recently, simulation studies were conducted to evaluate plume stabilization in this region~\citep{BURTONKELLY2025104456}. This study involved a single site model, and the effect of pressure interference was not evaluated.
In contrast, \citet{CHELLAL2027139835} investigated pressure interference effects in the same Williston Basin; however, their study focused on the Deadwood Formation, for which only one CCS project has been approved to date~\citep{NDIC2021EastTundraDeadwood}. Therefore, although a large number of CCS projects have been approved in the Broom Creek Formation and site-by-site modeling work has been performed, we are not aware of any studies focusing on inter-project pressure interference for this important storage region.

An obvious challenge with basin-scale modeling is that large simulation models are required. Even with modest levels of grid resolution, the number of cells can easily be many tens of millions, which leads to excessive computational demands, especially for uncertainty quantification, data assimilation, or optimization.
Several studies have therefore explored upscaling strategies to reduce computational cost while maintaining accuracy in critical regions. In the context of well placement optimization, for example, \citet{Zou2023Dissertation} employed a low-fidelity upscaled model in which near-well regions were resolved using LGR. Similarly, \citet{JIANG2024104124} introduced LGR regions around injectors for data assimilation applications. These studies involved single-project systems, and only a single LGR level was considered. 
Using a giant aquifer model inspired by the Santos Basin in Brazil,~\citet{KREUTZERDTMANN2026104660} evaluated several multiscale strategies, including effective-value upscaling, LGR, and an Algebraic Dynamic Multilevel (ADM) treatment. 
\citet{TVEIT2025213733} and \citet{Sandve2025MultiScaleSS} proposed a hierarchical two-stage workflow in which regional-scale simulations provide dynamic pressure or flux boundary conditions for detailed site-scale models. \citet{landamarban2025} applied transmissibility upscaling. These studies demonstrate the importance of developing efficient modeling strategies that can accurately capture both regional pressure interference and complex site-scale heterogeneity for basin-scale CCS simulations.

In this work, we develop a unified basin-scale model for the entire Broom Creek Formation in North Dakota by integrating publicly available geological information with data from existing and approved CO$_2$ storage projects. The resulting model provides a consistent framework for evaluating the performance of individual projects and basin-scale pressure interference under geological uncertainty. To address the computational challenges associated with simulations at this scale, we introduce a new multilevel treatment that combines nested LGRs, defined through the solution of a formal constraint-satisfaction problem, with power averaging for permeability upscaling in coarsened regions. The power-average exponents are determined via an optimization procedure. The resulting coarse model is shown to provide high accuracy -- within a few percent for the QoIs considered -- relative to the reference fine model solution. We then apply the resulting framework to quantify the effects of uncertainty in the geological realization and in associated model parameters, which include overburden and underburden permeability, pore volume of the surrounding regions, permeability anisotropy ratio, and relative permeability parameters. This sensitivity analysis requires 4400 simulation runs along with 29~fine-scale validation runs for selected models. Finally, we quantify the impact of inter-project interference by comparing predictions from the unified basin-scale model with those from standalone simulations for individual storage projects.

This paper proceeds as follows. An overview of the existing and planned CO$_2$ storage projects, and the construction of the geological model for the Broom Creek Formation, are described in Section~\ref{sec:geological model}. The numerical simulation methodology, including the setup for the fine-scale model, and the constraint satisfaction problem for LGR definition and power averaging for permeability for the coarse-scale model, are described in Section~\ref{sec:methodology}. In that section we also evaluate the resulting coarse model against the fine-scale reference solution. The formulation and results from the global sensitivity analysis, along with an evaluation of the robustness of the upscaling workflow for selected models, are presented in Section~\ref{sec:SensitivityAnalysis}. In Section~\ref{sec:single-project}, we quantify the effects of inter-project pressure interference by comparing unified basin-scale simulations with standalone simulations for each storage project. Finally, in Section~\ref{sec:conclusion}, we summarize our main findings and suggest directions for future research. Additional details on many aspects of this study are provided in online Supplementary Material.

In this study we use actual data in our modeling. It is, however, important to highlight how our setup differs from actual operations in North Dakota. First, individual operators will, of course, have proprietary data (e.g., 2D and/or 3D seismic survey data, relative permeability measurements) that are not available to us, so our simulation models and results may differ from theirs. Second, the time frames we simulate are much longer than those prescribed in the initial permits issued to Broom Creek Formation operators. The interference effects we observe would be expected to be weaker in shorter-duration operations. Finally, we need to make assumptions regarding the start times of approved but not-yet-operating projects. These assumed start times may differ from the actual start times, and this will impact the applicability of our results. In total, though we believe our models and results to be realistic under the provisos noted above, we are not suggesting that these models ``preempt'' those submitted to North Dakota regulatory authorities by individual operators. Rather, we believe our models should be considered in conjunction with the models from individual operators.

\section{Basin-scale Geological Model}
\label{sec:geological model}
In this section, we first provide an overview of the planned and operating storage projects. We then describe the geological setting of the Broom Creek Formation. Finally, the detailed geological model is presented. In this study, we treat the structural framework as deterministic, but the permeability realizations and many of the key model parameters are treated as uncertain. The base-case model employs median (or commonly used) parameters. In the uncertainty assessment, presented in Section~\ref{sec:SensitivityAnalysis}, we consider multiple realizations and sample from ranges for several of the key parameters.

\subsection{Storage Projects and Geological Setting}
\label{sec:geol_setting}
Eight storage facility permit (SFP) applications targeting the Broom Creek Formation in North Dakota have been approved, as summarized in Table~\ref{table:project}~\citep{nddmr2026classvi}.
Since the DCC East Center (DCCE) and DCC West Center (DCCW) projects are operated by the same entity, they are collectively referred to as DCC in this study. Similarly, Summit Carbon Storage Projects \#1, 2 and 3 (SCS1--3) are grouped and referred to as SCS. Consequently, the eight permitted SFP applications are represented by five projects in the analyses presented in this paper.

\begin{table}[h]
    \centering
    \caption{Permitted Broom Creek Formation CO$_2$ storage projects in North Dakota, USA (modified from~\citet{BURTONKELLY2025104456}). Storage facility permit (SFP) applications are referenced by Case ID on the North Dakota Department of Mineral Resources (DMR) website~\citep{nddmr2026classvi}. DCCE project is also referred to as Minkota Center MRYS Broom Creek Storage Facility\#1 on the DMR website.}
    \label{table:project}
    \begin{tabular}{p{4cm} p{4cm} c c c c }
    \hline
    Operator & Project & Project ID & Label & DMR Case ID \\ 
    \hline
    Gevo, Inc. & Red Trail Richardton Ethanol Broom Creek Storage Facility \#1 & RTE & RTE & 28848 \\ 
    Blue Flint Sequester Company, LLC & Blue Flint Underwood Broom Creek Storage Facility \#1 & BF & BF & 29888 \\ 
    Dakota Gasification Company & DGC Beulah Broom Creek Storage Facility \#1 & DGC & DGC & 29450\\ 
    DCC East Project LLC & DCC East Center Broom Creek Storage Facility \#1  & DCCE & DCC & 29029 \\ 
    DCC West Project LLC & DCC West Center Broom Creek Storage \#1 & DCCW & & 30122 \\ 
    Summit Carbon Storage \#1, LLC & TB Leingang Broom Creek Storage Facility & SCS1 & SCS & 30869 \\
    Summit Carbon Storage \#2, LLC & BK Fischer Broom Creek Storage Facility & SCS2 & & 30873 \\
    Summit Carbon Storage \#3, LLC & KJ Hintz Broom Creek Storage Facility & SCS3 & & 30877 \\ 
    \hline
    \end{tabular}
\end{table}

The Williston Basin is a large intracratonic sedimentary basin and has been identified as a favorable setting for long-term geological storage of CO$_2$~\citep{SORENSEN20092833}. The Broom Creek Formation, a Permian-age unit within the upper Minnelusa Group, is the target injection interval. Repeated episodes of marine transgression and regression resulted in substantial lateral variability and vertical heterogeneity in lithology and reservoir properties~\citep{Rygh1990}. The formation contains aeolian sandstone and associated dolostone and anhydrite deposits that act as baffles~\citep{BURTONKELLY2025104456}. The overlying Opeche Formation acts as the primary sealing unit, and the underlying Amsden Formation provides the lower seal~\citep{PECK201947}. More detailed descriptions of the lithostratigraphy, depositional history, and reservoir characteristics of the Broom Creek Formation and adjacent units can be found in~\citet{Rygh1990}, \citet{SORENSEN20092833}, \citet{osti_1606011}, and \citet{POTHANA2026214229}.

\subsection{Structural Framework}
\label{sec:structural_framework}
The areal extent of the Broom Creek Formation was originally defined by \citet{Rygh1990}. This definition has been widely used in CCS studies~\citep{osti_1606011}, though it was recently refined in the context of the BF project~\citep{NDIC2023BF}, located at the northeastern margin, based on new well control data. This areal boundary captures the pinch-out characteristics of the Broom Creek Formation toward the north and east. In this study, the modified areal extent developed in the BF project is used to define the northern and eastern boundaries of the structural framework. The western and southern boundaries are defined by the state boundaries with Montana and South Dakota, respectively. Our detailed geomodel includes only the North Dakota portion of the Broom Creek Formation, although the formation is expected to extend west into Montana and south into South Dakota. The formation pore volumes within Montana and South Dakota are treated as uncertain parameters in the sensitivity analysis described in Section~\ref{sec:SensitivityAnalysis}.

For horizon construction, formation top data for the Broom Creek and Amsden Formations were collected from three data sources. First, a compiled formation top dataset, derived from well log interpretations of existing wells by the North Dakota Industrial Commission (NDIC), is used. Second, well reports submitted from operators to the Department of Mineral Resources, North Dakota Industrial Commission, often include formation-top picks interpreted by the operator. This information, which is used where the NDIC formation-top dataset is sparse, is especially important for the Amsden Formation, as the NDIC dataset does not include Amsden Formation top information. Additional formation-top picks are interpreted from well-log data acquired from 17~wells drilled for CCS-related projects, which are listed in Table~S1 in Supplementary Material. For these wells, the raw well-log data were independently reviewed, and the interpretations cross-validated against the operator-reported picks to ensure consistency. In total, the dataset consists of 2970 Broom Creek and 221 Amsden Formation-top picks. The corresponding well locations are shown in Figure~S1 in Supplementary Material.

\begin{figure}[h!]
    \centering
    \includegraphics[width=0.5\textwidth]{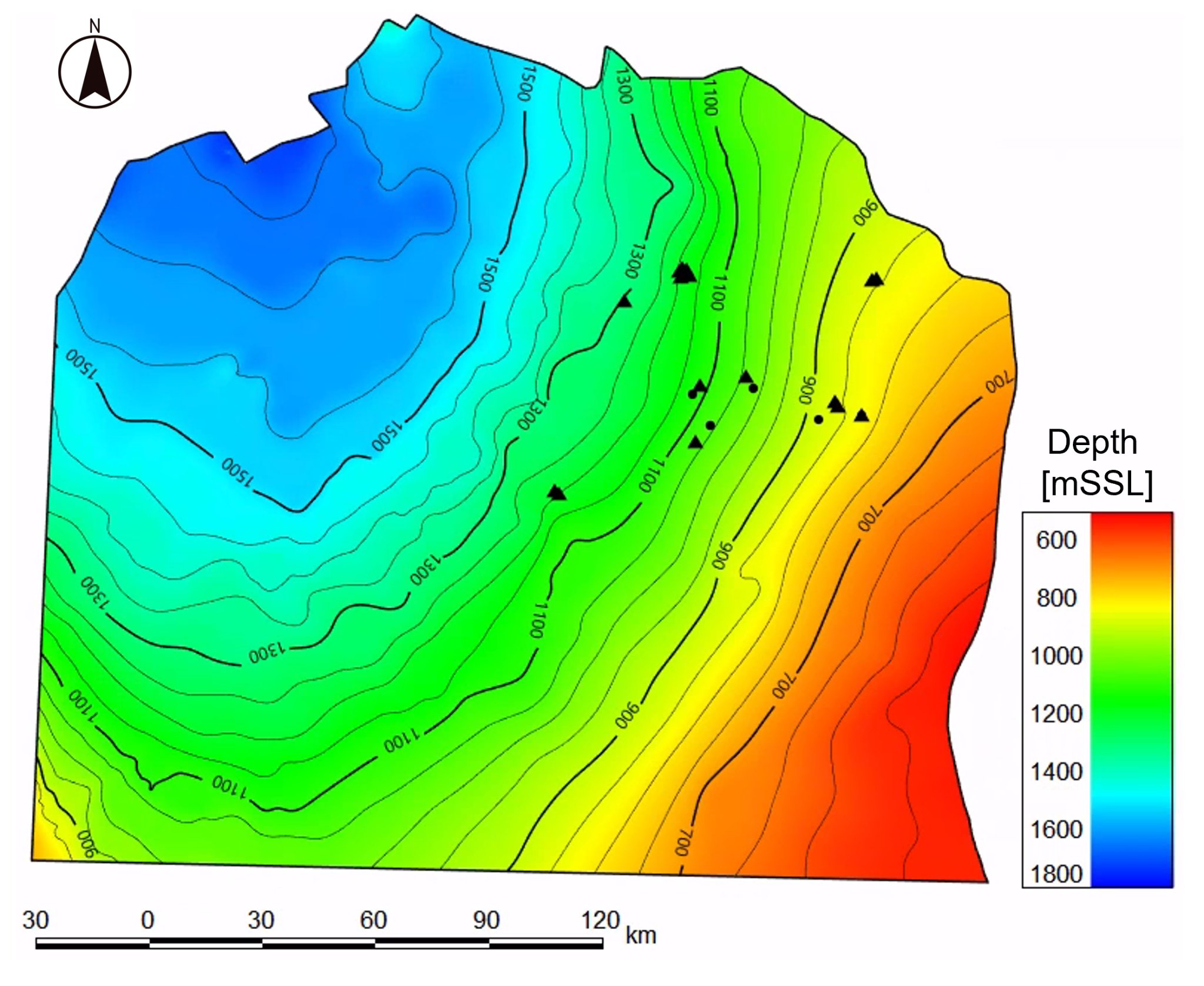}
    \caption{Horizon of the top of the Broom Creek Formation (mSSL). Black symbols indicate CCS-related wells, with triangles and circles representing drilled and planned wells, respectively.}
    \label{fig:BC_top_contours}
\end{figure}

The horizon of the top of the Broom Creek Formation, shown in Figure~\ref{fig:BC_top_contours}, is constructed using the minimum curvature method with the well formation-top data described above. The top of the Amsden Formation, shown in Figure~S2 in Supplementary Material, is constructed using the formation-top data with the areal extent data as an additional constraint. The minimum-curvature interpolation method is applied while honoring both datasets to define the horizon geometry and reproduce the observed pinch-out behavior within the study area. 

\begin{figure}[h!]
    \centering
    \includegraphics[width=0.5\textwidth]{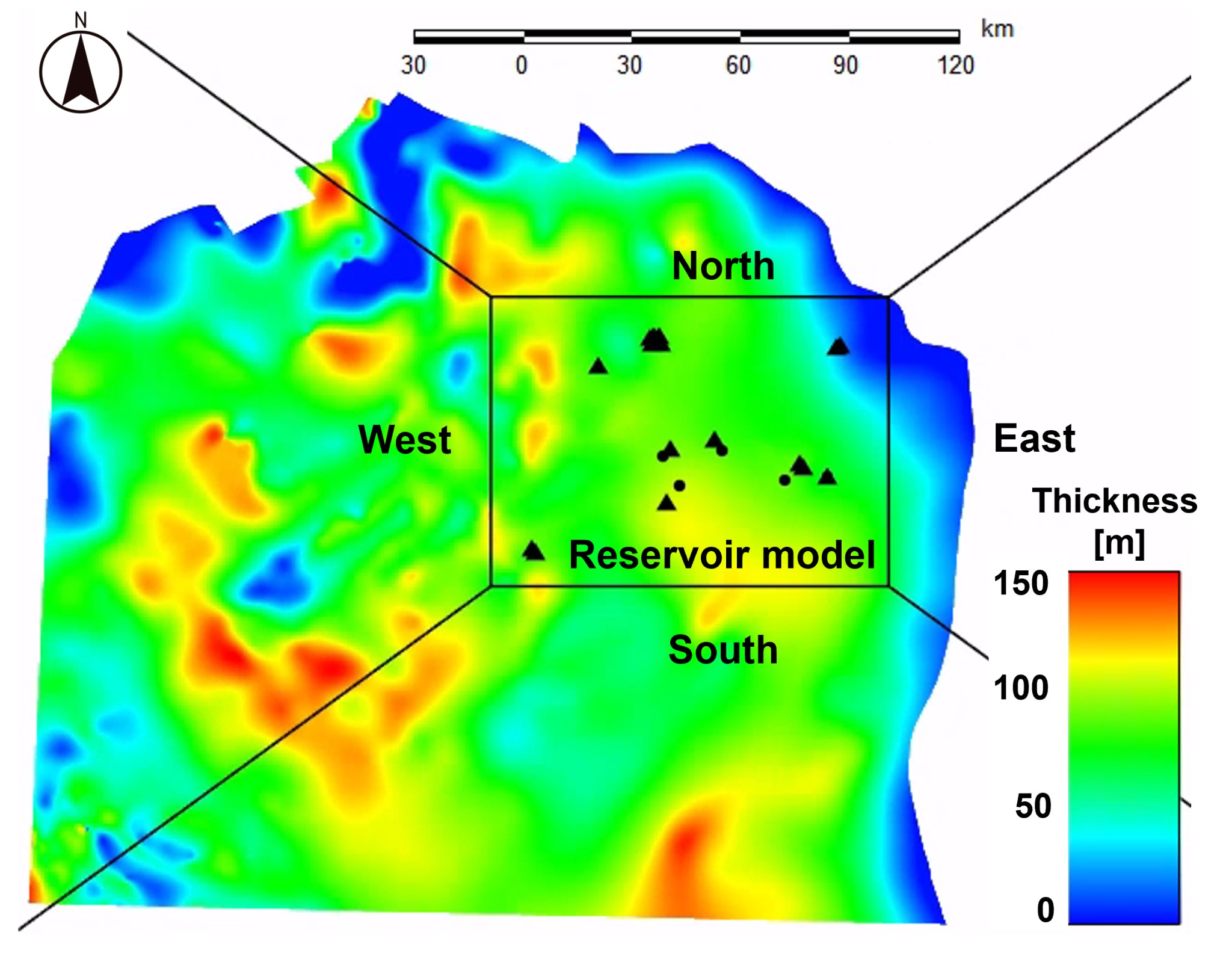}
    \caption{Thickness of the Broom Creek Formation. Black symbols indicate CCS-related wells, with triangles and circles representing drilled and planned wells, respectively.}
    \label{fig:BC_thickness}
\end{figure}

A map depicting the reservoir thickness is shown in Figure~\ref{fig:BC_thickness}. A minimum reservoir thickness of 3.05~m (10~ft) is enforced. The mean, median, and maximum thickness are 73.8~m, 76.6~m, and 152.1~m, respectively. The formation exhibits pinch-out behavior toward the north and east. Both the overlying and underlying formations are specified to be 1000~m thick. Although multiple stratigraphic units exist above and below the Broom Creek Formation, they are not modeled separately but are instead grouped into single overlying and underlying regions.

\subsection{Model Domain and Lithology}\label{sec:model_domain}
The geomodel spans 262~km $\times$ 230~km and captures the entire extent of the Broom Creek Formation in North Dakota (Figure~\ref{fig:BC_top_contours}). The total bulk volume is $3.67\times10^{12}$~m$^{3}$ (Table~\ref{table:BulkVolume}). Because geostatistical modeling and dynamic numerical simulation over the entire domain are computationally expensive, a smaller model, indicated by the black rectangle in Figure~\ref{fig:BC_thickness}, is extracted. This model includes all approved CCS projects within the study area and is hereafter referred to as the reservoir model, which is used for the numerical simulations in this study. The bulk volumes of each region are given in Table~\ref{table:BulkVolume}. 

\begin{table}[h]
\centering
\caption{Bulk volumes of the Broom Creek Formation and fractional contributions of each region. Regions are identified in Figure~\ref{fig:BC_thickness}.}
\label{table:BulkVolume}
\begin{tabular}{lcc}
\hline
Region & Bulk volume [m$^3$] & Fraction [-] \\
\hline
Reservoir model & $7.12 \times 10^{11}$ & 0.194 \\
North            & $2.90 \times 10^{11}$ & 0.079 \\
South            & $1.36 \times 10^{12}$ & 0.370 \\
East             & $4.14 \times 10^{10}$ & 0.011 \\
West             & $1.27 \times 10^{12}$ & 0.346 \\
\midrule
Total            & $3.67 \times 10^{12}$ & 1.000 \\
\hline
\end{tabular}
\end{table}

Previous investigators have defined four lithofacies in the Broom Creek Formation -- sandstone, dolomitic sandstone, dolostone, and anhydrite~\citep{osti_1606011, NDIC2024SCS1}. In this study, these four lithofacies are consolidated into two. Specifically, sandstone and dolomitic sandstone are classified as reservoir rock, and dolostone and anhydrite as nonreservoir rock. This treatment is similar to that in the RTE project, which uses only two sandstone and dolostone lithofacies~\citep{NDIC2021RTE}. In addition, a single lithofacies is assigned to both the overlying and underlying formations.

Facies logs for 17~wells are constructed using publicly available well-log data. Information on the sources for these data is provided in Table~S1 in Supplementary Material. For 10~of these wells, facies logs appear in SFP reports, and these interpretations are used directly. For the remaining seven wells, which were drilled after project approval and for which no operator interpretation reports are available, facies classification is performed manually based on well-log data, with particular emphasis on resistivity, sonic, bulk density, and gamma ray logs. Examples of facies logs are shown in Figure~S3 in Supplementary Material. The net-to-gross ratio (N/G), defined as the proportion of reservoir rock within the Broom Creek Formation, is calculated from the facies logs as listed in Table~S1 in Supplementary Material. The average N/G value is 0.65, and this value is used in the geostatistical simulations.

Several variogram models have been used for facies and/or porosity geostatistical simulation in previous Broom Creek Formation studies~\citep{NDIC2021EastTundra, NDIC2022DGC,NDIC2021RTE}. The variogram parameters from the most recently approved SCS project~\citep{NDIC2024SCS1} are adopted in this work, with those for sandstone assigned to the reservoir rock and those for dolostone assigned to the nonreservoir rock. 
For the reservoir rock, the horizontal variogram ranges along the main direction and the direction orthogonal to the main direction are 1520~m (5000~ft) and 1370~m (4500~ft), respectively. The vertical variogram range is 9.14~m (30~ft), and the azimuth of the main anisotropy direction is 45$^\circ$. For the nonreservoir rock, the horizontal variogram ranges along the main and orthogonal directions are 2500~m (8200~ft) and 1830~m (6000~ft), respectively. The vertical variogram range is 10.7~m (35~ft), and the azimuth of the main anisotropy direction is 90$^\circ$.

\begin{figure}[h!]
    \centering

    \begin{subfigure}[b]{0.4\textwidth}
        \centering
        \includegraphics[width=\textwidth]{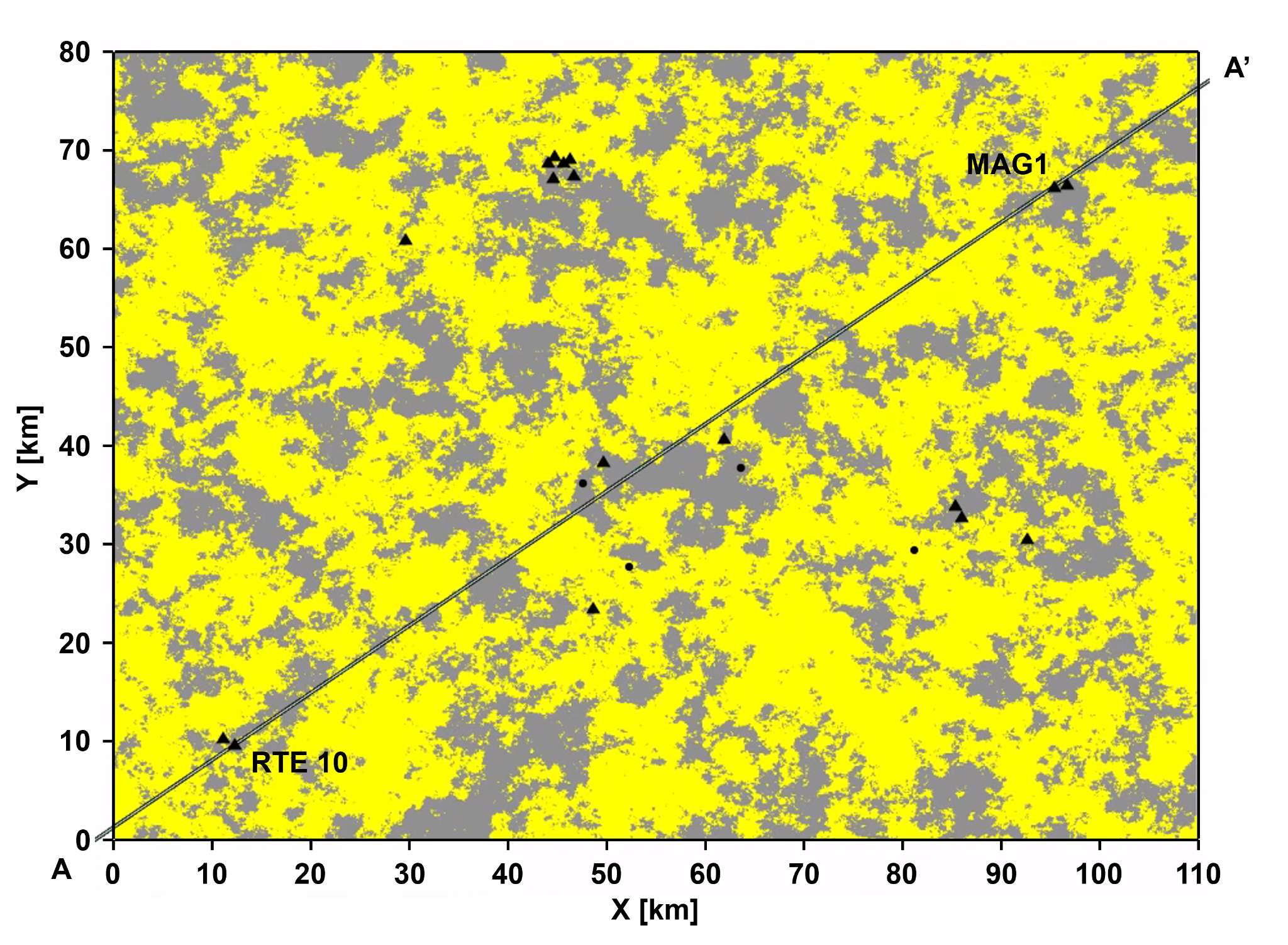}
        \caption{Top-layer facies distribution.}
    \end{subfigure}
    \hfill
    \begin{subfigure}[b]{0.58\textwidth}
        \centering
        \includegraphics[width=\textwidth]{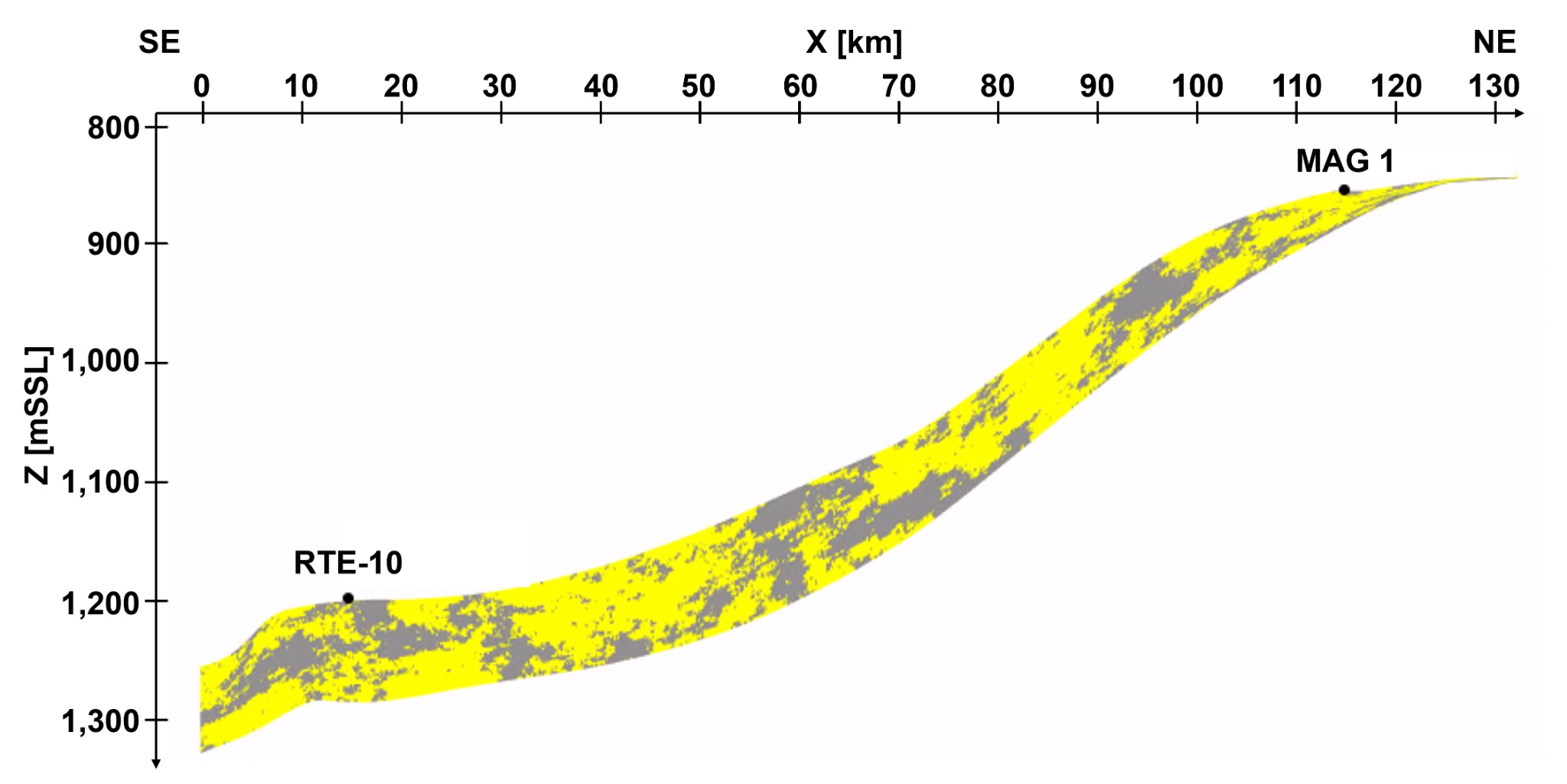}
        \caption{A--A$^\prime$ cross section. Vertical axis exaggerated by a factor of 119.}
    \end{subfigure}

    \caption{
    Facies distribution of the Broom Creek Formation. Reservoir and nonreservoir rocks are shown in yellow and gray, respectively. In the map view (left), triangles indicate drilled CCS-related wells used for well conditioning and circles represent planned wells. The A--A$^\prime$ cross section (right), at the location indicated in (a), intersects well RTE-10 in the RTE project and well MAG~1 in the BF project.}
    \label{fig:FaciesDistribution}
\end{figure}

The facies distributions are generated using Sequential Indicator Simulation (SIS), conditioned on facies logs from the 17~wells listed in Table~S1 in Supplementary Material. Geostatistical simulation inputs, including facies logs, N/G, and variogram parameters, were provided earlier. An exponential variogram model is applied, and the Geology Designer module in tNavigator is used to perform the geostatistical simulations. Multiple realizations will be generated for the sensitivity analysis presented in Section~\ref{sec:SensitivityAnalysis}.
The facies distribution in the Broom Creek Formation (in map and cross-sectional views) for the base case is shown in Figure~\ref{fig:FaciesDistribution}. 

\subsection{Porosity and Permeability Specifications}\label{sec:PorosityPermeability}
\begin{figure}[h!]
    \centering
    \includegraphics[width=0.5\textwidth]{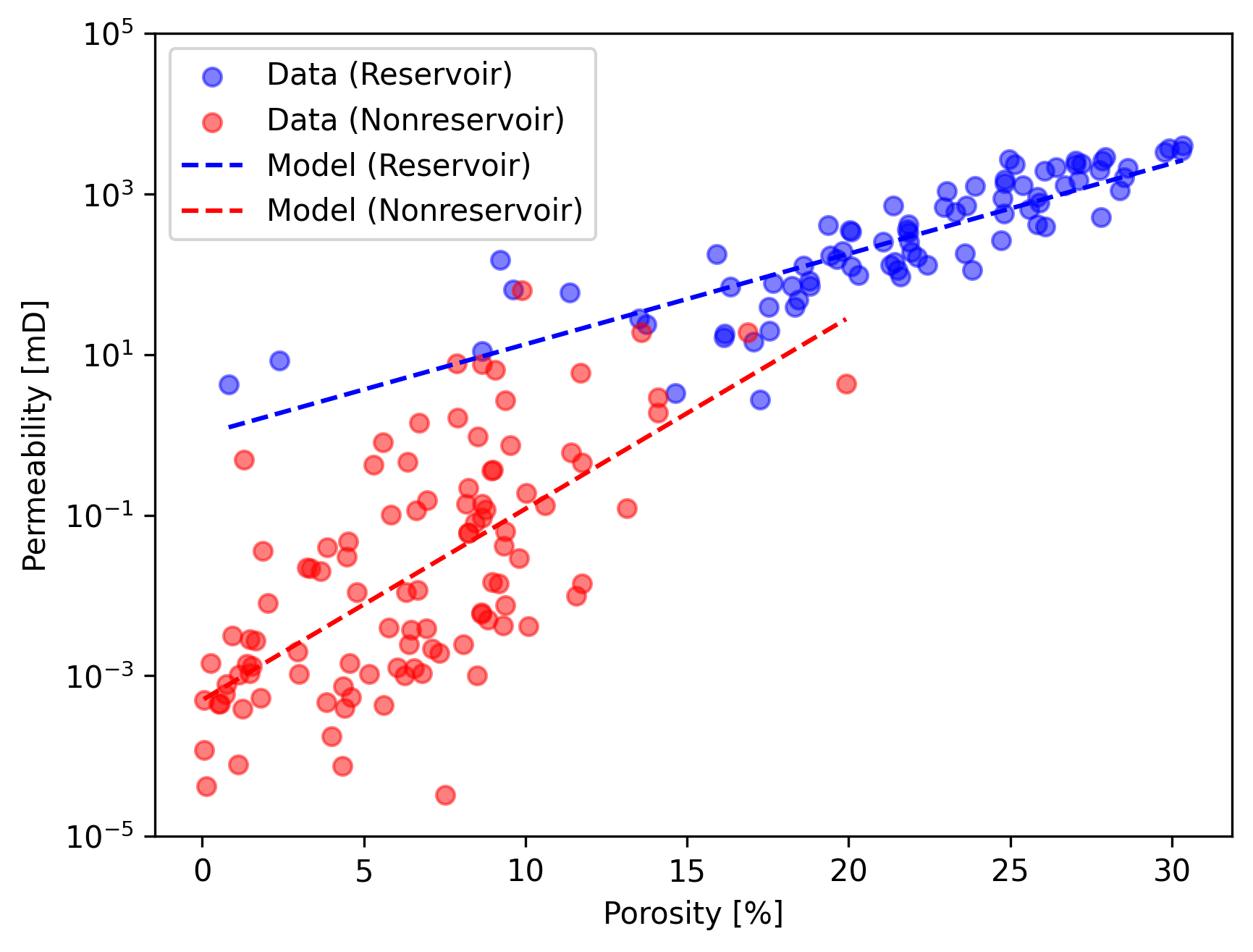}
    \caption{\centering Porosity and Klinkenberg permeability of the Broom Creek Formation from RCA data measured between 12.4~MPa (1800~psi) and 16.5~MPa (2400~psi).}
    \label{fig:RCA-Phi-K-Res}
\end{figure}

The statistical characteristics of porosity and permeability, as well as their inter-relationship, are derived using publicly available routine core analysis (RCA) data from the North Dakota Oil and Gas Division dataset. Figure~\ref{fig:RCA-Phi-K-Res} shows the porosity and Klinkenberg permeability of reservoir and nonreservoir rock from the Broom Creek Formation. Here, rock types of each sample are assigned based on facies logs, as described earlier. These data support our two-facies treatment of the Broom Creek Formation (as opposed to considering four distinct facies). Tables~\ref{table:StatisticsPorosity} and \ref{table:StatisticsPermeability} summarize some of the porosity and permeability statistics, as determined from the RCA data. 

\begin{table}[h!]
    \centering
    \caption{Summary of porosity derived from RCA data. SD denotes standard deviation.}
    \begin{tabular}{ccccc}
    \toprule
        Rock type & Mean [-] & SD [-] & Min [-] & Max [-]\\
    \midrule
        Reservoir & 0.215 & 0.059 & 0.082 & 0.303 \\
        Nonreservoir & 0.065 & 0.067 & 0.0005 & 0.199 \\
        Overburden & 0.079 & 0.035 & 0.002 & 0.154 \\
        Underburden & 0.049 & 0.032 & 0.001 & 0.122 \\
    \bottomrule
    \end{tabular}
    \label{table:StatisticsPorosity}
\end{table}

\begin{table}[h!]
    \centering
    \caption{Summary of permeability derived from RCA data.}
    \begin{tabular}{ccccccc}
    \toprule
        Rock type & Median [mD] & Min [mD] & Max [mD] \\
    \midrule
        Reservoir & 334 & 2.75 & 3990 \\
        Nonreservoir & $1.11\times10^{-2}$ & $3.26\times10^{-5}$ & 62.9 \\
        Overburden & $7.78\times10^{-3}$ & $4.39\times10^{-5}$ & 3.09 \\
        Underburden & $1.06\times10^{-3}$ & $8.00\times10^{-5}$ & 4.32 \\
    \bottomrule
    \end{tabular}
    \label{table:StatisticsPermeability}
\end{table}

The porosity and permeability relationship is established for each facies by fitting exponential regression curves to the RCA data:

\begin{equation}
    k = k_0 e^{a\phi},
    \label{eq:k-phi-noise}
\end{equation}
where \(k\) is the permeability in mD, \(\phi\) is porosity, and $k_0$ and $a$ are the model parameters. The fitted parameters for the reservoir rock are $k_0 = 1.0$ and $a = 26.0$, whereas those for the nonreservoir rock are $k_0 = 4.97\times10^{-4}$ and $a = 54.9$. There is a clear trend in the measured data for the reservoir rock in the Broom Creek Formation, which is captured by the regression line shown in Figure~\ref{fig:RCA-Phi-K-Res}. There is more scatter in the regression for the nonreservoir rock, though a trend is evident.

The permeability values discussed above refer to the horizontal ($x$ and $y$) components of permeability. Permeability anisotropy is quantified through the ratio $k_v/k_h$, where $k_h$ and $k_v$ are the horizontal and vertical permeabilities. Only six RCA measurements are available to derive $k_v/k_h$. These data, shown in Figure~S4 in Supplementary Material, involve air permeability measured at 5.52~MPa (800~psi). Based on these data, a median value of 0.045 is applied uniformly to all rock types in the base case. Because of the limited amount of available data, the anisotropy ratio is considered as an uncertain parameter in the sensitivity analysis.

\begin{figure}[h!]
    \centering
    \includegraphics[width=0.7\textwidth]{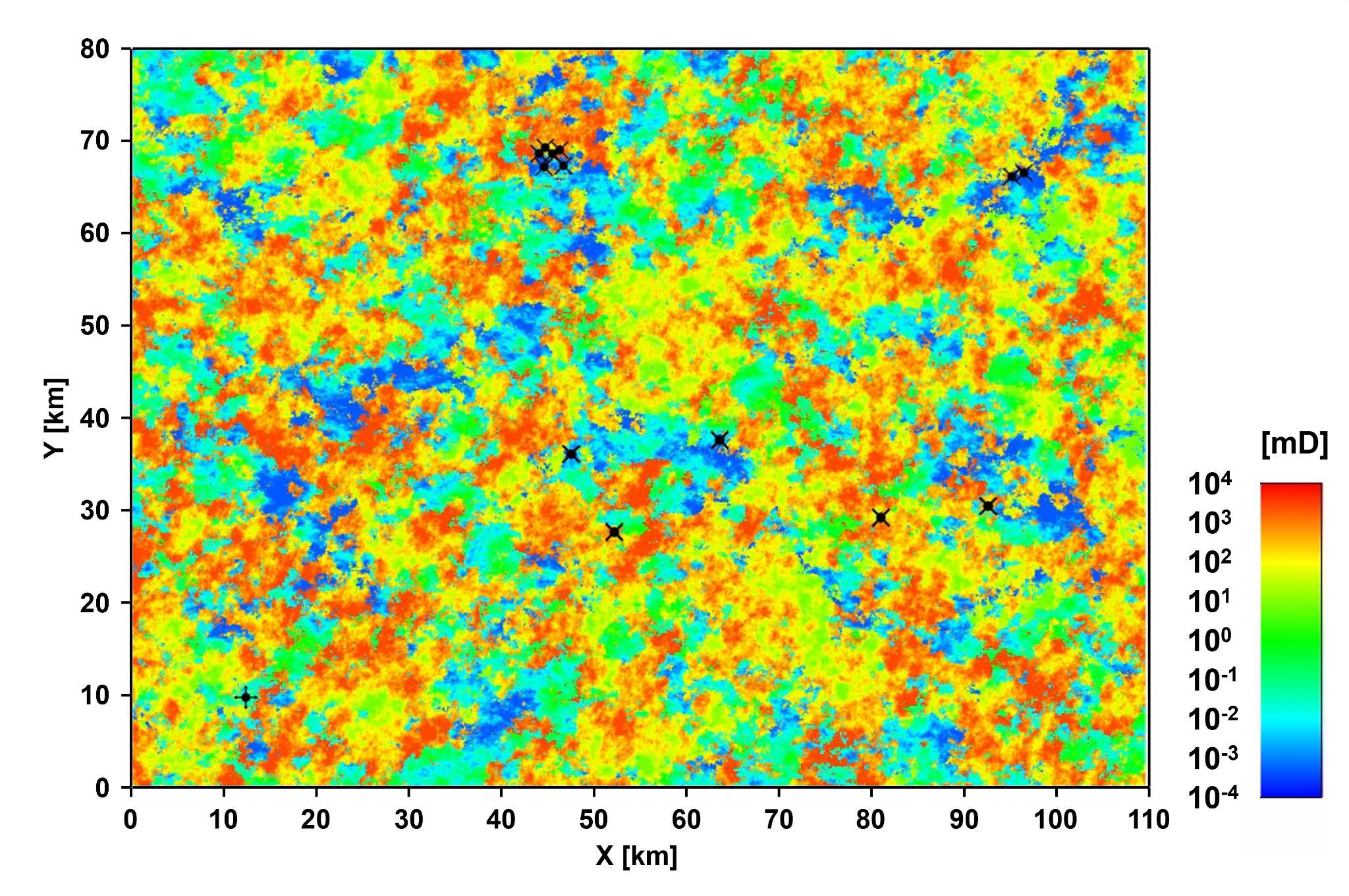}
    \caption{Permeability distribution of the top layer of the Broom Creek Formation. }
    \label{fig:K-Distribution}
\end{figure}

The porosity distribution for the Broom Creek Formation is generated using Sequential Gaussian Simulation (SGS). The same variogram model used for the facies distribution, described in Section~\ref{sec:model_domain}, is applied. The mean and standard deviation derived from the RCA data, listed in Table~\ref{table:StatisticsPorosity}, are used as input parameters. The resulting field for the top layer of the Broom Creek Formation is shown in Figure~S5 in Supplementary Material.
Grid-block permeability values are then generated using Eq.~\ref{eq:k-phi-noise}. 
Figure~\ref{fig:K-Distribution} shows the permeability distribution for the top layer of the Broom Creek Formation.
The geostatistically simulated porosity and permeability values for each facies are clipped to the minimum and maximum values observed in the RCA, which are given in Tables~\ref{table:StatisticsPorosity} and~\ref{table:StatisticsPermeability}.

For the overlying and underlying formations, homogeneous porosity and permeability values, corresponding to the mean (porosity) and median (permeability) values listed in Tables~\ref{table:StatisticsPorosity} and~\ref{table:StatisticsPermeability}, are specified. 
Since the permeabilities of the overlying and underlying formations span a wide range, as shown in Table~\ref{table:StatisticsPermeability}, they will be treated as uncertain parameters in the sensitivity analysis in Section~\ref{sec:SensitivityAnalysis}. A median $k_v/k_h$ value of 0.045, derived from the RCA data across all rock types, is assigned to the overlying and underlying formations.

\section{Numerical Methodology}\label{sec:methodology}

In this section we present the setup for the numerical simulations. We then describe the two methods used in the construction of coarse-scale models -- nested local grid refinement (LGR) to define the regions of high, low, and intermediate resolution, and power averaging to compute permeability in coarsened regions. Finally, the performance of the resulting coarse model is evaluated through comparison against the fine model. Most of the parameters and settings provided in this section are for the base case. We will indicate the parameters to be varied in the sensitivity analysis.

\subsection{Numerical Simulation Setup}\label{sec:SimSetting}
As discussed in Section~\ref{sec:model_domain}, the aquifer model is defined by extracting a subdomain containing the approved projects plus surrounding regions (Figure~\ref{fig:BC_thickness}). The full grid for the fine-scale model contains 1100 $\times$ 800 $\times$ 50~cells, for a total of $44\times10^6$~cells. These cells are of size 100~m~$\times$~100~m in the $x$ and $y$-directions throughout the domain. The cell thickness varies from region to region. The storage aquifer (Broom Creek Formation) is gridded with 40~layers of average thickness of 1.92~m. The overburden and underburden are each of total thickness 1000~m. These are gridded with cells of increasing thickness as we move away from the Broom Creek Formation. The specific cell thicknesses are 32.3~m (in the layers just above and just below the storage aquifer), 64.5~m, 129.0~m, 258.1~m, and 516.1~m.

All flow simulations are performed using tNavigator~\citep{RFD2023}. The simulation models involve two fluid phases (water/brine and gas) and three components (water, CO$_2$, and salt). The initial pressure is 132.4~bar at 536.4~mSSL, and the initial gas saturation is zero. The system is isothermal, at a temperature of 58~$^\circ$C. Salinity and rock compressibility are 69,025~ppm and $7.12 \times 10^{-5}$~bar$^{-1}$. Observed values of reservoir pressure and salinity are provided in Figures~S6(b) and S7 in Supplementary Material.

To preserve the total volume of the original structural framework in North Dakota, pore-volume multipliers are applied to the lateral boundary cells of the extracted reservoir model. The pore-volume multipliers are calculated from the bulk volumes of the corresponding regions listed in Table~\ref{table:BulkVolume}. The structural framework described in Section~\ref{sec:structural_framework} is truncated at the state boundaries with Montana and South Dakota, although the Broom Creek Formation is expected to extend beyond these boundaries. The pore volumes outside the modeled domain may allow for additional pressure dissipation and storage capacity, thereby mitigating pressure interference among storage projects. To account for this uncertainty, the boundary bulk volumes at the southern and western edges are treated as uncertain parameters in the sensitivity analysis in Section~\ref{sec:SensitivityAnalysis}. No-flow boundary conditions are specified on the outer boundaries of the overall model. 

Water and gas relative permeabilities ($k_{rw}$ and $k_{rg}$) are modeled using the Brooks--Corey formulation, and capillary pressure ($P_c$) is modeled using the van Genuchten formulation. The specific expressions are as follows:
\begin{equation}
    k_{rw} = (\hat S_w)^{n_w},
    \label{eq:Krw}
\end{equation}
\begin{equation}
    k_{rg} = k_{rg,\mathrm{max}}(1-\hat S_w)^{n_g},
    \label{eq:Krg}
\end{equation}
\begin{equation}
    \hat S_w = \frac{S_w-S_{wi}}{1-S_{wi}},
    \label{eq:Sw_star}
\end{equation}
\begin{equation}
    P_{c} = P_e\left(\hat S_w^{-\frac{1}{n}}-1\right)^{\frac{1}{m}}.
    \label{eq:Pc}
\end{equation}
Here $S_w$ is the water saturation, $\hat S_w$ is a rescaled water saturation, $P_e$ is capillary entry pressure, and $n_w$, $n_g$, $S_{wi}$, $k_{rg,\mathrm{max}}$, $m$, and $n$ are model parameters. The parameter $n$ is taken to be $n=1-\frac{1}{m}$. Values of the various parameters for the four rock types considered in the base case are summarized in Table~\ref{tab:kr_pc_parameters}. The $n_w$ and $n_g$ parameters for reservoir rock ($n_{w,\mathrm{res}}$ and $n_{g,\mathrm{res}}$) are treated as uncertain parameters in the sensitivity analysis. The base case parameters in Table~\ref{tab:kr_pc_parameters} are derived from the SFP application for the DCCE project~\citep{NDIC2021EastTundra}.

\begin{table}[htbp]
    \centering
    \caption{Relative permeability (Brooks--Corey) and capillary pressure (van Genuchten) parameters for the base case.}
    \label{tab:kr_pc_parameters}
    \begin{tabular}{lccccccc}
    \hline
    \multirow{2}{*}{Rock type} & \multicolumn{4}{c}{Relative permeability} & \multicolumn{3}{c}{Capillary pressure} \\
    \cline{2-5} \cline{6-8}
     & $n_w$ & $n_g$ & $S_{wi}$ & $k_{rg,\mathrm{max}}$ & $P_e$ [Pa] & $m$ & $S_{wi}$ \\
    \hline
    Reservoir      & 5.8 & 2.2  & 0.025 & 0.91 & 3{,}792   & 5.5 & 0.01 \\
    Nonreservoir  & 5.0 & 2.25 & 0.175 & 0.97 & 75{,}686  & 2.1 & 0.08 \\
    Overburden     & 5.0 & 2.25 & 0.175 & 0.97 & 344{,}738 & 2.1 & 0.08 \\
    Underburden    & 5.0 & 2.25 & 0.025 & 0.91 & 11{,}032  & 2.3 & 0.00 \\
    \hline
    \end{tabular}
\end{table}

Our model contains 18 CO$_2$ injection wells. These correspond to wells in projects already operating and to planned wells in projects that have been approved (though some of these wells have not yet been drilled). Detailed project information, including injection rates and injection start years, is summarized in Table~\ref{tab:injection_schedule}. The injection rates range from 0.18~MTPA to 3.24~MTPA based on information from the project applications. The RTE, BF, and DGC projects commenced CO$_2$ injection in 2022, 2023, and 2024, respectively, and the injection start year is set accordingly. The SCS and DCC projects, by contrast, have not yet commenced operation, and their injection start times remain uncertain. For purposes of this study, wells in these projects are assumed to start injection between 2026 and 2030, as indicated in Table~\ref{tab:injection_schedule}. 

\begin{table}
    \centering
    \caption{Target injection rates and injection start years. MAG~2 is a monitoring well. Start times for DCC and SCS wells are assumed for purposes of this study.}
    \label{tab:injection_schedule}
    \begin{tabular}{llrr}
    \toprule
    Project & Well & Rate [MTPA] & Injection start \\
    \midrule
    RTE & RTE-10    & 0.18 & 2022 \\
    \midrule
    BF  & MAG~1     & 0.20 & 2023 \\
    BF  & MAG~2     & -- & --   \\
    \midrule
    DGC & Coteau~1  & 0.36 & 2024 \\
    DGC & Coteau~2  & 0.36 & 2024 \\
    DGC & Coteau~3  & 0.36 & 2024 \\
    DGC & Coteau~4  & 0.36 & 2024 \\
    DGC & Coteau~5  & 0.36 & 2024 \\
    DGC & Coteau~6  & 0.36 & 2024 \\
    \midrule
    DCC & Liberty~1 & 1.92& 2026 \\
    DCC & Unity~1   & 1.92& 2026 \\
    DCC & IIW-N     & 3.07 & 2027 \\
    DCC & IIW-S     & 3.07 & 2027 \\
    \midrule
    SCS & KJ Hintz~1      & 3.11 & 2028 \\
    SCS & KJ Hintz~2      & 3.11 & 2028 \\
    SCS & TB Leingang~1   & 2.46 & 2029 \\
    SCS & TB Leingang~2   & 2.46 & 2029 \\
    SCS & BK Fischer~1    & 3.24 & 2030 \\
    SCS & BK Fischer~2    & 3.24 & 2030 \\
    \bottomrule
    \end{tabular}
\end{table}

The CO$_2$ injection simulations are conducted over a 50-year period starting in 2022. Thus, the injection durations range from 42 to 50~years. It is important to emphasize that these injection durations are substantially longer than the 12–20~year injection periods for the currently approved CO$_2$ storage projects. The simulations presented in this study thus represent an extended-duration scenario, so some of the effects modeled in this work will not be observed in the 12–20~year time frame, though they could be observed if operations are continued.

All wells are assumed to be vertical and to fully penetrate the Broom Creek Formation. A maximum bottom-hole pressure ($\mathrm{BHP}_{\max}$) constraint is specified to prevent pressure from exceeding the fracture pressure, i.e.,

\begin{equation}
    \mathrm{BHP}_{\max} = 0.9 (d_z \times g_{\mathrm{frac}}),
    \label{eq:BHPmax}
\end{equation}
where $d_z$ is the depth of the Broom Creek Formation, $g_{\mathrm{frac}}$ is the fracture pressure gradient, and the 0.9 multiplier represents a safety factor. The minimum observed value of $g_{\mathrm{frac}}$ in the Broom Creek Formation, 0.147~bar/m (0.65~psi/ft), is used for all wells. Observed values of $g_{\mathrm{frac}}$ are provided in Figure~S6(a) in Supplementary Material. If $\mathrm{BHP}_{\max}$ is reached, the well switches from rate control to BHP control, which results in a reduced injection rate.

\subsection{Nested Local Grid Refinement (LGR)}
\label{sec:NestedLGR}
The model presented above, which will be referred to as the fine-scale model, contains $44\times10^6$~cells and is time-consuming to run (run times will be given later). To enable uncertainty quantification, which entails many simulation runs, and for later use in optimization, a coarsened model that runs much faster is required. Our coarsening strategy involves the use of nested local grid refinement regions along with permeability upscaling. We now describe these two procedures in turn.

\begin{figure}[h!]
    \centering
    \begin{subfigure}{0.35\textwidth}
        \centering
        \includegraphics[height=4.8cm]{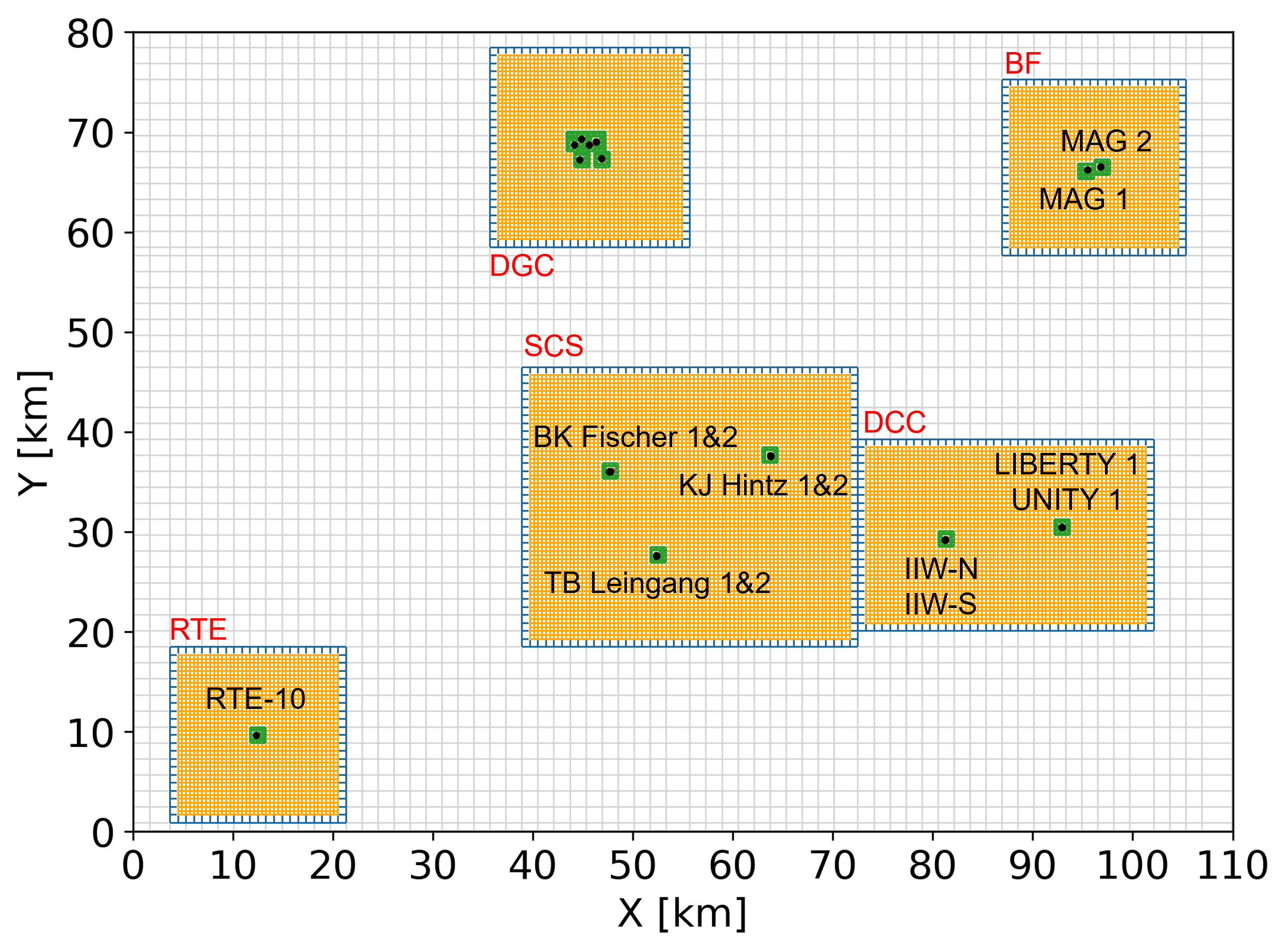}
        \caption{Overall simulation domain}
    \end{subfigure}
    \hfill
    \begin{subfigure}{0.29\textwidth}
        \centering
        \includegraphics[height=4.8cm]{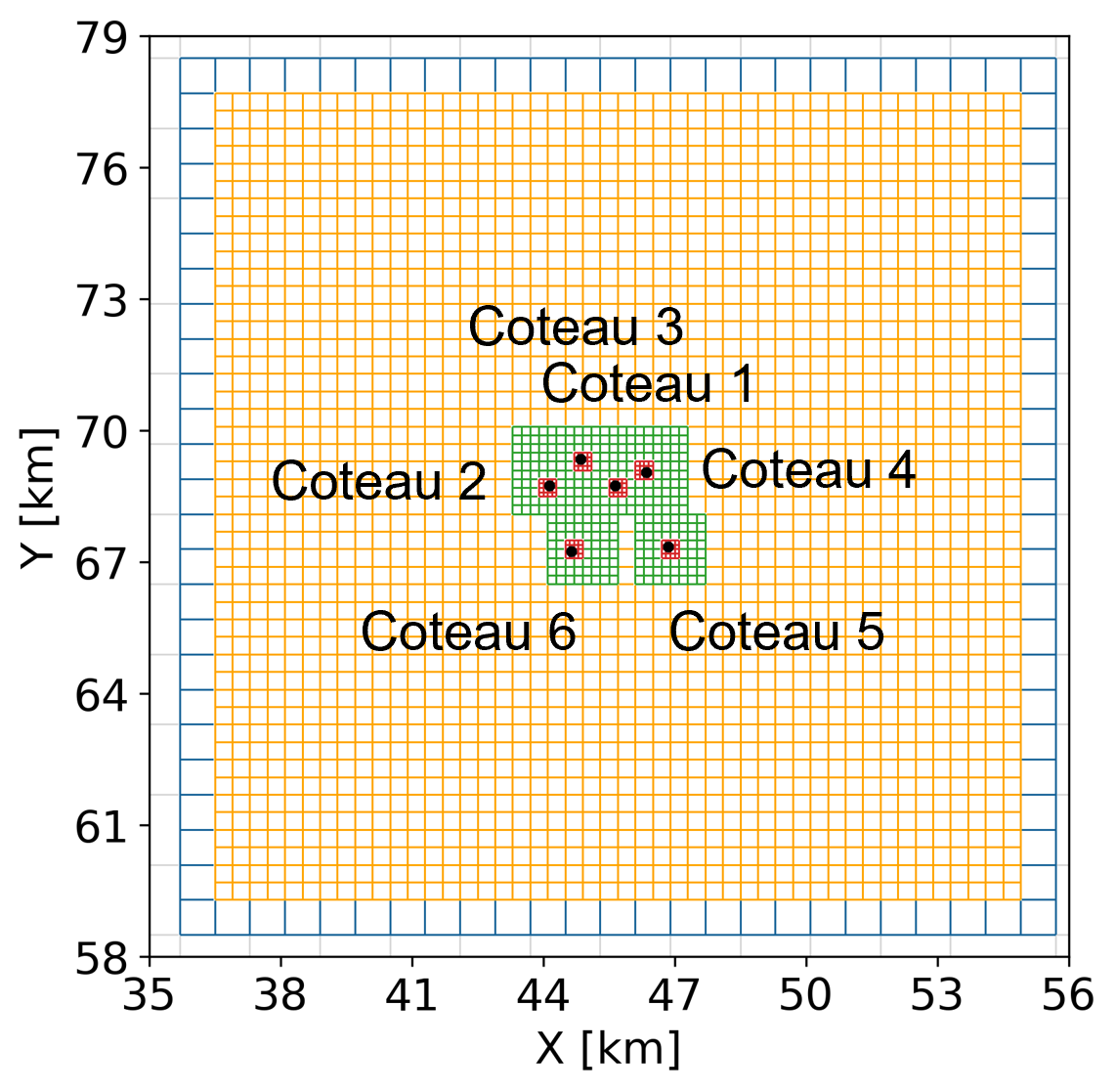}
        \caption{DGC project area}
    \end{subfigure}
    \begin{subfigure}{0.29\textwidth}
        \centering
        \includegraphics[height=4.8cm]{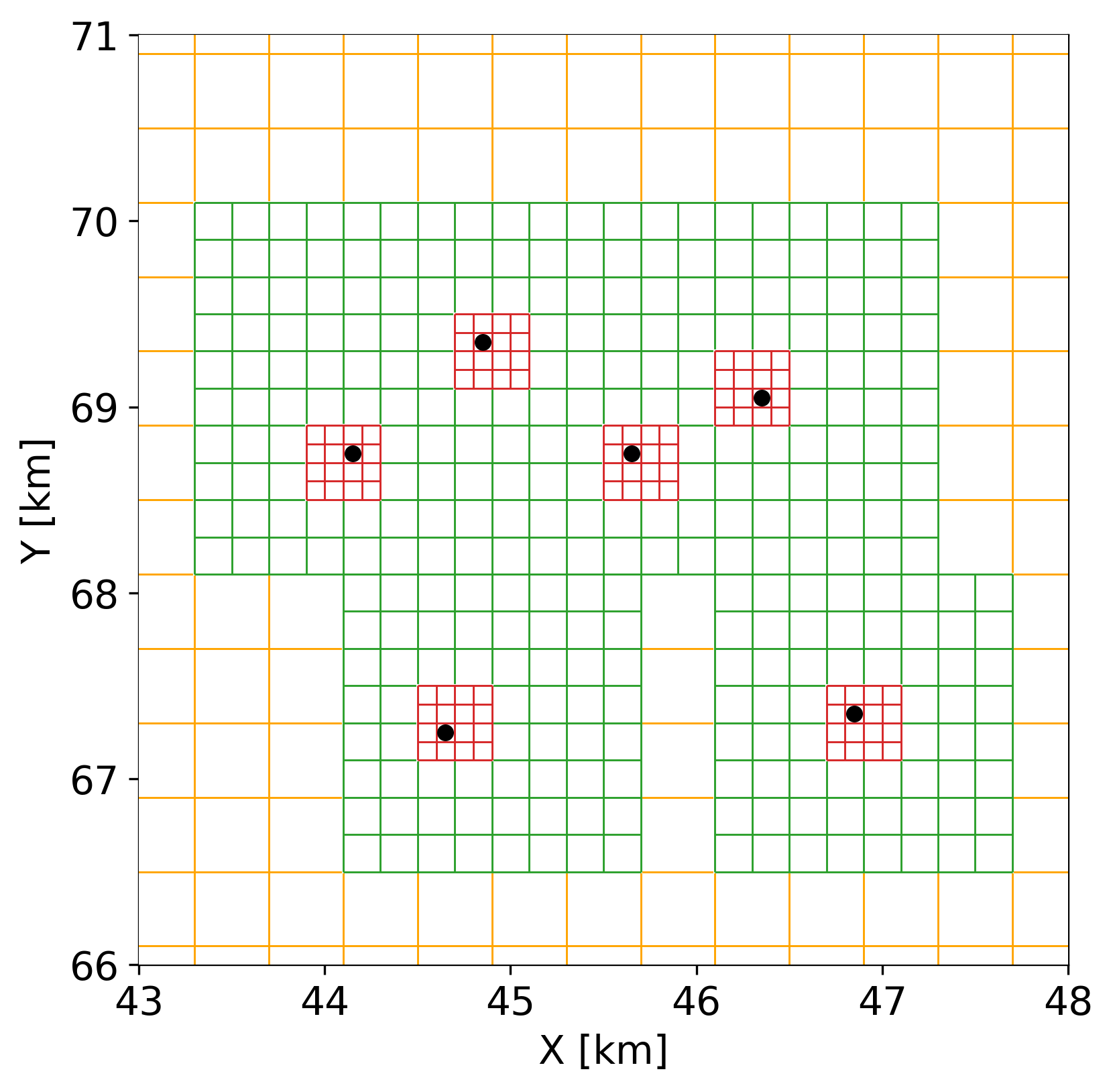}
        \caption{Near-well regions}
    \end{subfigure}
    \caption{Example of gridding with four levels of nested LGRs. Gray, blue, orange, green, and red lines indicate the global domain and the level-1, 2, 3, and 4 LGR regions, respectively. Black circles and labels indicate well locations and names. Red labels denote project names.}
    \label{fig:Nested LGR}
\end{figure}

Nested LGRs, shown in Figure~\ref{fig:Nested LGR}, allow us to introduce different levels of grid resolution over different areas of the coarse model. The finest cells are used around the wells to enable the accurate representation of injectivity. The grid is progressively coarsened away from the wells. LGR regions are in all cases taken to be rectangular. In the $x$-$y$ plane, they are fully defined by the grid indices in the lower-left corner $(i_{\min},j_{\min})$ and the upper-right corner $(i_{\max},j_{\max})$ of the refined region. LGR regions are defined only within the Broom Creek Formation, not in the overburden or underburden. The region outside the LGRs is referred to as the global region, which is depicted with gray grid lines in Figure~\ref{fig:Nested LGR}(a).

In a nested LGR hierarchy, a region that contains one or more other LGRs is referred to as the parent, and the enclosed LGRs are referred to as children. For example, in Figure~\ref{fig:Nested LGR}(b), the level-1 LGR (blue) is the parent of the level-2 LGR (orange). The outermost LGRs are direct children of the global region. LGRs that share the same parent are referred to as siblings. The four level-4 LGRs shown in red in the upper half of Figure~\ref{fig:Nested LGR}(c) are all enclosed within the same level-3 LGR (green) and are therefore sibling LGRs. Although the green area in Figure~\ref{fig:Nested LGR}(c) is a single connected region, it is treated as three separate rectangular LGR regions. 
These three level-3 LGR regions are sibling LGRs that share the same parent level-2 LGR, shown in orange.

The grid indices of the two opposite corners of a (rectangular) LGR region are given by
\begin{equation}
    i_{\min} = i_{\mathrm{well}} - n_{i} + 1 + \Delta_{i}, \ \ j_{\min} = j_{\mathrm{well}} - n_{j} + 1 + \Delta_{j},
    \label{eq:dmin}
\end{equation}
\begin{equation}
    i_{\max} = i_{\min} + 2n_{i}, \ \  j_{\max} = j_{\min} + 2n_{j},
    \label{eq:dmax}
\end{equation}
where $(i_{\mathrm{well}}, j_{\mathrm{well}})$ indicates the cell containing the well, $n_i$ and $n_j$ denote the half-widths measured in grid cells, in $x$ and $y$, of the LGR region, and $\Delta_i$ and $\Delta_j$ are shifts.
In our convention, cell ($i,j$) is bounded by grid lines with indices $i$ and $i+1$ in the $x$-direction, and $j$ and $j+1$ in the $y$-direction. 
Because coarsening factors are restricted to powers of~2, $n_i$ and $n_j$ are constrained to be even integers. Consequently, the spans of the LGR regions, $2n_i$ and $2n_j$, are always multiples of~4. As a result, the well block is not located exactly at the center of the LGR region. In this study, the default is to place the well block one cell to the left of the center in the $i$-direction and one cell below the center in the $j$-direction. This accounts for the $+1$ offset in Eq.~\ref{eq:dmin}. The shift terms $\Delta_i$ and $\Delta_j$ are used to satisfy the conformance constraint, described below. 

LGR regions are initially defined independently for each well. When two or more LGR regions at the same level overlap, they are merged into a single rectangular region with $x$-boundaries determined by the minimum $(i_{\min})$ and the maximum $(i_{\max})$ $i$-indices for all overlapping regions, and similarly for the $y$-boundaries and $j$-indices. This merging is performed in several places in Figure~\ref{fig:Nested LGR}(a). Specifically, although there are originally 19 level-3 LGR regions (one for each of the 19~wells, counting the monitoring well), only five level-3 LGR regions (orange) are present in Figure~\ref{fig:Nested LGR}(a). This is because the level-3 LGR regions associated with wells in the same project are merged.

Nested LGRs are defined by specifying the upscaling factors $l_i$ and $l_j$, half-widths $n_i$ and $n_j$, and shifts $\Delta_i$ and $\Delta_j$ for each LGR region. As described below, $\Delta_i$ and $\Delta_j$ are treated as decision variables that can be varied to satisfy the conformance requirement as described below. Therefore, the upscaling factors and half-widths constitute the primary input parameters for defining the LGRs. The parameter values used in this study are listed in Table~\ref{tab:lgr_upscaling_factor}. Although they can be specified independently for each well, identical values are adopted for all wells for simplicity.

The above discussion applies only to the definition of LGR regions in the $x$-$y$ plane. In the $z$-direction, coarsening is applied uniformly across the Broom Creek Formation domain. Coarsening factors, $l_k$, of 8, 4, 2, or 1, as listed in Table~\ref{tab:lgr_upscaling_factor}, are applied over all 40 original layers of the fine model. Because these $l_k$ values all divide evenly into 40, the resulting LGRs do not violate the conformance requirement described below.

\begin{table}[htbp]
    \centering
    \caption{Upscaling factors and half-widths of nested LGR}
    \label{tab:lgr_upscaling_factor}
    \begin{tabular}{lcccccc}
    \toprule
    \multirow{2}{*}{Level} & \multicolumn{3}{c}{Upscaling factor} & & \multicolumn{2}{c}{Half-width} \\
    \cline{2-4} \cline{6-7}
     & $l_i$ & $l_j$ & $l_k$ & & $n_i$ & $n_j$ \\
    \midrule
    Global & 16 & 16 & 8 & & --  & --  \\
    1      & 8  & 8  & 4 & & 88  & 88  \\
    2      & 4  & 4  & 2 & & 80  & 80  \\
    3      & 2  & 2  & 2 & & 8   & 8   \\
    4      & 1  & 1  & 1 & & 2   & 2   \\
    \bottomrule
    \end{tabular}
\end{table}

The upscaling factors $l_i$ and $l_j$ for the global region given in Table~\ref{tab:lgr_upscaling_factor} are treated as target values, with local adjustments made where necessary to preserve grid conformity. 
The details of defining the global region are provided in Section~S2.1 in Supplementary Material. Note that the global region covers the entire model and therefore does not require specification of $n_i$ and $n_j$.

\begin{figure}[h!]
    \centering
    \begin{subfigure}{0.45\textwidth}
        \centering
        \includegraphics[height=6cm]{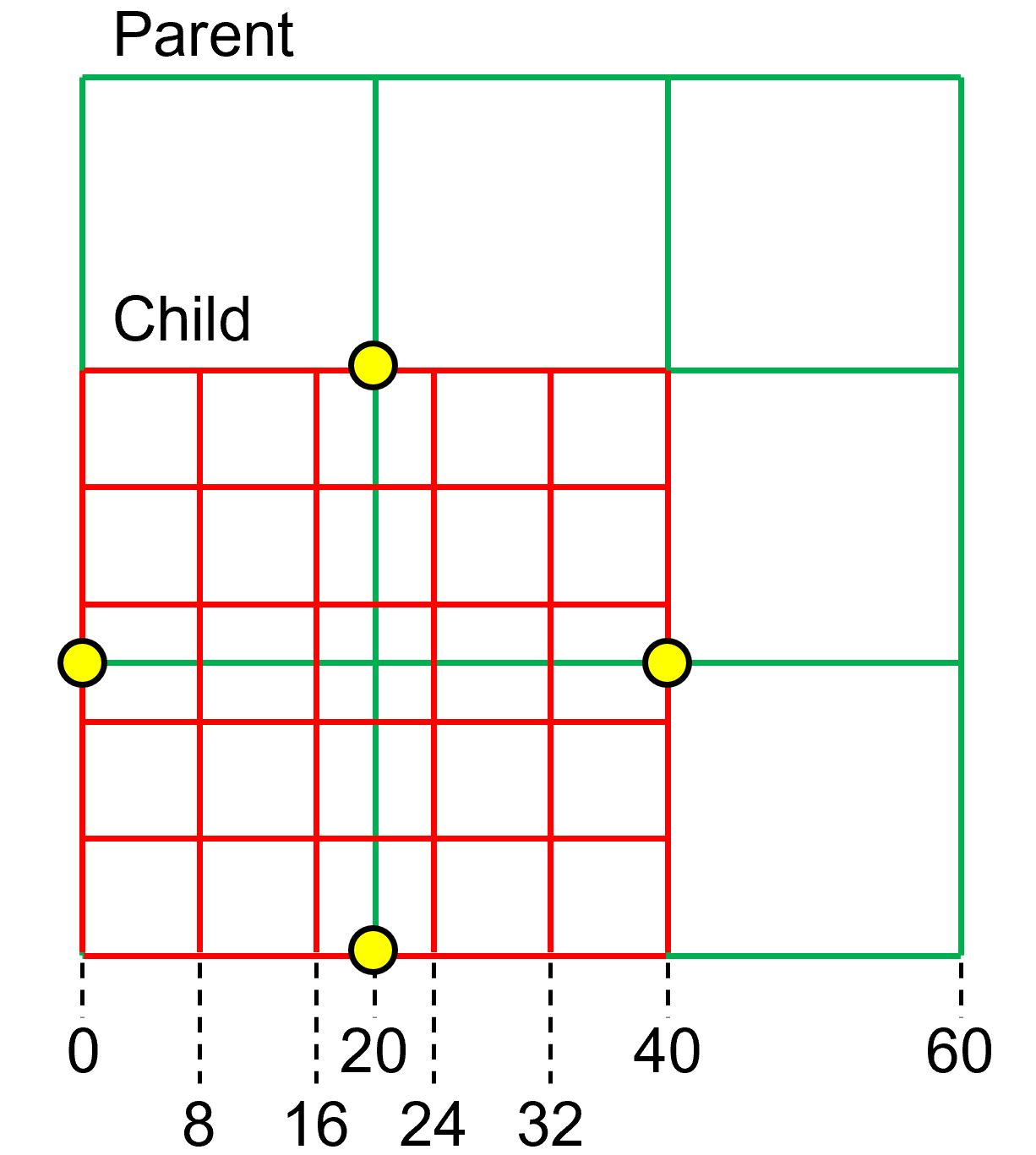}
        \caption{Nonconforming example}
    \end{subfigure}
    \hfill
    \begin{subfigure}{0.45\textwidth}
        \centering
        \includegraphics[height=6cm]{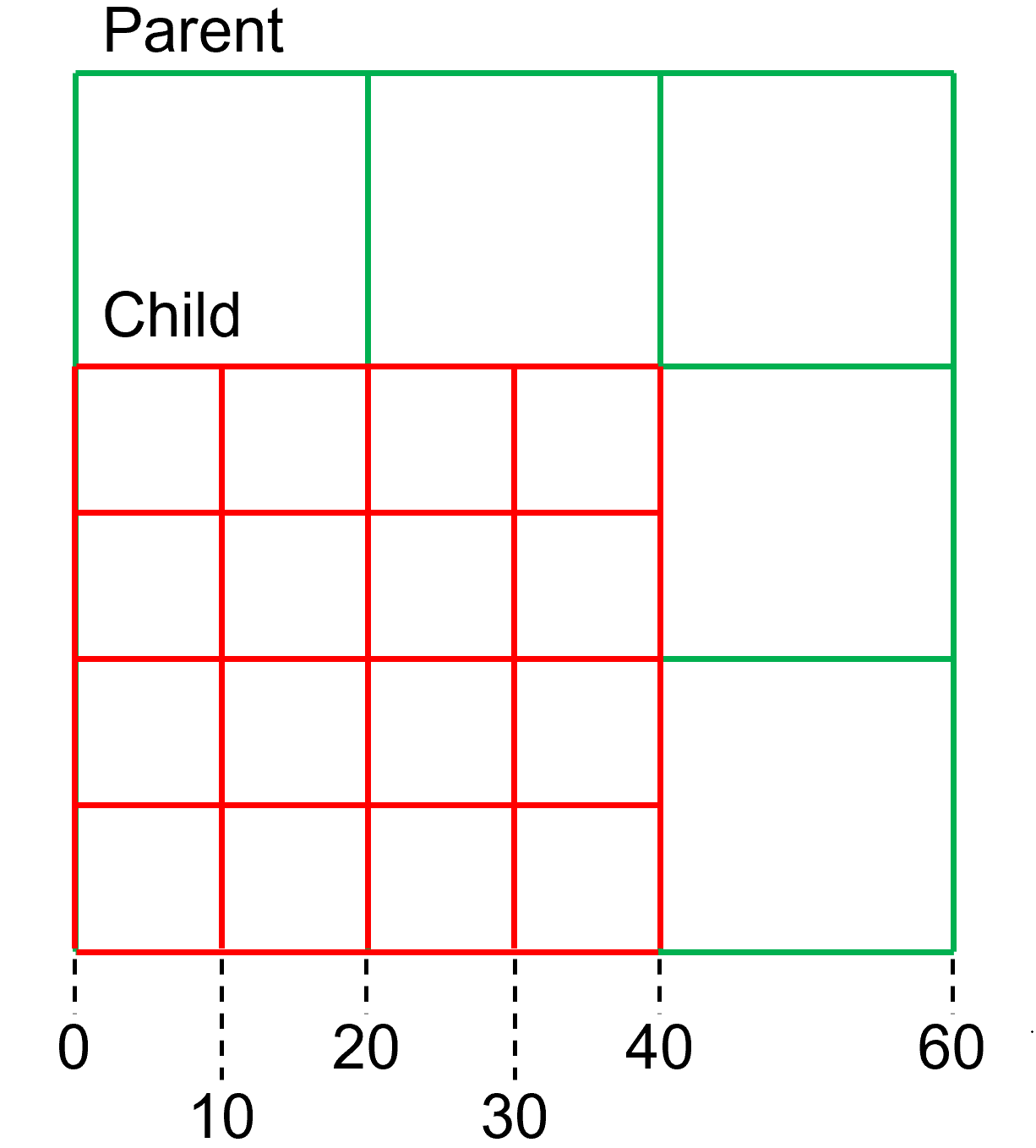}
        \caption{Conforming example}
    \end{subfigure}
    \caption{First example of grid conformance: requirement that every parent-grid index coincide with a child-grid index.}
    \label{fig:csp-conformance-example1}
\end{figure}

Simulators typically require adjacent regions with different resolutions to be conforming. By conforming, we mean that certain nodes (cell corner points) must be shared by the adjacent cells. To illustrate the LGR conformance requirements, we present two representative examples. The first example, shown in Figure~\ref{fig:csp-conformance-example1}, involves the relationship between a parent grid (green) and child grid (red). In Figure~\ref{fig:csp-conformance-example1}(a), the set of parent ($p$) and child ($c$) $i$-direction grid indices are given by $\mathbf{i}^p=[i_1^p,\dots,i_4^p]^T=[0,20,40,60]^T$ and $\mathbf{i}^c=[i_1^c,\dots,i_6^c]^T=[0,8,16,24,32,40]^T$. These grid index values are all referenced to the fine-scale grid. The corresponding scaling factors of the parent and child regions are $l_i^p=20$ and $l_i^c=8$, respectively. Within the child grid, i.e., from $i_{\min}^c=i_1^c=0$ to $i_{\max}^c=i_6^c=40$, one of the LGR conformance requirements is that every parent-grid index must coincide with one of the child-grid indices. 
However, as highlighted by the yellow circles, the parent-grid index $i^p_2=20$ does not correspond to a child-grid index. This results in a nonconforming interface, which violates the LGR requirements.

This condition can be satisfied by restricting the upscaling factor of each LGR region to be a power of~2. As a result, the ratio of the upscaling factors between any two adjacent regions is also a power of~2, thereby ensuring the first conformance requirement. Figure~\ref{fig:csp-conformance-example1}(b) shows a conforming example in which the parent upscaling factor is exactly twice that of the child region ($l_i^p=20$ and $l_i^c=10$). Under this restriction, all parent-grid indices coincide with a child-grid index, and the nonconforming interface point is eliminated.

\begin{figure}[h!]
    \centering
    \begin{subfigure}{0.45\textwidth}
        \centering
        \includegraphics[height=6cm]{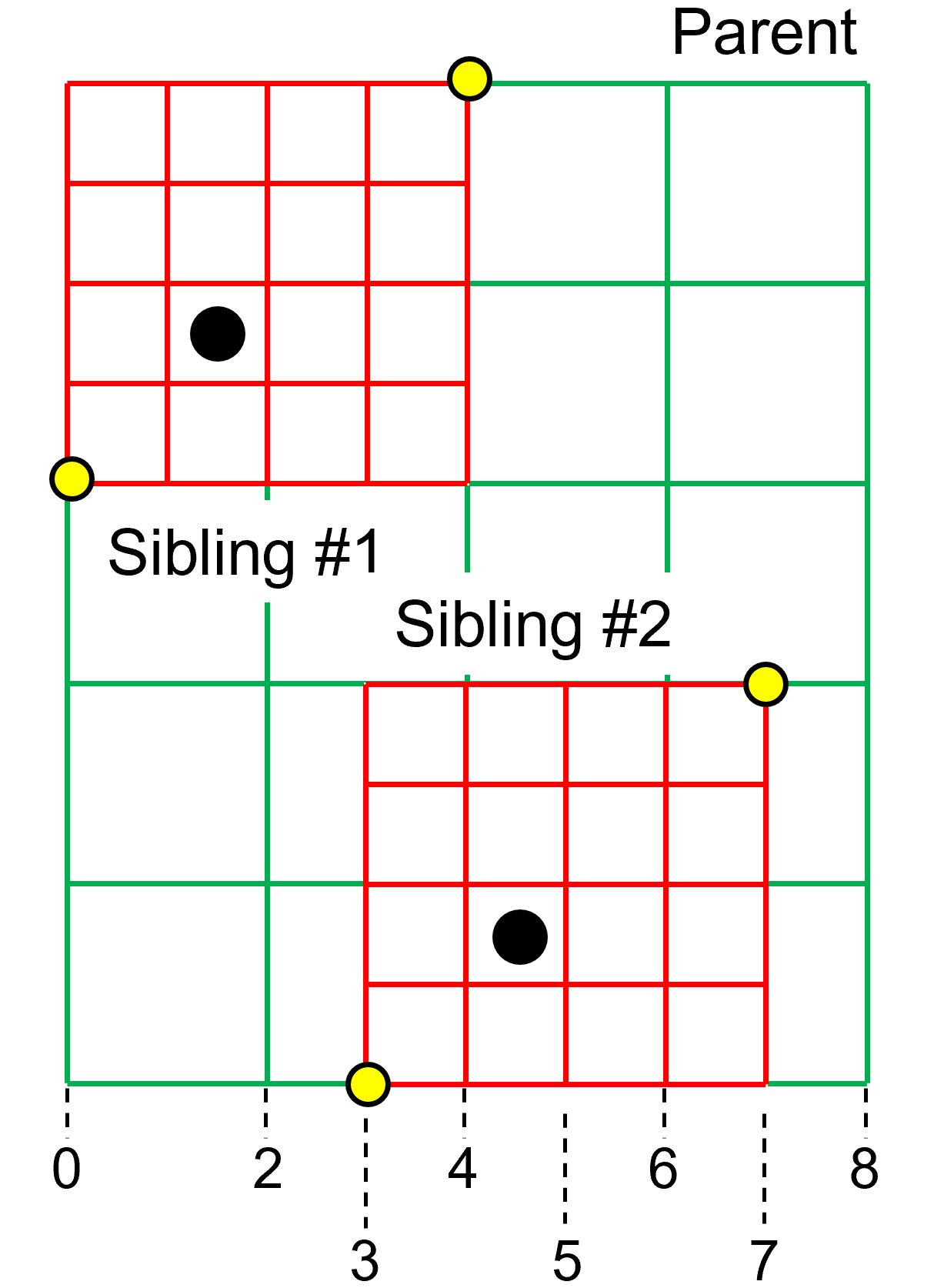}
        \caption{Nonconforming example}
    \end{subfigure}
    \hfill
    \begin{subfigure}{0.45\textwidth}
        \centering
        \includegraphics[height=6cm]{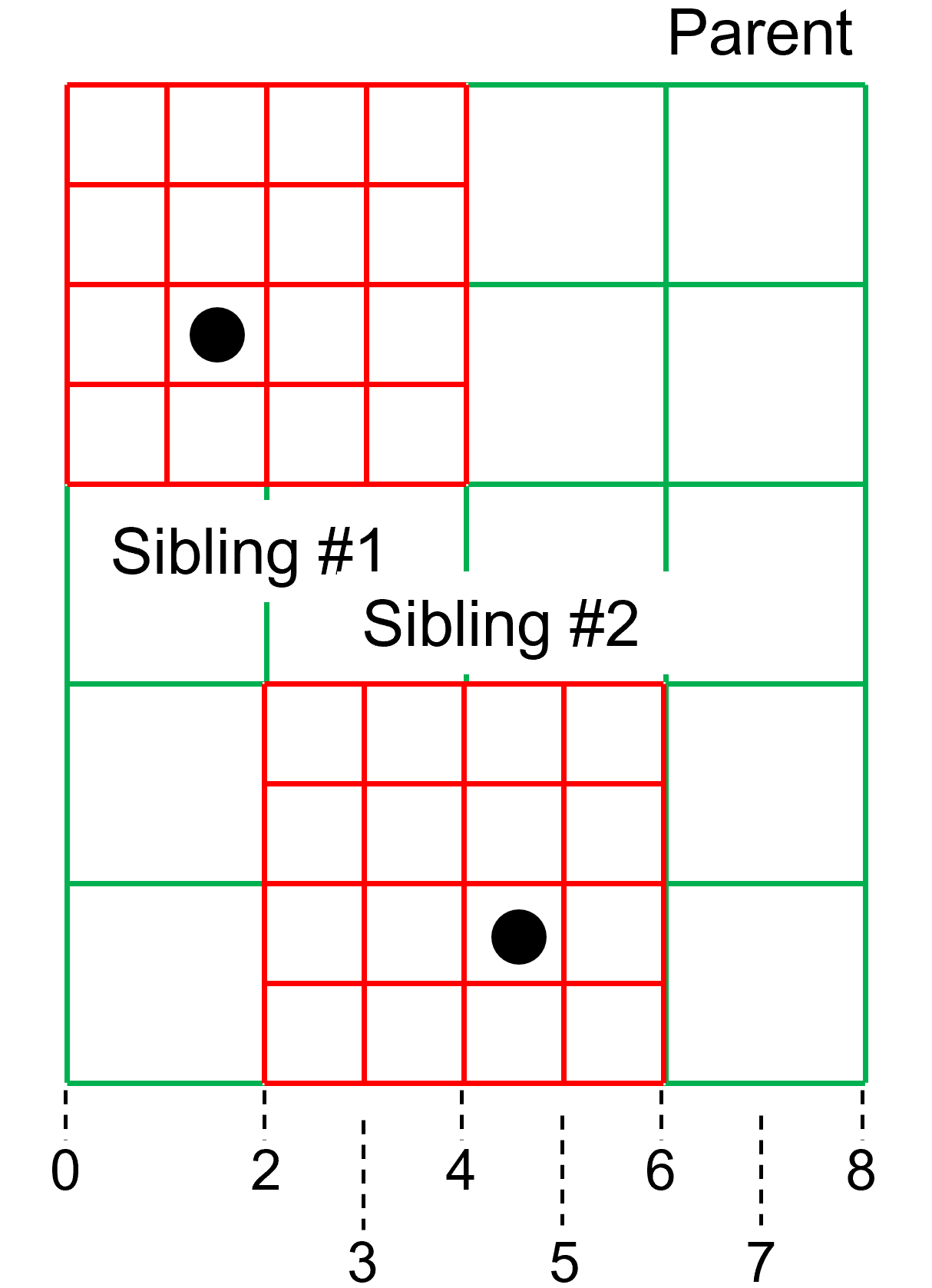}
        \caption{Conforming example}
    \end{subfigure}
    \caption{Second example of grid conformance: requirement that child-region corner points coincide with parent nodes.}
    \label{fig:csp-conformance-example2}
\end{figure}

A second conformance requirement is that both corner points of a child region must coincide with nodes in its parent grid. This case is illustrated in Figure~\ref{fig:csp-conformance-example2}. The upper-left sibling region (Sibling~\#1) in Figure~\ref{fig:csp-conformance-example2}(a) satisfies the conformance requirement because its two corner points, highlighted by the yellow circles, coincide with parent-grid nodes. Sibling~\#2, however, violates this requirement because its two corner points do not coincide with any parent-grid nodes. 

This nonconformance can be corrected by adjusting the shift parameters ($\Delta_i$ and $\Delta_j$) of the individual LGR regions. In other words, the locations of the LGR regions are shifted, when necessary, such that all region boundaries become conforming. Figure~\ref{fig:csp-conformance-example2}(b) illustrates this adjustment. Here, the shift for Sibling~\#2, denoted $\Delta_i^{s2}$, is set to -1 ($\Delta_i^{s2}=0$ in Figure~\ref{fig:csp-conformance-example2}(a)). It is apparent that the Sibling~\#2 LGR now satisfies this second conformance requirement. Sibling~\#1 is unaffected by this adjustment.

Although the adjustment performed in Figure~\ref{fig:csp-conformance-example2} appears straightforward, satisfying the conformance requirements can be challenging in cases with multiple wells and many nested LGRs. In such scenarios, the number of design variables, $\Delta_i$ and $\Delta_j$, and the number of the possible combinations and interactions increase rapidly with the number of wells and LGR levels. 
Thus, to determine the shift variables such that the conformance requirements are satisfied, we formulate the problem as a Constraint Satisfaction Problem (CSP):
\begin{equation}
    \boldsymbol{\Delta}_{\rm opt} = \mathrm{arg}\min_{\boldsymbol{\Delta}} \; \|\boldsymbol{\Delta}\|_1 \quad \text{subject to} \quad g_m(\boldsymbol{\Delta}) \leq 0, \; m = 1, \dots, M ,
    \label{eq:csp}
\end{equation}
where $\boldsymbol{\Delta}$ is the vector of decision variables containing all the $\Delta_i$ and $\Delta_j$ shifts for all LGR regions, at every well and at every LGR level. The vector $\boldsymbol{\Delta}_{\rm opt}$ represents the optimal solution and $g_m$ denotes constraint $m$. Note that a simple CSP formulation would only seek a $\boldsymbol{\Delta}_{\rm opt}$ that satisfies all constraints $g_m$. This can result in LGR regions in which the wells are far from the center, which is undesirable in our setting. To avoid this, the problem is instead formulated as a minimization problem where we minimize the $\ell_1$-norm of the shifts. This results in wells being positioned as close as possible to the center of their respective LGR regions. A complete definition of the constraints corresponding to $g_m$ in Eq.~\ref{eq:csp} is given in Section~S2.2 in Supplementary Material. Eq.~\ref{eq:csp} is solved using the CP-SAT solver provided by Google OR-Tools~\citep{cpsatlp}.

\subsection{Permeability Upscaling Procedure}
\label{sec:PowerAveraging}

Many methods for upscaling permeability have been proposed -- see \citet{durlofsky2005upscaling} for a discussion of several of the existing approaches. The most accurate (single-phase parameter) upscaling procedures typically entail global transmissibility upscaling, where coarse-scale transmissibilities are computed from the solution of the global fine-scale pressure equation, with flow driven by the actual wells. Methods of this type can also be formulated as local-global approaches that avoid global fine-scale solutions~\citep{wen_spej_2006}, which are useful with very high resolution fine models such as that considered here. Although it is possible to combine a global or local-global transmissibility upscaling procedure with simple LGR structures (this was in fact accomplished by \citet{Zou2023Dissertation}), the computations become complicated and can lose accuracy in cases with multiple nested LGR regions. In addition, some simulators may not allow the specification of transmissibility in such regions. Because our models do in fact contain many nested LGR regions, and because the grid resolution and LGR nesting are constructed such that the low-resolution regions experience limited dynamics, we apply a much simpler empirical power averaging approach in this study. This strategy will be especially useful in optimization settings (considered in later work), where a new coarse model must be constructed for each candidate well configuration.

Power averaging entails the computation of upscaled grid-block permeabilities through application of~\citep{durlofsky2005upscaling}
\begin{equation}
    k_i^*=\left(\frac{1}{V_b}\int_{V_b}[k_i(\mathbf{y})]^{\omega_i}dV\right)^{\frac{1}{\omega_i}}.
\end{equation}
Here $k_i^*$ is the upscaled permeability component in the coordinate direction $i\in \{x, y, z\}$, $V_b$ is the coarse block bulk volume, $k_i(\mathbf{y})$ denotes component $i$ of the fine-grid permeability in a fine cell within the target coarse block, and $\omega_i \in [-1, 1]$ is the power averaging exponent for direction $i$. Note that, with this approach, both the fine- and coarse-scale permeabilities are assumed to be diagonal tensors. Commonly used averages are readily recovered by this procedure, i.e., the arithmetic average corresponds to $\omega_i=1$, the harmonic average to $\omega_i=-1$, and the geometric average to the limit $\omega \to 0$. In this study, we take $\omega_x=\omega_y$, which maintains isotropy in the $x$-$y$ plane. 

The values of $\omega_x$ ($=\omega_y$) and $\omega_z$ are typically determined empirically by minimizing a measure of the mismatch between fine and coarse solutions. Technically, we can vary both the LGR structure and the $\omega_i$ values to determine the optimal model, but here we instead identify them sequentially. The LGR structure is determined through limited numerical experimentation, using estimates for $\omega_x$ and $\omega_z$, with the goal of balancing computational efficiency and accuracy. Then, once the gridding is established, we proceed with the determination of optimal values for $\omega_x$ and $\omega_z$.

The three quantities of interest (QoIs) considered here and in our subsequent assessments are (1)~total CO$_2$ injected over 50~years from all five projects, (2)~total plume area at 50~years, and (3)~time-varying injection rate at each well. The error quantity (denoted $E$) we seek to minimize through the determination of optimal values of $\omega_x$ and $\omega_z$ is given by
\begin{equation}
    E= \left| \frac{Q^{f} - Q^{c}}{Q^{f}} \right|
    + \left| \frac{A^{f} - A^{c}}{A^{f}} \right|
    + \frac{1}{N_t N_w} \sum_{t=1}^{N_t} \sum_{i=1}^{N_w} \left| \frac{q^{f}_{t,i} - q^{c}_{t,i}}{q^{f}_{t,i}} \right|.
    \label{eq:objective_function}
\end{equation}
In this expression, $Q$ is the total CO$_2$ injected over 50~years [MT], $A$ is the total plume area [km$^2$] at 50~years, and $q_{t,i}$ is the injection rate of well $i$ at year $t$ [MTPA]. The superscripts $f$ and $c$ indicate that the corresponding quantity is obtained from the fine model or the coarse model, and $N_t$ and $N_w$ are the total number of time steps and wells. Weightings can be introduced in Eq.~\ref{eq:objective_function} to emphasize a particular quantity, though this is not done here. The minimization of $E$ is performed only for the base case. The resulting $\omega_x$ and $\omega_z$ values are then used for all the sensitivity cases. In Section~\ref{sec:SensitivityAnalysis} we will show that these values are indeed applicable for a wide range of cases.

The plume area ($A^f$ and $A^c$ in Eq.~\ref{eq:objective_function}) can be computed in different ways. Some studies, e.g.,~\citet{TANG2025104404}, have defined it in terms of the area of the minimum rectangle that encompasses the plume in the $x$--$y$ plane. Here we use a procedure similar to that described by~\citet{BURTONKELLY2025104456}, which is also consistent with the approach used in SFP applications for CCS projects in North Dakota~\citep{NDIC2022DGC}. Specifically, for each $(i,j)$ cell (in the $x$--$y$ plane), we scan vertically in the $z$-direction. If any cell along the $z$-direction has a gas saturation greater than 0.01, we count that $(i,j)$ cell as contributing to the plume area. The sum of all contributing $(i,j)$ cell areas is the total plume area. This treatment is well-suited for the discontinuous CO$_2$ saturation fields resulting from widely spaced injection wells, as we have in our basin-scale model.  

Porosity in coarse-scale grid blocks is computed via bulk-volume averaging, which acts to preserve pore volume exactly. Facies type is assigned to each coarse cell based on the dominant (in terms of volume fraction) facies over the corresponding set of fine-scale cells. The relative permeability and capillary pressure curves are unchanged between the fine and coarse models.

\subsection{Assessment of Coarse Model Accuracy}\label{sec:UpscalingResult}

The coarse model is gridded using the nested LGR method described in Section~\ref{sec:NestedLGR}. The resulting grid with four levels of nested LGR regions is shown in Figure~\ref{fig:Nested LGR}, and the corresponding grid parameters are summarized in Table~\ref{tab:lgr_upscaling_factor}. The LGR regions are successfully generated, with no constraint violations, by solving the CSP given in Eq.~\ref{eq:csp}. As a result, the wells are positioned near the centers of their LGR regions, as is evident in Figure~\ref{fig:Nested LGR}(b) and (c). 

The upscaling procedure reduces the number of grid cells from $44\times10^6$ to 394,610, corresponding to a 99\% reduction. The time required to simulate the fine model is 51.6~hours on 32 cores of an AMD EPYC 7502 processor. The time to simulate the coarse model, on the same 32 cores, is 605~seconds. This corresponds to a $307\times$ speedup. 

The optimal $\omega_x$ and $\omega_z$ are determined by minimizing the objective function $E$ in Eq.~\ref{eq:objective_function}. This optimization is performed using the modified Powell algorithm~\citep{Powell1964} as implemented in SciPy~\citep{Virtanen2020}, which is a derivative-free optimization method. The resulting optimal values are $\omega_x=\omega_y=0.61$ and $\omega_z=0.31$. With these parameters, the relative error in the total CO$_2$ injected over 50~years is 0.88\%, the relative error in total plume area at 50~years is 2.62\%, and the average relative error in the yearly injection rate at each well is 1.23\%. These results demonstrate that the use of nested LGRs combined with power averaging for permeability upscaling is able to provide a coarse model of reasonable accuracy and greatly reduced computational cost.

\begin{figure}[h!]
    \centering
    \begin{subfigure}{0.55\textwidth}
        \centering
        \includegraphics[height=6cm]{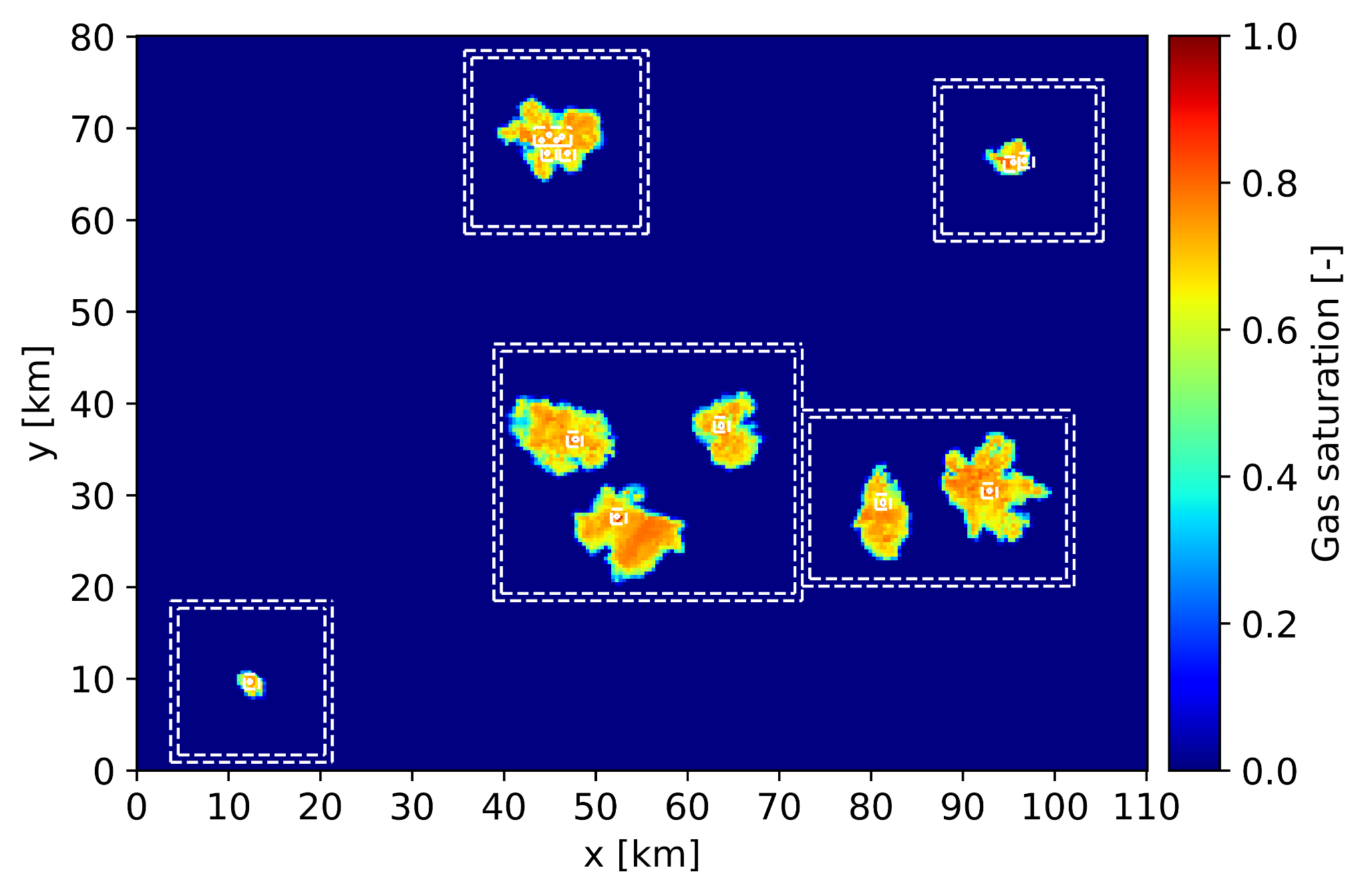}
        \caption{Overall domain for the fine model}
    \end{subfigure}
    \begin{subfigure}{0.4\textwidth}
        \centering
        \includegraphics[height=6cm]{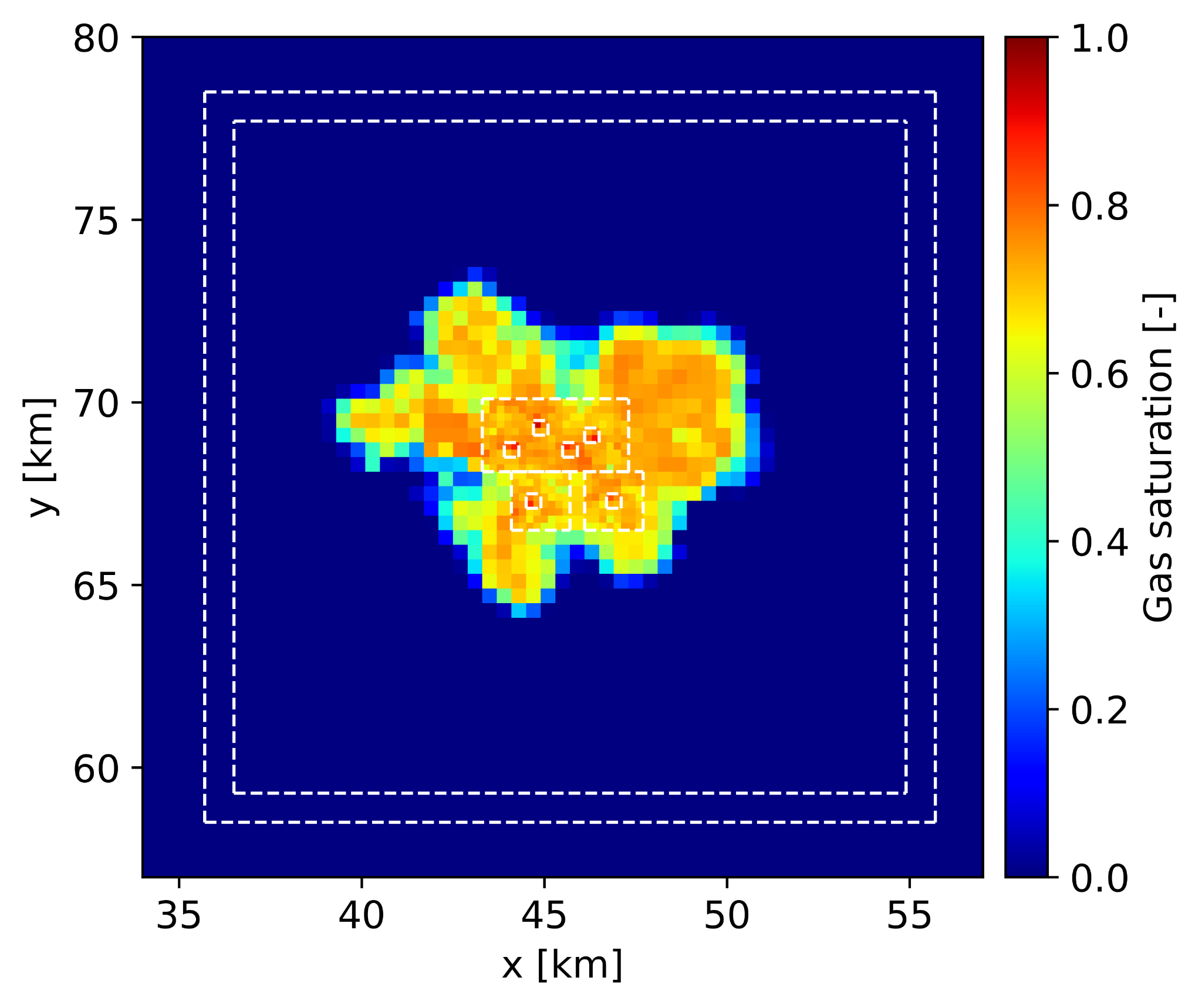}
        \caption{DGC project area in the fine model}
    \end{subfigure}
    \vspace{0.5em}
    \begin{subfigure}{0.55\textwidth}
        \centering
        \includegraphics[height=6cm]{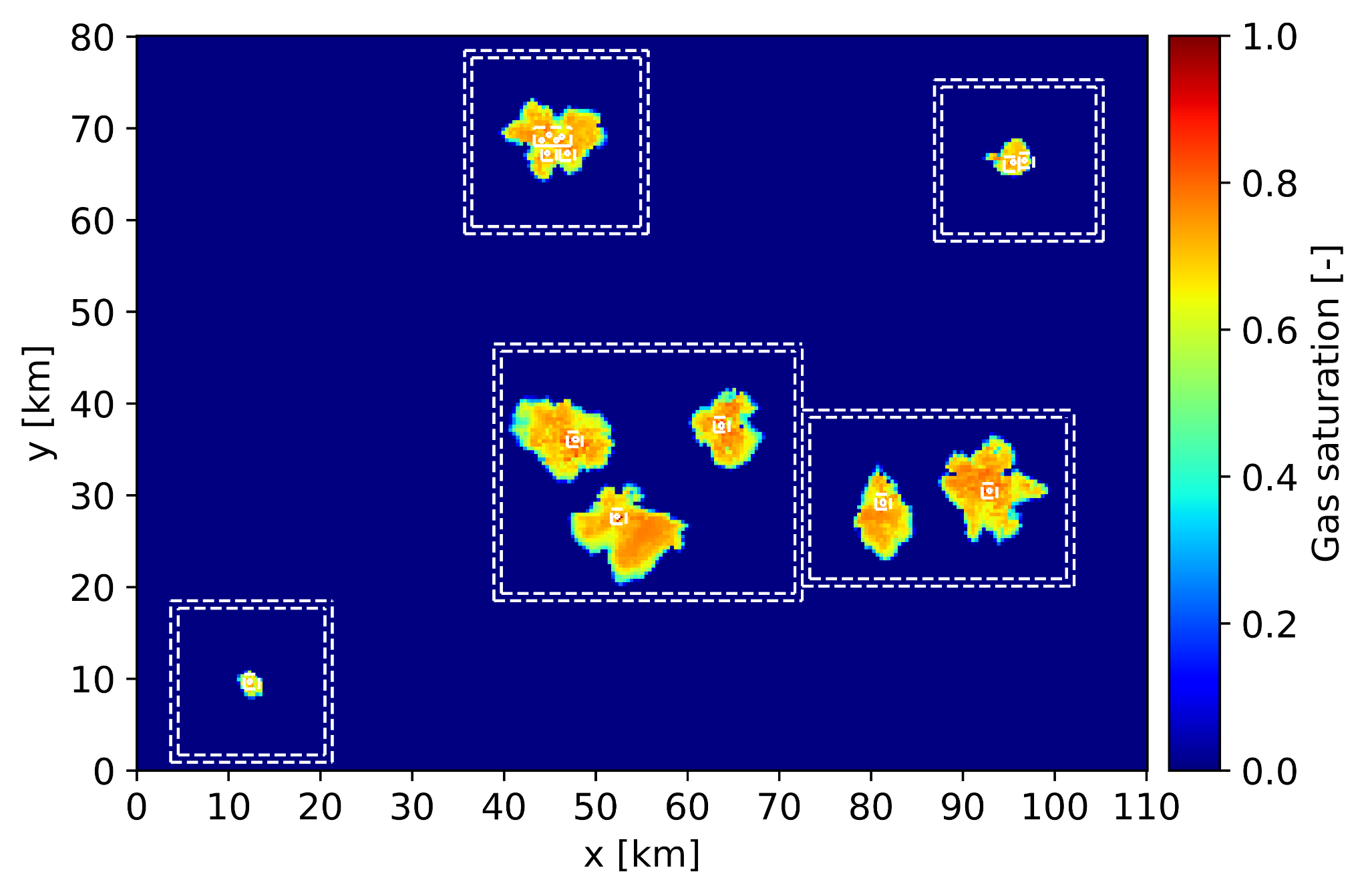}
        \caption{Overall domain for the coarse model}
    \end{subfigure}
    \begin{subfigure}{0.4\textwidth}
        \centering
        \includegraphics[height=6cm]{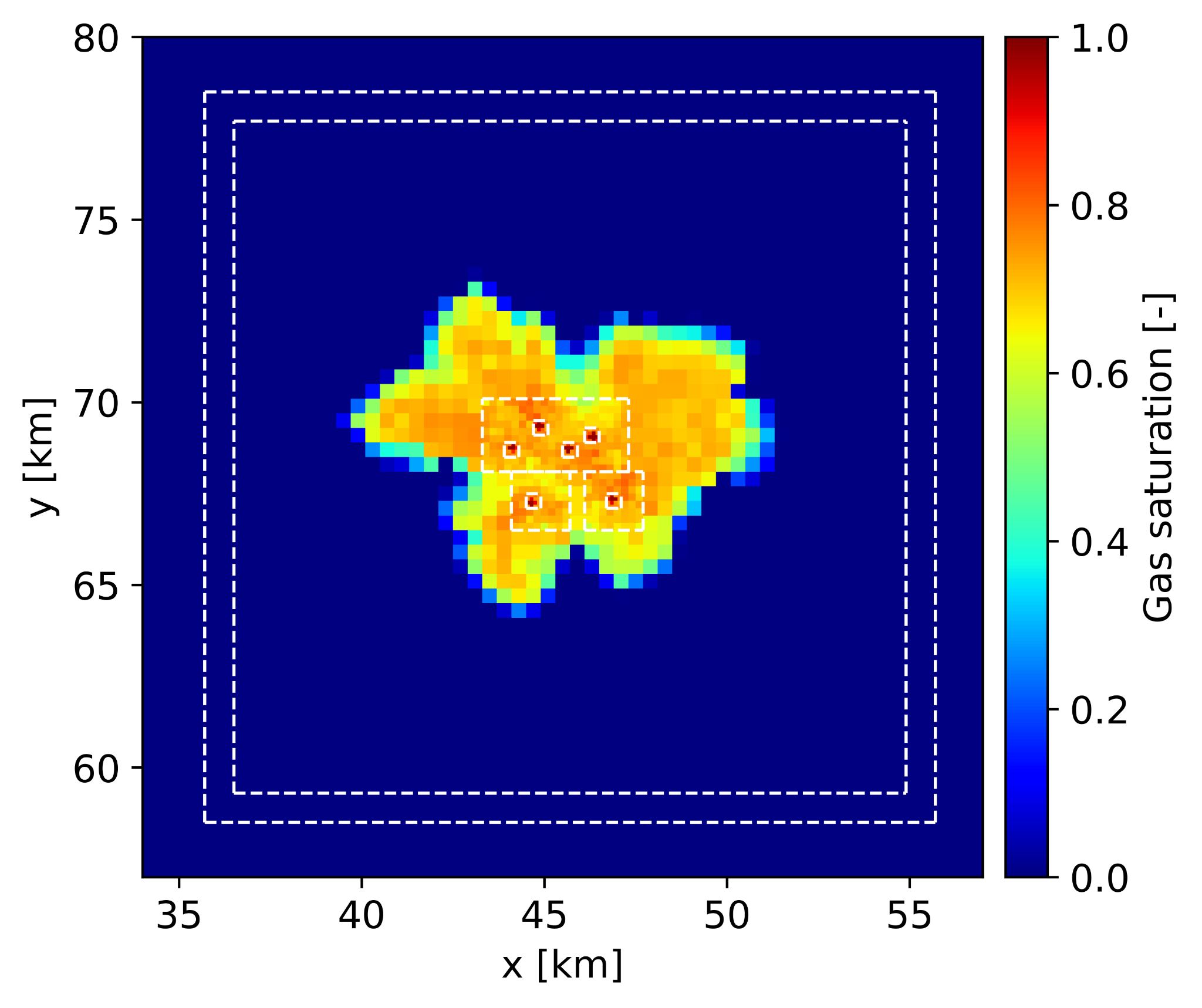}
        \caption{DGC project area in the coarse model}
    \end{subfigure}
    \caption{Comparison of gas saturation distributions at 50~years between the fine and coarse models. White dashed lines indicate the boundaries of the nested LGR regions in the coarse model.}
    \label{fig:Sg comp for upscaling err}
\end{figure}

We now present some comparisons of fine and coarse-scale responses. To enable a direct comparison of saturation results, the gas saturation for the fine model is first pore-volume averaged onto the coarse-model grid. Figure~\ref{fig:Sg comp for upscaling err} shows the maximum gas saturation along the $z$-axis for cells in the $x$-$y$ plane for both models at 50~years. It is apparent from Figure~\ref{fig:Sg comp for upscaling err}(a) and (c) that the plume distribution over all five projects is captured by the coarse model. The white dashed lines indicate the boundaries of the nested LGR regions in the coarse model. The outer level-1 and level-2 LGR regions encompass the two-phase flow regions.
The inner level-3 and level-4 LGR regions are confined to the vicinity of the injection wells. Enlarged views of the DGC project area are presented in Figure~\ref{fig:Sg comp for upscaling err}(b) and (d). We see that the coarse model accurately captures the main features evident in the high-resolution model, though some of the very fine detail is lost.

\begin{figure}[h!]
    \centering
    \begin{subfigure}{0.5\textwidth}
        \centering
        \includegraphics[height=5.4cm]{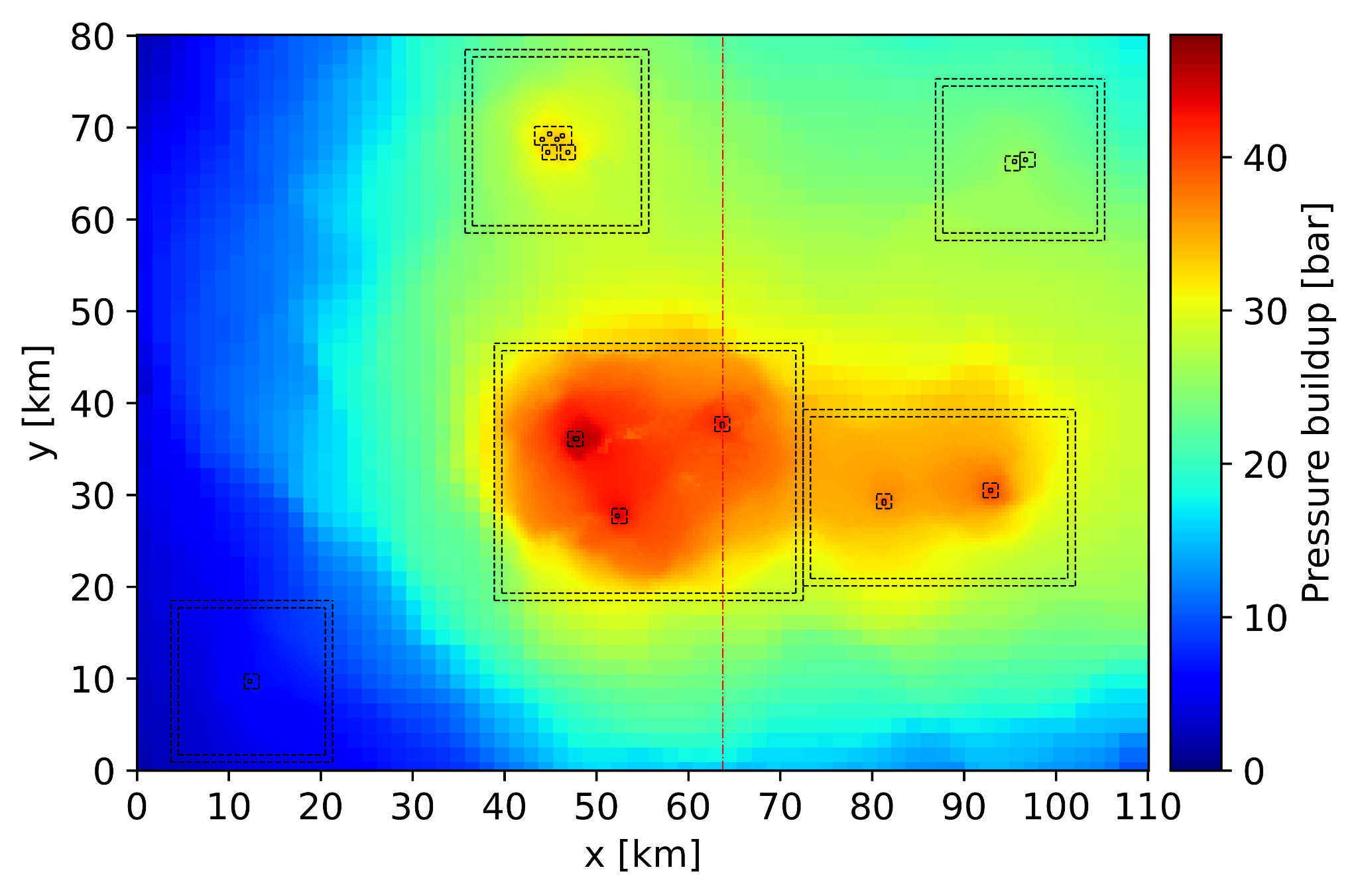}
        \caption{Pressure buildup in the coarse model}
    \end{subfigure}
    \begin{subfigure}{0.44\textwidth}
        \centering
        \includegraphics[height=5.4cm]{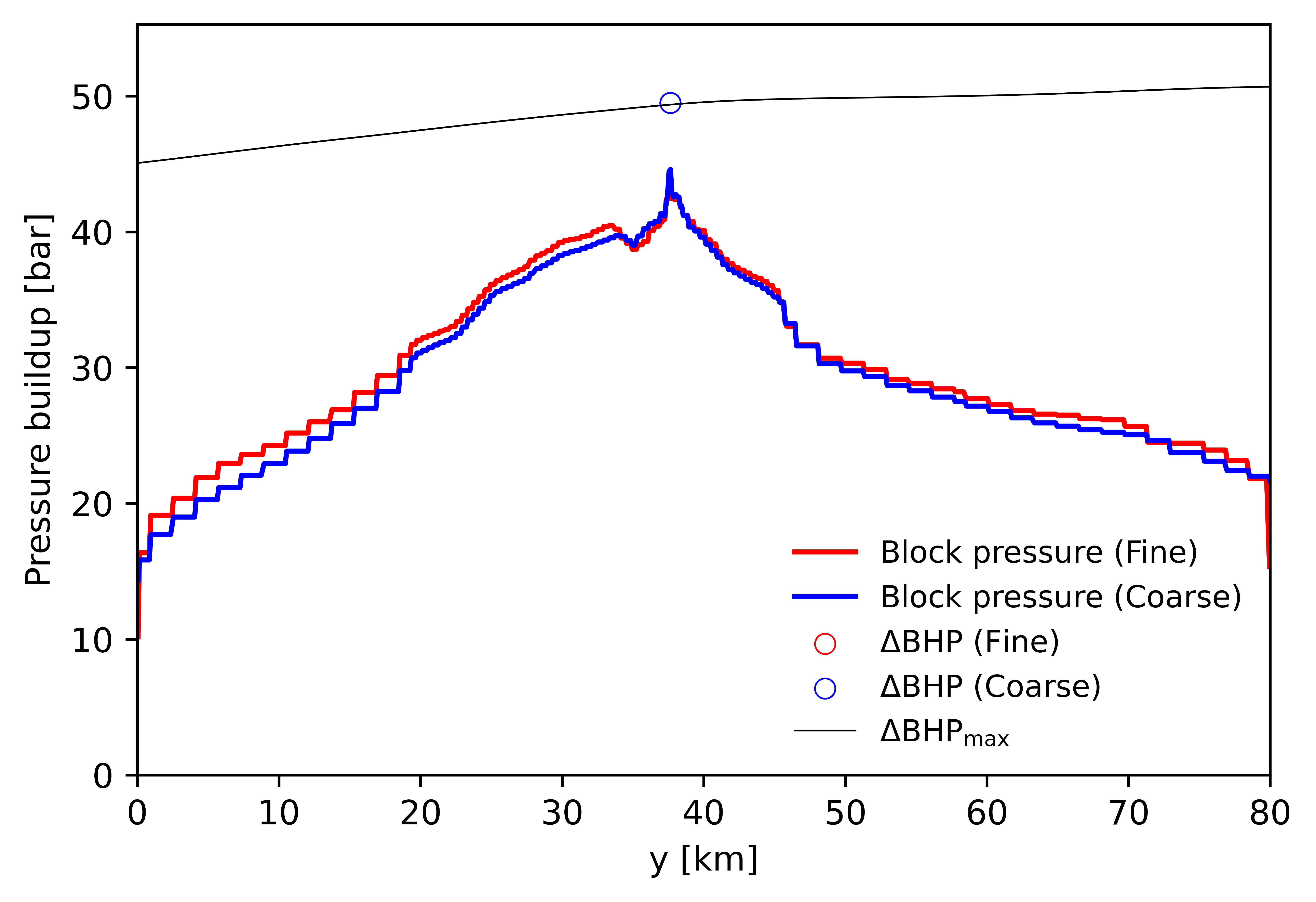}
        \caption{Cross-sectional pressure and BHP buildup profiles}
    \end{subfigure}
    \caption{Vertically averaged reservoir pressure buildup distribution at 50~years in the Broom Creek Formation. In (a), the black dashed lines indicate the boundaries of LGR regions at each level, and the red dashed line (passing through well KJ Hintz~1 in the SCS project) indicates the cross-sectional location shown in (b). Red and blue lines in (b) represent block pressure buildup for the fine and coarse models, and red and blue circles (which overlap) indicate BHP buildups. The black line represents the maximum allowable BHP buildup.}
    \label{fig:DP comp for upscaling error}
\end{figure}

The pressure response is shown in Figure~\ref{fig:DP comp for upscaling error}. Coarse-scale results for pressure buildup, defined as the difference between the initial pressure and the pressure at 50~years, appear in Figure~\ref{fig:DP comp for upscaling error}(a). This map is constructed by averaging the pressure buildup in the $z$-direction over the Broom Creek Formation (via bulk-volume weighting) at each $x$-$y$ location. Larger pressure buildup is observed in the SCS and DCC project areas than in the RTE and BF regions because of the higher injection rates for the SCS and DCC projects. In fact, the pressure effect of the RTE-10 well (in the lower left) is barely visible for this reason. Pressure effects in most cases, however, extend well beyond the individual plume areas, as is evident by comparing Figure~\ref{fig:DP comp for upscaling error}(a) to Figure~\ref{fig:Sg comp for upscaling err}(a) or (c). The global pressure effects apparent in Figure~\ref{fig:DP comp for upscaling error}(a) suggest that pressure interference between projects may be important in this system.

Next, in Figure~\ref{fig:DP comp for upscaling error}(b), we compare the average pressure buildup along the cross section indicated by the red dashed line at $x=63.75$~km in Figure~\ref{fig:DP comp for upscaling error}(a). The blue line in Figure~\ref{fig:DP comp for upscaling error}(b) represents the vertically averaged pressure buildup of the coarse model along this cross section, and the red line represents the corresponding result for the fine model. To enable a direct comparison, the fine-model pressure buildup is first bulk-volume averaged onto the coarse grid (and then averaged vertically).
The pressure buildups for the fine and coarse models are seen to be in very close agreement over the full cross section. 
The thin black line shows the maximum BHP buildup at the top of the Broom Creek Formation (BHP results do not involve any vertical averaging). The maximum BHP buildup varies with location due to depth variation across the formation, as shown in Figure~\ref{fig:BC_top_contours}. The red and blue circles, which overlap, represent the BHP buildup for well KJ Hintz~1 in the fine and coarse models. The fact that the two results coincide and lie on the maximum BHP buildup line indicates that the well reaches the maximum BHP (defined in Eq.~\ref{eq:BHPmax}) in both models.  

\begin{figure}[h!]
    \centering
    \begin{subfigure}{0.48\textwidth}
        \centering
        \includegraphics[width=\textwidth]{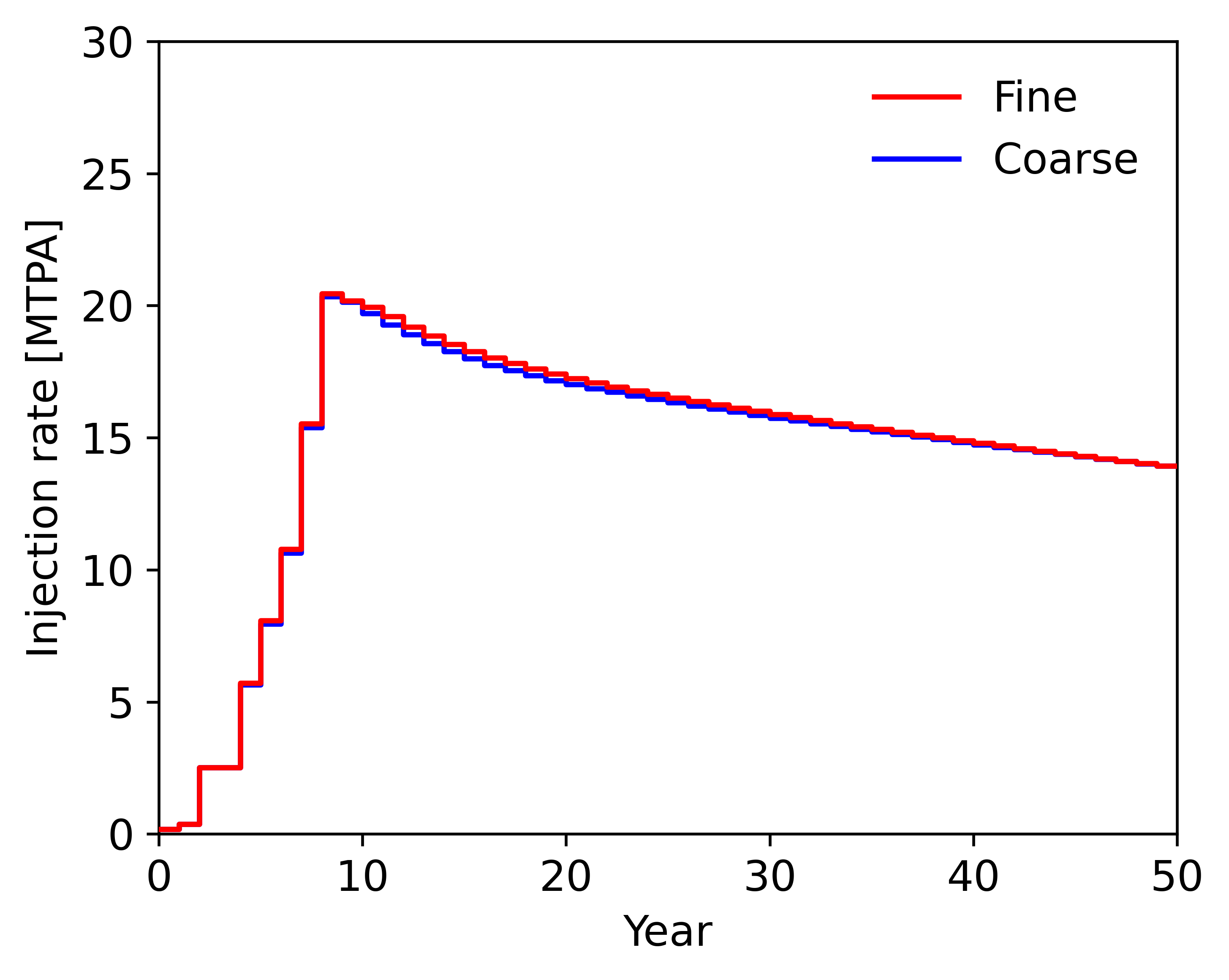}
        \caption{Basin total}
    \end{subfigure}
    \begin{subfigure}{0.48\textwidth}
        \centering
        \includegraphics[width=\textwidth]{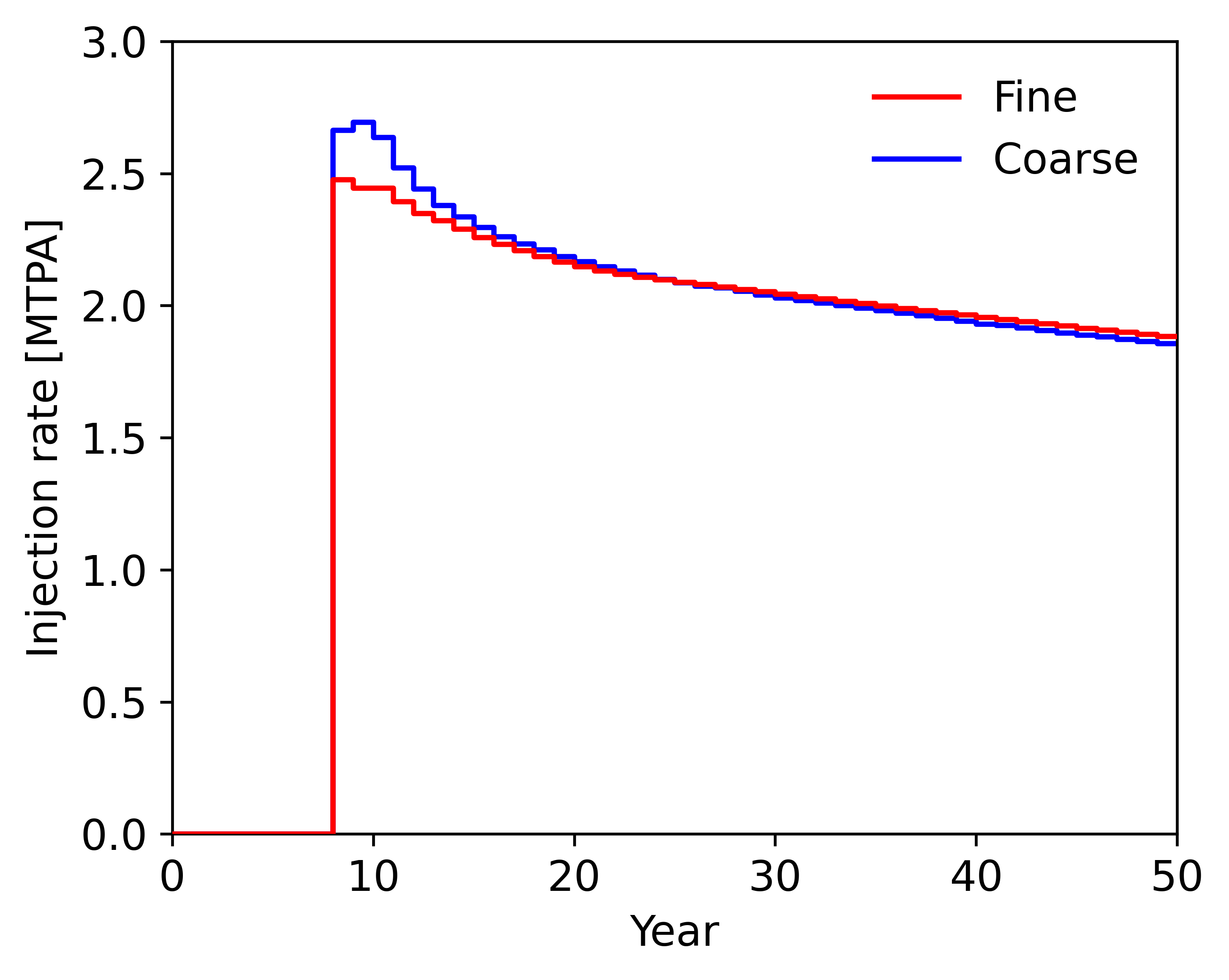}
        \caption{Well BK Fischer 1}
    \end{subfigure}
    \begin{subfigure}{0.48\textwidth}
        \centering
        \includegraphics[width=\textwidth]{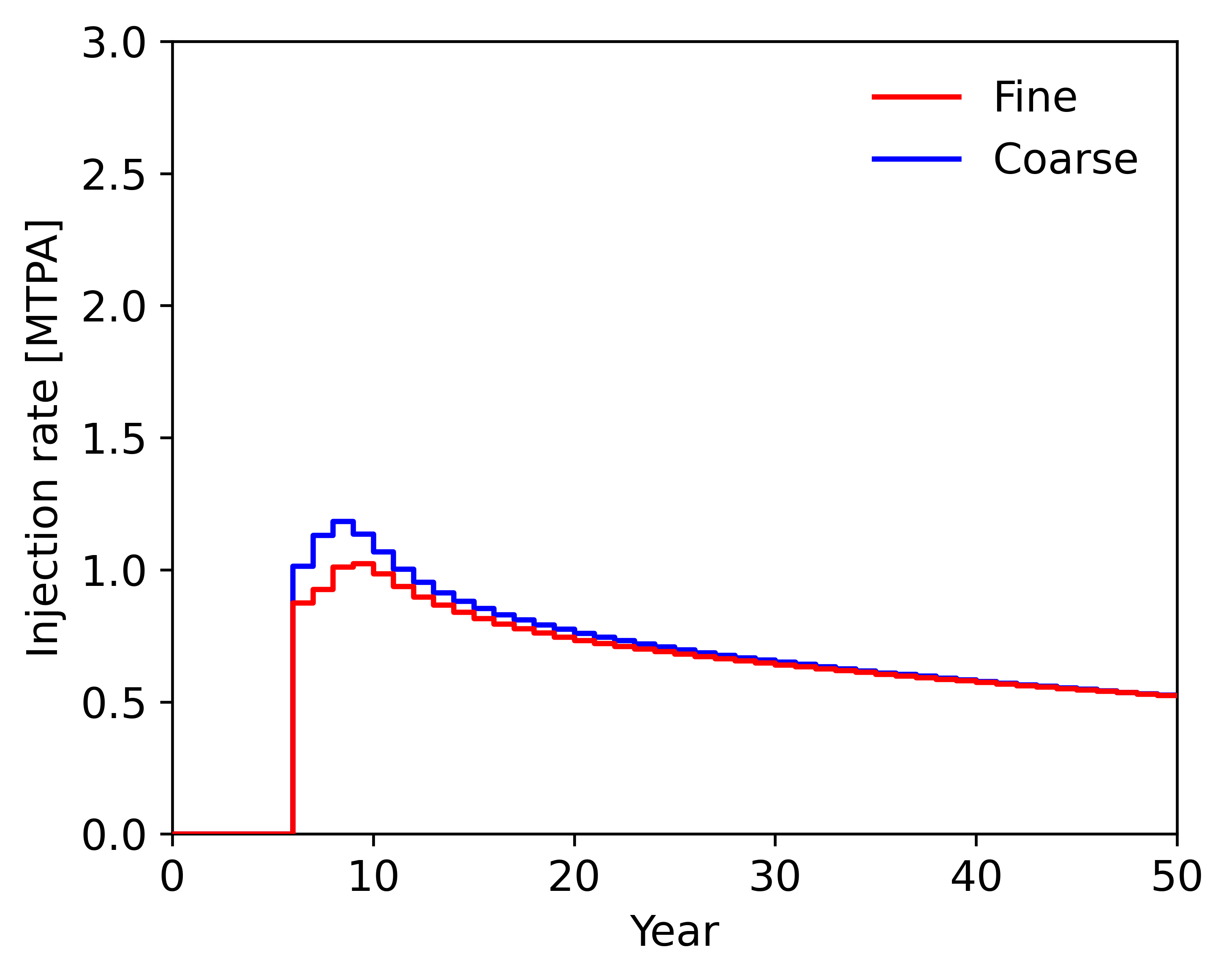}
        \caption{Well KJ Hintz 1}
    \end{subfigure}
    \begin{subfigure}{0.48\textwidth}
        \centering
        \includegraphics[width=\textwidth]{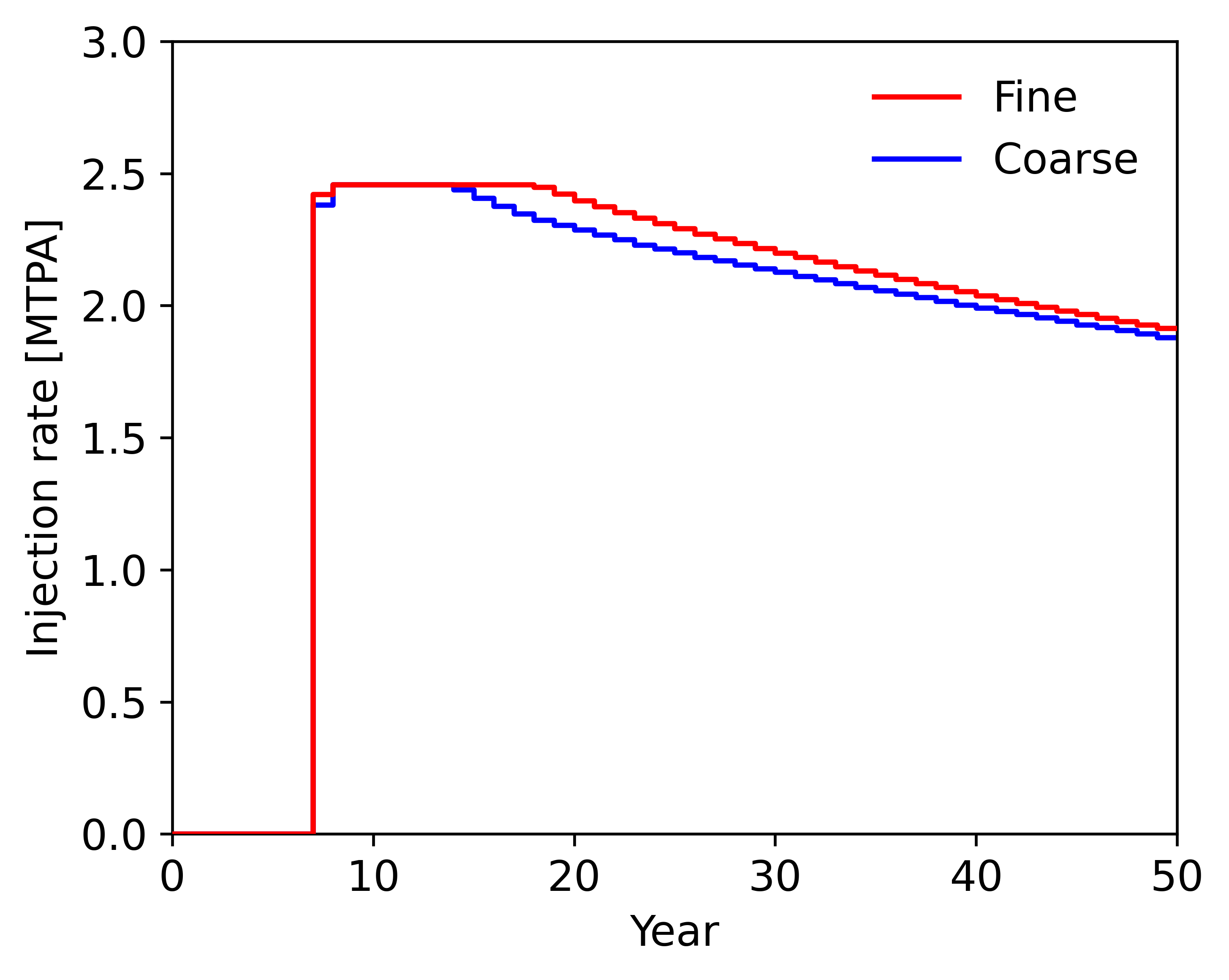}
        \caption{Well TB Leingang 1}
    \end{subfigure}
    \caption{Comparison of injection rates for the fine and coarse models. Basin-scale results and results for three injection wells from the SCS project are shown.}
    \label{fig:Rate comp for upscaling error}
\end{figure}

Finally, we compare injection rates from the fine and coarse-scale models. Results for the entire basin and for three individual wells are shown in Figure~\ref{fig:Rate comp for upscaling error}. The basin-wide results (Figure~\ref{fig:Rate comp for upscaling error}(a)) show excellent overall agreement. The relative error in the total CO$_2$ injected over 50~years is only 0.88\%, as noted above. The injection rates for the three wells (Figure~\ref{fig:Rate comp for upscaling error}(b)--(d)) show larger differences. These discrepancies, though relatively small, are due to errors in predicting the time for the switch from rate to BHP control and to errors in injection rate when the well is under BHP control. As we see from Figure~\ref{fig:Rate comp for upscaling error}(a), the well-by-well errors tend to cancel at the basin scale.

\section{Sensitivity Analysis}\label{sec:SensitivityAnalysis}

We now perform sensitivity analyses to quantify the uncertainties in the key flow responses discussed in Section~\ref{sec:UpscalingResult}. First, a global sensitivity analysis is performed using an ensemble generated by Latin Hypercube Sampling (LHS) to assess uncertainty in the QoIs. Next, the robustness of the model coarsening (nested LGRs and upscaling) procedure is assessed by comparing QoIs for several coarse and fine models. Finally, a surrogate model is trained using the ensemble employed in the global sensitivity analysis. A variance-based global sensitivity analysis is then conducted to quantify the contributions of the parameters and geological realizations to the variability of the QoIs. 

\subsection{Sensitivity Analysis Using Coarse Models}
\label{sec:sensitivity analysis using coarse model}
In this section, we perform a global sensitivity analysis using the coarsening strategy described earlier. We introduce geological uncertainty by considering 60 facies realizations (and corresponding porosity and permeability realizations) and by introducing uncertainty ranges for seven of the key parameters. The geostatistical realizations are generated using the procedures described in Sections~\ref{sec:model_domain} and~\ref{sec:PorosityPermeability} for the base case, i.e., within tNavigator, we use Sequential Indicator Simulation for the facies distribution and Sequential Gaussian Simulation for the porosity distribution. All realizations are conditioned to the same well data as the base case. 

The set of parameters now treated as uncertain, given in Table~\ref{tab:uncertainty_parameters} (along with their ranges) and referred to collectively as $\mathbf{h}$, are
\begin{equation} \label{eq:h}
    \mathbf{h} = [BV_\mathrm{south}, BV_\mathrm{west},k_\mathrm{over},k_\mathrm{under},(k_v/k_h)_{\mathrm{res}},n_{w,\mathrm{res}},n_{g,\mathrm{res}}]^T.
\end{equation}
Here ${BV_\mathrm{south}}$ and $BV_\mathrm{west}$ denote the bulk volumes of the boundary cells along the southern (South Dakota) and western (Montana) model boundaries. As described in Section~\ref{sec:SimSetting}, the bulk volumes in the base case represent the total volume of the original structural framework in North Dakota. However, because the structural framework is truncated at the North Dakota state boundary (Section~\ref{sec:structural_framework}), the base case does not include the extensions of the formation into South Dakota and Montana. To evaluate the impact of these regions, ${BV_\mathrm{south}}$ and $BV_\mathrm{west}$ are treated as uncertain parameters. Their minimum values are set equal to the base-case values because the uncertainty in the modeled North Dakota portion is considered to be relatively small. The upper bounds, by contrast, are chosen to represent the possible contribution from the formation extensions outside North Dakota. Because previous regional studies covering Montana and South Dakota commonly combine the Broom Creek Formation with other stratigraphic units and treat them collectively as the Minnelusa Group~\citep{GELMAN2023106390}, the volume of the Broom Creek Formation outside North Dakota is subject to substantial uncertainty. Therefore, we set the upper bounds to be 100~times the base-case values to represent large extensions outside North Dakota.

The quantities $k_\mathrm{over}$ and $k_\mathrm{under}$ denote the permeabilities of the overlying and underlying formations. Median values are used for base-case values as listed in Table~\ref{table:StatisticsPermeability}. Here we treat these parameters as uncertain because the core analysis data span a wide range. The upper and lower bounds of these parameters are determined from the core analysis results summarized in Table~\ref{table:StatisticsPermeability}. However, the measured maximum and minimum permeabilities are not directly used as the parameter bounds.
Given the existence of layering in the overlying and underlying formations, the effective permeability of these regions would not correspond to the highest-measured permeability. The upper bound is, therefore, conservatively set to $10^{-2}$~mD. This value is lower than the highest-measured permeability in the overlying and underlying formations, and corresponds to the highest permeability reported for caprock in previous CCS studies~\citep{https://doi.org/10.1002/wrcr.20197}. 
The minimum measured permeabilities are $4.39\times10^{-5}$~mD and $8.0\times10^{-5}$~mD for the overlying and underlying formations, respectively. However, several core permeabilities reached the lowest measurable limit of the laboratory analyses.
Consequently, the reported minimum permeabilities are likely constrained by the measurement limits rather than representing the true permeability. Therefore, a lower bound of $10^{-6}$~mD is used for both formations.

The base-case value of the anisotropy ratio of the reservoir rock, $(k_v/k_h)_\mathrm{res}$, is taken as the median of the available core analysis, as described in Section~\ref{sec:PorosityPermeability}. However, only a total of six measurements of permeability anisotropy are available, and only three of these are for reservoir rock (see Figure~S4 in Supplementary Material). This limited dataset indicates considerable uncertainty in the anisotropy ratio. Here we take the lower and upper bounds to be the minimum and maximum measured values.

Lastly, $n_{w,\mathrm{res}}$ and $n_{g,\mathrm{res}}$ denote the model parameters for relative permeability curves for the reservoir rock. The base-case values are taken from sandstone parameters adopted in the Storage Facility Permit (SFP) reports of four projects, which correspond to the reservoir rock considered in this study~\citep{NDIC2022DGC, NDIC2023BF, NDIC2023WestTundra, NDIC2024SCS1}. Although the same parameter values were employed across all four projects, this does not necessarily imply that the uncertainty associated with these parameters is small. For example, the DGC project study also involved numerical simulations using relative permeability curves derived from site-specific core samples. The parameter values used in the DCCE project, for which an SFP had already been granted~\citep{NDIC2022DGC}, were ultimately adopted, however. This was because the site-specific DGC relative permeability curves resulted in smaller predicted CO$_2$ plume extents, so the DCCE parameter set (which resulted in larger plumes) was adopted as a more conservative basis for permitting. This example suggests that the widespread use of the base-case parameter values may partly reflect a preference for conservative plume predictions in regulatory applications. Numerous studies have reported the influence of relative permeability curves on plume migration~\citep{SARKARFARSHI201461, YOSHIDA2016161}, so these are important model parameters. Since site-specific data for the Broom Creek Formation appear to be insufficient to constrain their uncertainty, the upper and lower bounds reported by~\citet{Gao2024GHGT}, which were derived from a compilation of datasets from multiple fields, are used here.

A total of 4400 models are generated using Latin Hypercube Sampling (LHS)~\citep{McKay1979}, as implemented in SciPy~\citep{Virtanen2020}. The parameters $k_{\mathrm{over}}$, $k_{\mathrm{under}}$, and $(k_v/k_h)_\mathrm{res}$ are sampled uniformly in logarithmic space, and all other parameters are sampled uniformly in linear space. Each sample of $\mathbf{h}$ is then combined with one of the 60 (randomly sampled) facies and porosity realizations. The coarse model for each sample--realization pair is then simulated. As noted earlier, the nested LGR configuration and the upscaling parameters determined for the base case ($\omega_x=\omega_y=0.61$, $\omega_z=0.31$) are applied in all 4400 simulations. The QoIs considered in Section~\ref{sec:UpscalingResult}, namely (1)~total CO$_2$ injected over 50~years from all five projects, (2)~total plume area at 50~years, and (3)~time-varying injection rate at each well, are then evaluated.

\begin{table}[htbp]
    \centering
    \caption{Ranges of uncertain parameters used in global sensitivity analysis.}
    \label{tab:uncertainty_parameters}
    \begin{tabular}{lccc}
    \toprule
    Parameter & Base case & Min & Max \\
    \midrule
    $BV_{\mathrm{south}}$ & $1.36 \times 10^{12}\ \mathrm{m^3}$ & $1.36 \times 10^{12}\ \mathrm{m^3}$ & $1.36 \times 10^{14}\ \mathrm{m^3}$ \\
    
    $BV_{\mathrm{west}}$ & $1.27 \times 10^{12}\ \mathrm{m^3}$ & $1.27 \times 10^{12}\ \mathrm{m^3}$ & $1.27 \times 10^{14}\ \mathrm{m^3}$ \\
    
    $k_{\mathrm{over}}$ & $7.78 \times 10^{-3}\ \mathrm{mD}$ & $1.0 \times 10^{-6}\ \mathrm{mD}$ & $1.0 \times 10^{-2}\ \mathrm{mD}$ \\
    
    $k_{\mathrm{under}}$ & $1.06 \times 10^{-3}\ \mathrm{mD}$ & $1.0 \times 10^{-6}\ \mathrm{mD}$ & $1.0 \times 10^{-2}\ \mathrm{mD}$ \\
    
    $(k_v/k_h)_\mathrm{res}$ & $4.5 \times 10^{-2}$ & $1.5 \times 10^{-3}$ & $6.41 \times 10^{-1}$ \\
    
    $n_{w,\mathrm{res}}$ & 5.8 & 3.5 & 7.0 \\
    $n_{g,\mathrm{res}}$ & 2.2 & 1.5 & 3.0 \\
    \bottomrule
    \end{tabular}
\end{table}

\begin{figure}[h!]
    \centering
    \begin{subfigure}{0.45\textwidth}
        \centering
        \includegraphics[width=\textwidth]{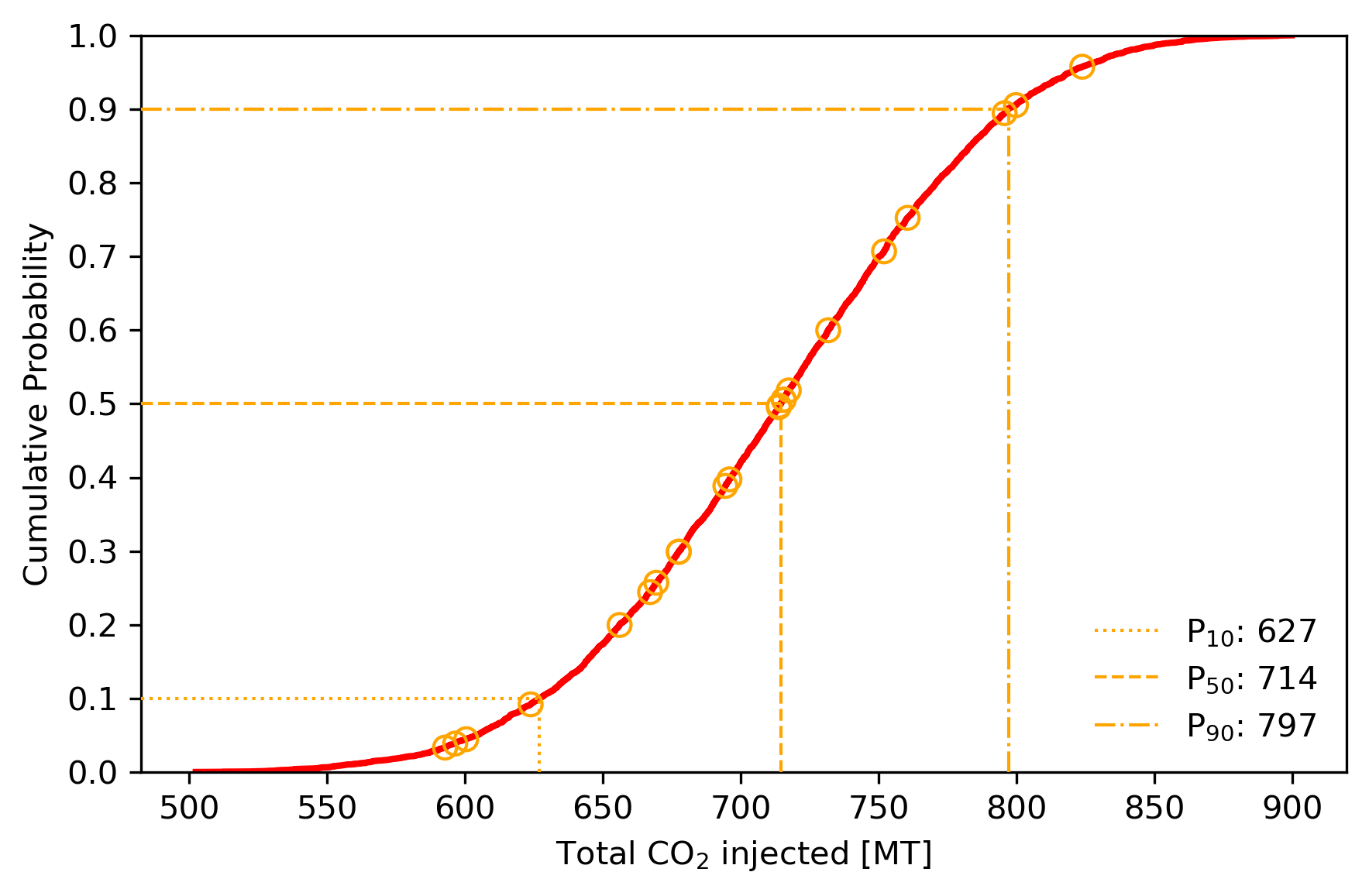}
        \caption{Total CO$_2$ injected}
    \end{subfigure}
    \begin{subfigure}{0.45\textwidth}
        \centering
        \includegraphics[width=\textwidth]{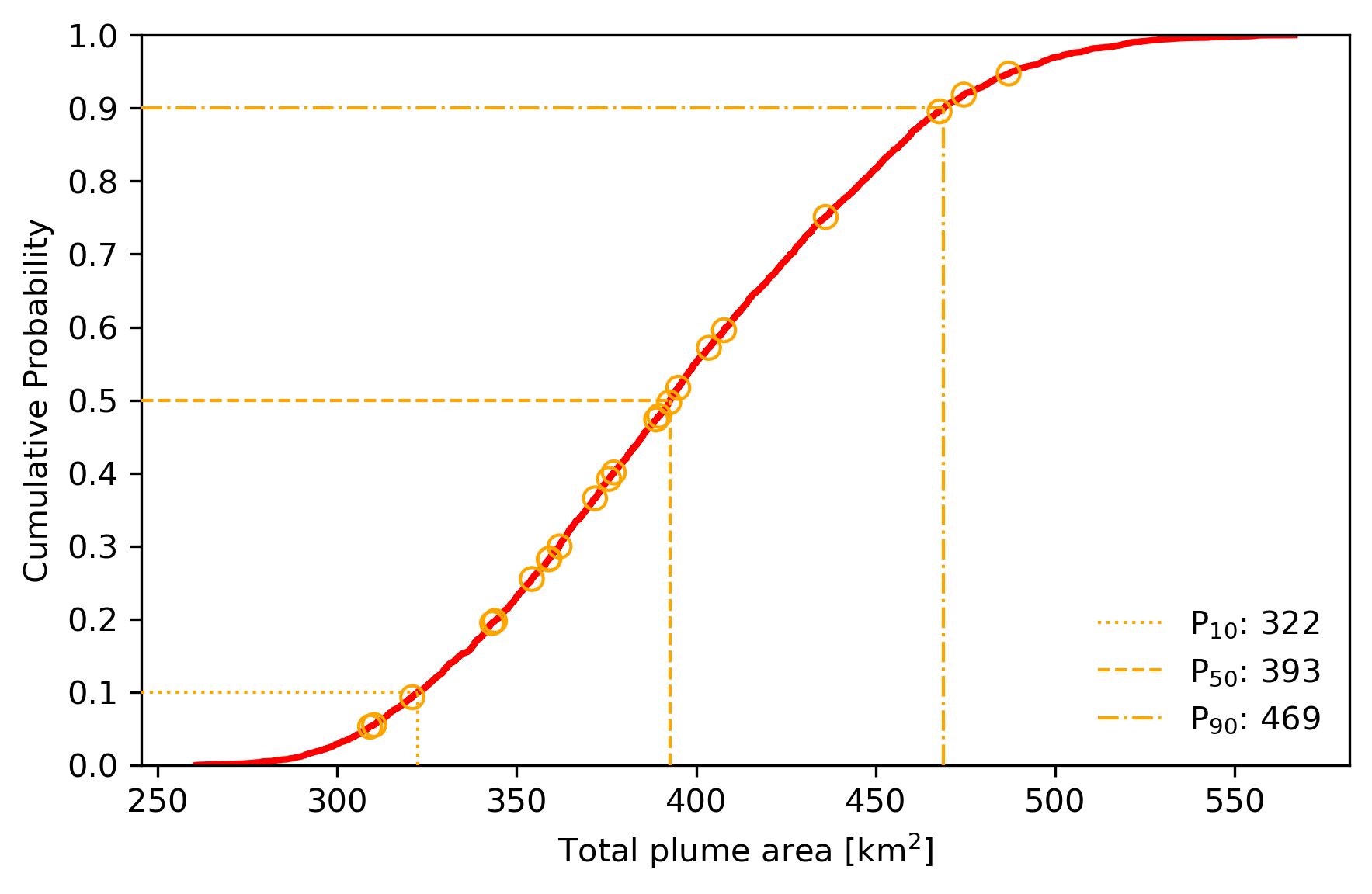}
        \caption{Total plume area}
    \end{subfigure}
    \caption{Cumulative distribution functions for total CO$_2$ injected and plume area at 50~years.  Circles show the cases for which fine-scale simulations are conducted to evaluate upscaling error (shown in Figures~\ref{fig:Sg comp sensitivity} and~\ref{fig:upscaling error}).}
    \label{fig:CDF sensitivity}
\end{figure}

We now present the results from this global sensitivity assessment. In Figure~\ref{fig:CDF sensitivity} we show the cumulative probabilities of the total CO$_2$ injected and the plume area at 50~years. The dashed orange lines indicate the P$_{10}$, P$_{50}$, and P$_{90}$ percentiles, and the corresponding QoI values are noted on the plots. Summary statistics along with base-case results are listed in Table~\ref{tab:statistics sensitivity}. Substantial uncertainty ranges are observed for both QoIs. For the total CO$_2$ injected, the P$_{10}$--P$_{90}$ range is 170~MT. This range corresponds to 24\% of the median response (714~MT). The uncertainty is even greater for the total plume area, for which the P$_{10}$--P$_{90}$ range corresponds to 37\% of the median value. 

\begin{table}[t]
    \centering
    \caption{Statistical values of QoIs from the global sensitivity analysis.}
    \label{tab:statistics sensitivity}
    \begin{tabular}{lcc}
    \toprule
     & Total CO$_2$ injected & Total plume area \\
    \midrule
    Base case & 728.8 MT& 434.5 km$^2$\\
    Mean & 713.3 MT& 395.2 km$^2$\\
    Standard deviation & 65.7 MT & 55.4 km$^2$ \\
    P$_{10}$--P$_{90}$ range & 170.0 MT& 146.4 km$^2$ \\
    (P$_{90}$--P$_{10}$)/P$_{50}$ & 23.8\% & 37.3\% \\
    \bottomrule
    \end{tabular}
\end{table}

\begin{figure}[h!]
    \centering
    \begin{subfigure}{0.32\textwidth}
        \centering
        \includegraphics[height=3.6cm]{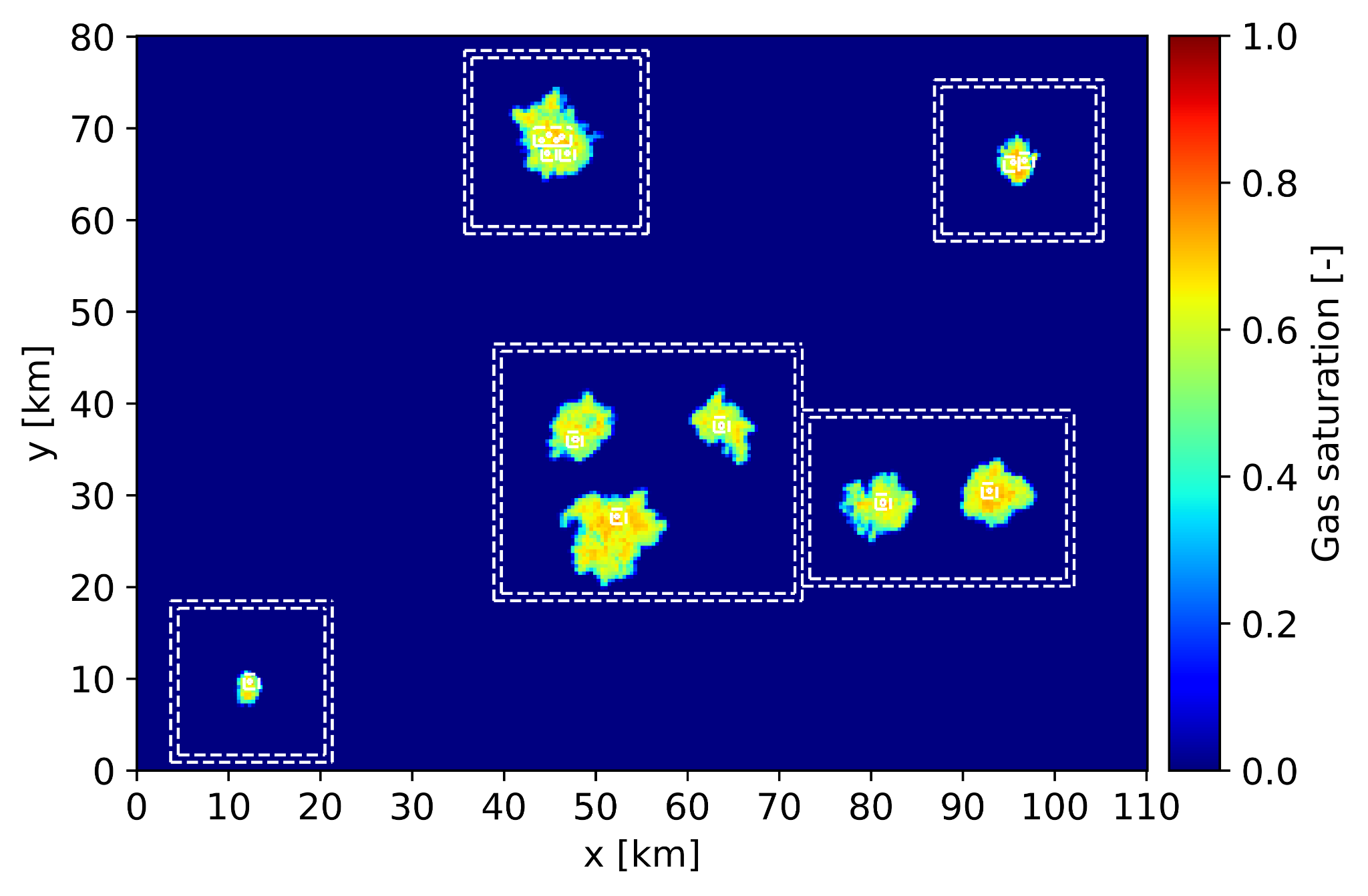}
        \caption{Fine model, P$_{10}$}
    \end{subfigure}
    \begin{subfigure}{0.32\textwidth}
        \centering
        \includegraphics[height=3.6cm]{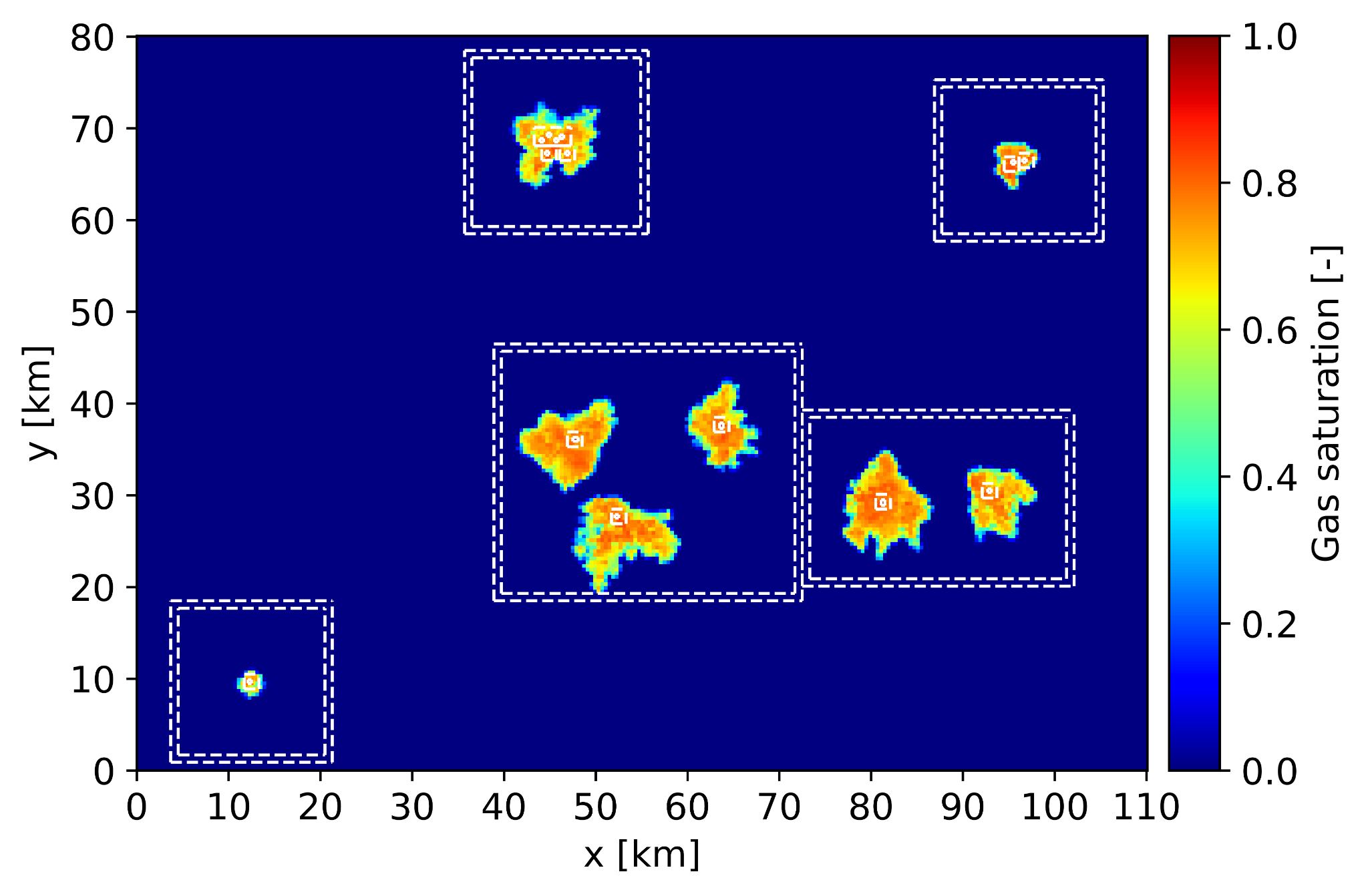}
        \caption{Fine model, P$_{50}$}
    \end{subfigure}
    \begin{subfigure}{0.32\textwidth}
        \centering
        \includegraphics[height=3.6cm]{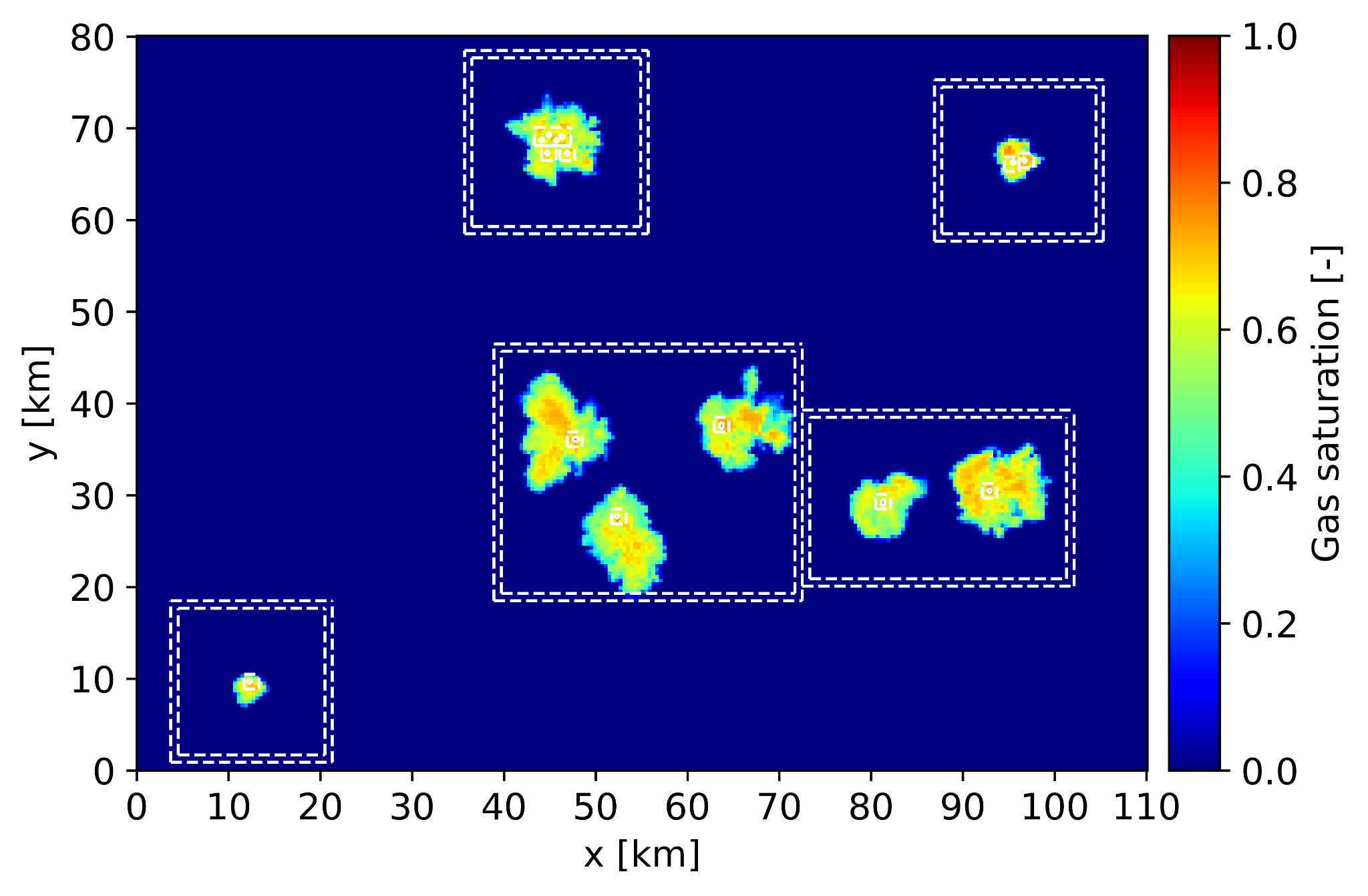}
        \caption{Fine model, P$_{90}$}
    \end{subfigure}
    \begin{subfigure}{0.32\textwidth}
        \centering
        \includegraphics[height=3.6cm]{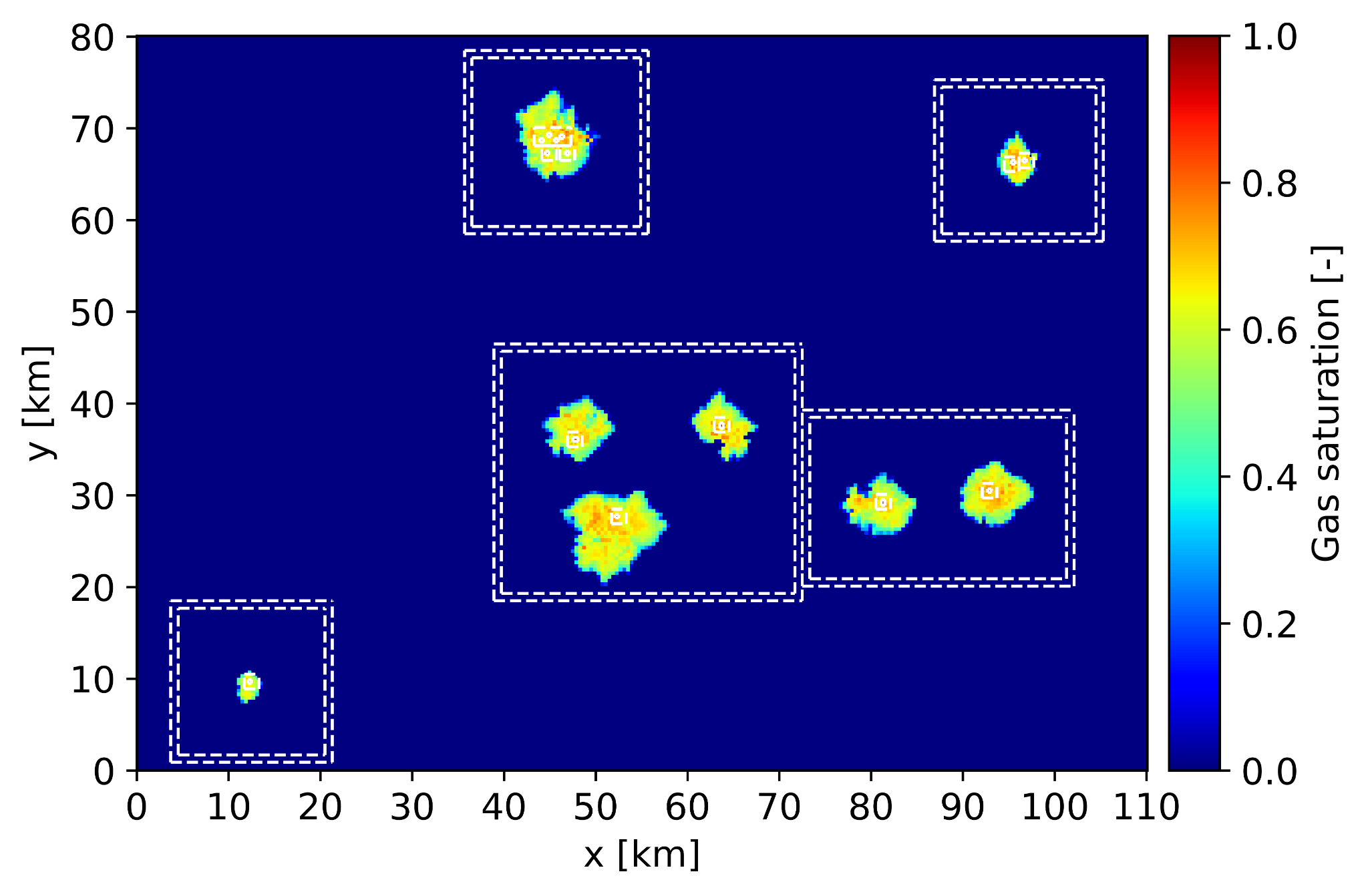}
        \caption{Coarse model, P$_{10}$}
    \end{subfigure}
    \begin{subfigure}{0.32\textwidth}
        \centering
        \includegraphics[height=3.6cm]{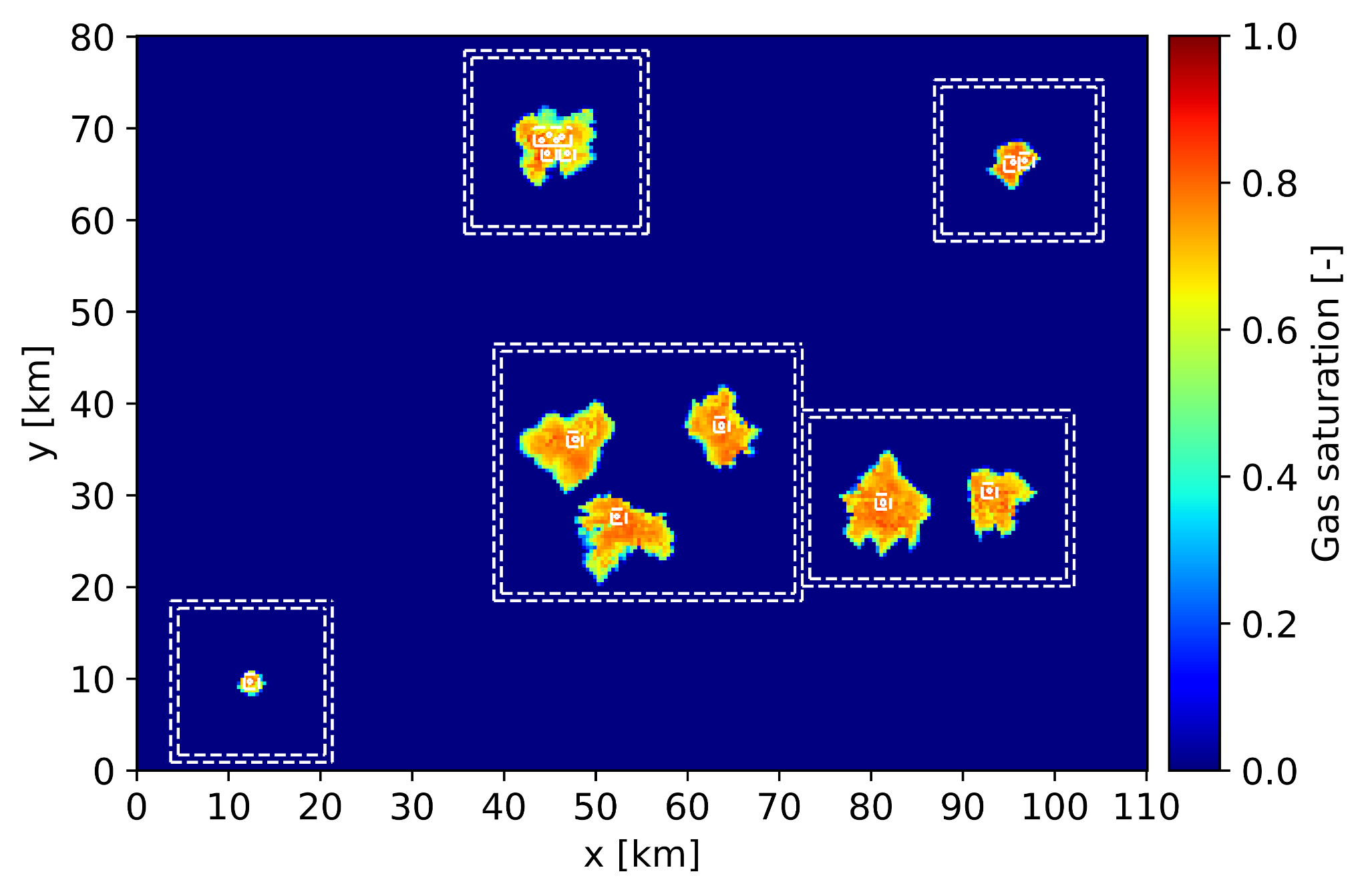}
        \caption{Coarse model, P$_{50}$}
    \end{subfigure}
    \begin{subfigure}{0.32\textwidth}
        \centering
        \includegraphics[height=3.6cm]{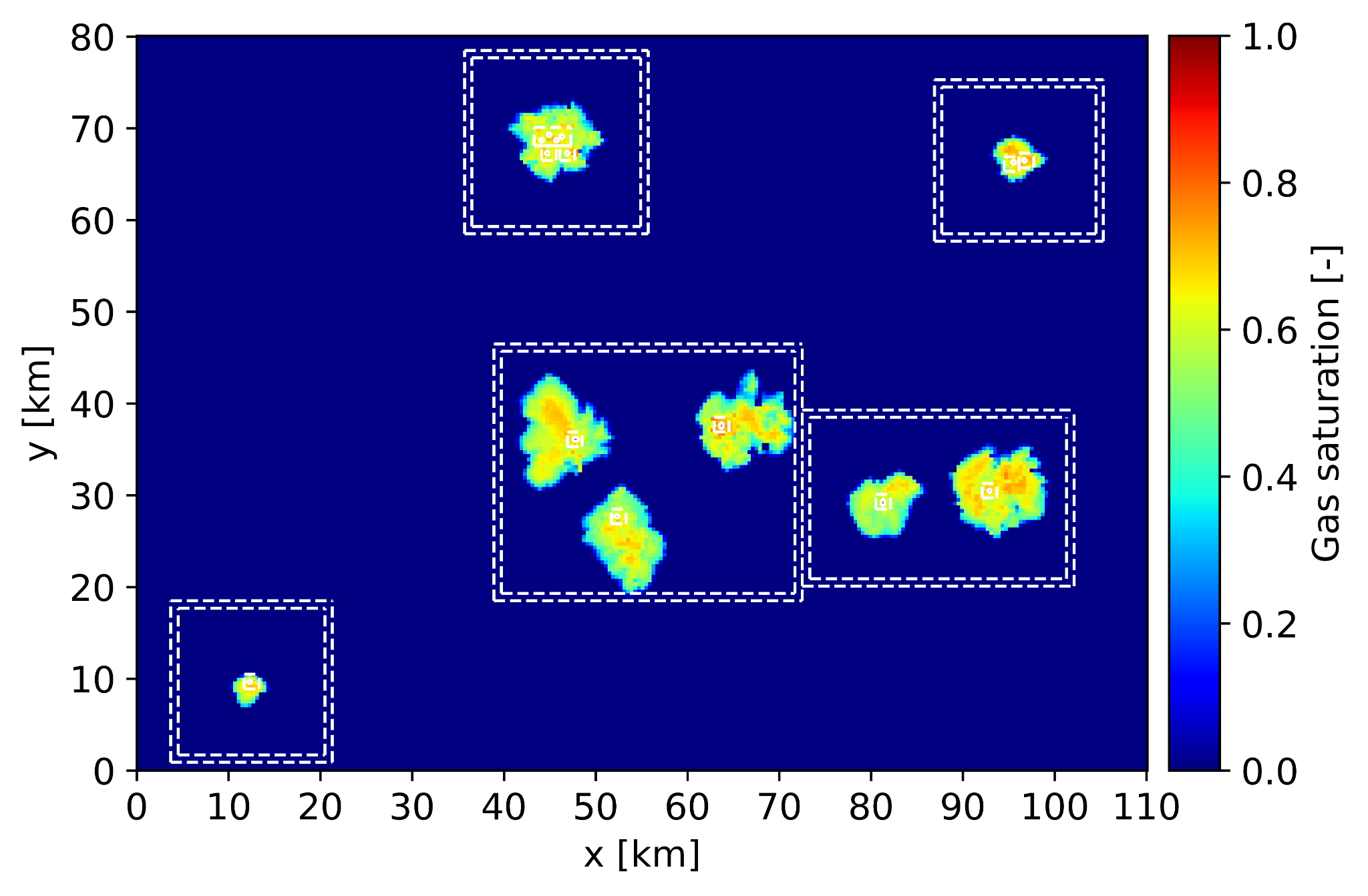}
        \caption{Coarse model, P$_{90}$}
    \end{subfigure}
    \caption{CO$_2$ plume distributions at 50~years for models selected near the P$_{10}$, P$_{50}$, and P$_{90}$ values of the total plume area shown in Figure~\ref{fig:CDF sensitivity}(b). Fine model results are on the top row, and coarse model results are on the bottom row.}
    \label{fig:Sg comp sensitivity}
\end{figure}

We now perform fine-scale simulations for a small subset of the 4400 cases considered in the sensitivity analysis. Recall that each fine-scale run requires more than 2~days of simulation on 32~CPU cores, so the models considered in this assessment should be selected carefully. To achieve representative results, we simulate a set of 29~models (indicated by the orange circles) that span the range of the QoI cumulative distribution functions (CDFs) shown in Figure~\ref{fig:CDF sensitivity}. In Figure~\ref{fig:Sg comp sensitivity} we compare the CO$_2$ plume distributions at 50~years, for fine and coarse models, for cases located near the P$_{10}$, P$_{50}$, and P$_{90}$ plume areas shown in Figure~\ref{fig:CDF sensitivity}(b). These figures are generated using the same procedure as was used for Figure~\ref{fig:Sg comp for upscaling err}. The plume geometries in Figure~\ref{fig:Sg comp sensitivity} differ from the base-case results in Figure~\ref{fig:Sg comp for upscaling err} due to differences in the model parameters and geological realizations. Nonetheless, in all three cases, the coarse models accurately reproduce the CO$_2$ distributions observed in the fine model.

\begin{figure}[h!]
    \centering
    \includegraphics[width=0.6\textwidth]{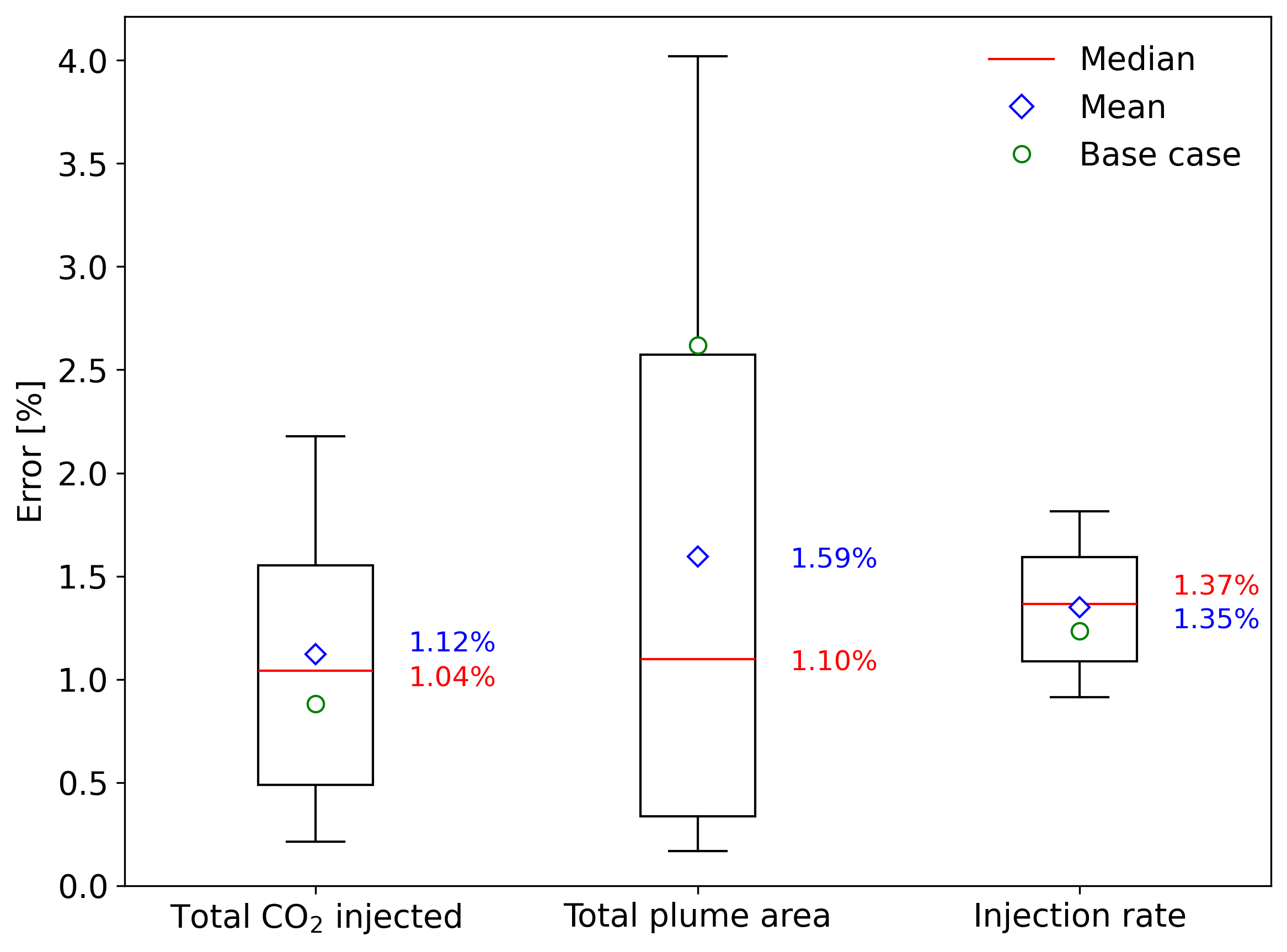}
    
    \caption{Box plots of the upscaling errors for QoIs: total CO$_2$ injected and total plume area at 50 years, and yearly injection rate over all wells. Errors correspond to the three terms in Eq.~\ref{eq:objective_function}. Boxes display P$_{10}$, P$_{25}$, P$_{50}$, P$_{75}$, and P$_{90}$ errors.}
    \label{fig:upscaling error}
\end{figure}

Upscaling errors for the three QoIs, computed from coarse and fine model results for the 29~cases, are now evaluated. Box plots for these errors, which correspond to the three terms in Eq.~\ref{eq:objective_function}, are shown in Figure~\ref{fig:upscaling error}. The open green circles represent the error for the base-case model, while the red lines and blue diamonds denote the median and mean errors across the 29 validation cases, respectively. For the three QoIs, both the mean and median upscaling errors range from 1.04\% to 1.59\%, indicating that the prediction errors introduced by the upscaling procedure remain consistently small. This is an important observation since the nested LGR structure and the upscaling parameters are fixed at the settings determined for the base case. In addition, the errors for the base case are well within the distribution of the errors for the full set. They are slightly below the median error for total CO$_2$ injected and injection rate, but well above the median for total plume area. In total, these error results demonstrate that the proposed nested LGR gridding and upscaling workflow is robust over a wide range of parameter combinations and geological realizations. This finding will be very useful in our future work on well placement optimization for new projects in the Broom Creek Formation.

\subsection{Global Sensitivity Analysis Using a Surrogate Model}\label{sec:sensitivity analysis using surrogate model}

We now conduct an additional global sensitivity analysis to quantify the contributions of individual parameters and geological realizations to QoI variability. Because this assessment requires a very large number of function evaluations, we first construct a surrogate model using the 4400 coarse-model simulation results. Variance-based global sensitivity analysis is then performed with this surrogate.

Here we consider two QoIs -- total CO$_2$ injected and total plume area at 50~years, denoted as $\mathbf{Y}\in \mathbb{R}^2$. These are obtained by performing flow simulations using the coarse model. We express this process as
\begin{equation}
    \mathbf{Y}=[Q^c,\;A^c]^T = f(\mathbf{h},r),
\end{equation}
where $f$ represents the flow simulation and $r$ denotes a particular realization consisting of a facies distribution and associated porosity and permeability fields. Because we consider a discrete set of realizations, we view $r$ as an index, with $r \in \{1, \ldots, 60\}$.

The surrogate model prediction, denoted $\hat{f}$, is expressed as
\begin{equation}
    \hat{\mathbf{Y}}=[\hat{Q}^c,\;\hat{A}^c]^T = \hat{f}(\mathbf{h},r;\mathbf{w}).
\end{equation}
Here $\hat{Q}^c$ and $\hat{A}^c$ denote the surrogate-predicted total CO$_2$ injected and plume area at 50~years, and $\mathbf{w}$ represents the neural network parameters for the surrogate model. We use a fully connected neural network, shown in Figure~S8 and further defined in Table~S2 in Supplementary Material, for the surrogate model. The input variables are the seven parameters in $\mathbf{h}$ and the realization index $r$. The index $r$ is first mapped to a continuous vector space through an embedding layer, after which the embedded representation is concatenated with $\mathbf{h}$ to form the input to the neural network. 

For training purposes, the QoIs are first standardized by subtracting the training set mean and dividing by the standard deviation. The surrogate model is then trained by minimizing the mean squared error (MSE) between the standardized predictions and targets.
The dataset (comprising the 4400~samples described above) is randomly divided into training and validation sets with a ratio of 80:20. The Adam optimizer~\citep{kingma2015adam} is used, with the initial learning rate set to $1.0\times10^{-2}$ and reduced by a factor of 0.5 when the validation loss does not improve for 150~consecutive epochs. The minimum learning rate is $1.0\times10^{-6}$. Training is terminated if the validation loss does not improve for 300~consecutive epochs. The neural network parameters $\mathbf{w}$ corresponding to the minimum validation loss are retained for the surrogate model. The training process is performed on an NVIDIA A100 GPU and requires less than 3~minutes for 2826~training epochs.

Figure~S9 in Supplementary Material compares the surrogate predictions with the corresponding coarse-model simulations. The mean relative errors for the validation set are 0.11\% for the total CO$_2$ injected and 0.32\% for the plume area. These results demonstrate that the surrogate model accurately reproduces the coarse-model responses.

Variance-based global sensitivity analysis~\citep{SOBOL2001271} is now performed using the surrogate model.
We follow the general framework of \citet{HAN2026114801}, though the treatment of geological uncertainty is different. In \citet{HAN2026114801}, principal component analysis (PCA) was used to parameterize the geological realizations, and the resulting latent variables were collectively treated as a single uncertain variable. In our case, by contrast, we denote realizations by the categorical variable $r \in \{1, \ldots, 60\}$.
The uncertainty space, denoted by $\mathbf{X}$, is therefore given by
\begin{equation}
    \mathbf{X} = [\mathbf{h},\ r]^T =[BV_\mathrm{south}, BV_\mathrm{west},k_\mathrm{over},k_\mathrm{under},(k_v/k_h)_\mathrm{res},n_{w,\mathrm{res}},n_{g,\mathrm{res}},r]^T. 
\end{equation}

The first-order sensitivity index~\citep{SALTELLI2010259}, denoted $S_{i,j}$, quantifies the direct effect of each uncertain variable $X_i$, $i=1, \ldots, 8$, on the prediction variance of each QoI ($Y_1$ and $Y_2$). It is defined as
\begin{equation}
    S_{i,j} = \frac{\mathrm{Var}_{X_i}\left(\mathrm{E}_{\mathbf{X}_{\sim i}}(Y_j|X_i)\right)}{\mathrm{Var}_{\mathbf{X}}(Y_j)},
\end{equation}
where $\mathbf{X}_{\sim i}$ is the set of all uncertain variables except $X_i$. The conditional expectation, $\mathrm{E}_{\mathbf{X}_{\sim i}}(Y_j|X_i)$, is obtained by averaging $Y_j$ over all possible values of $\mathbf{X}_{\sim i}$ while holding $X_i$ fixed. The numerator measures the variance of this conditional expectation as $X_i$ varies. The denominator represents the total variance of $Y_j$ over the entire uncertainty space. 

The total-effect sensitivity index, denoted $S_{Ti,j}$, quantifies the total contribution of $X_i$ to the variance of the QoIs. Unlike the first-order sensitivity index, $S_{Ti,j}$ accounts for both the direct effect of $X_i$ and higher-order interaction effects involving $X_i$. It is defined as 
\begin{equation}
    S_{Ti,j} = \frac{\mathrm{E}_{\mathbf{X}_{\sim i}}\left(\mathrm{Var}_{X_i} \left(Y_j|\mathbf{X}_{\sim i}\right) \right)}{\mathrm{Var}_{\mathbf{X}}(Y_j)}.
\end{equation}
Here, $\mathrm{Var}_{X_i} \left(Y_j|\mathbf{X}_{\sim i}\right)$ denotes the variance of $Y_j$ over all possible values of $X_i$, holding $\mathbf{X}_{\sim i}$ fixed. The expectation, $\mathrm{E}_{\mathbf{X}_{\sim i}}\left(\mathrm{Var}_{X_i} \left(Y_j|\mathbf{X}_{\sim i}\right)\right)$, is obtained by averaging this conditional variance over all possible values of $\mathbf{X}_{\sim i}$.

A total of 655,360 samples of the set of uncertain variables $\mathbf{X}$ are generated using the Sobol sequence-based Saltelli sampling scheme~\citep{SALTELLI2010259}. For each sample, the total CO$_2$ injected and plume area are predicted using the surrogate model. This sample size is necessary to ensure convergence of the sensitivity indices. Convergence is considered to be achieved when the average absolute change in sensitivity indices, averaged over all uncertain variables, is less than 0.001.

\begin{figure}[h!]
    \centering
    \begin{subfigure}{0.47\textwidth}
        \centering
        \includegraphics[width=\textwidth]{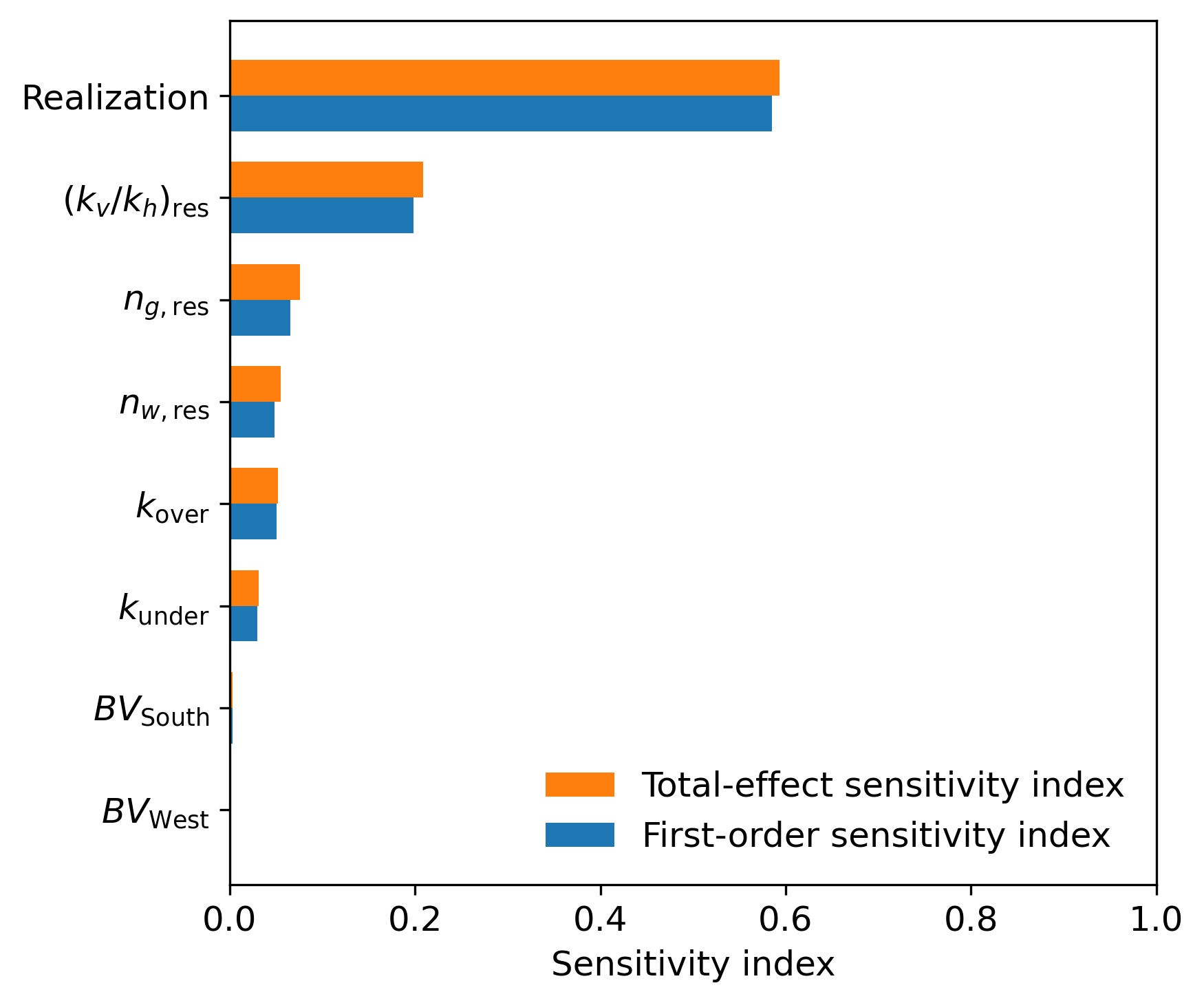}
        \caption{Total CO$_2$ injected}
    \end{subfigure}
    \begin{subfigure}{0.47\textwidth}
        \centering
        \includegraphics[width=\textwidth]{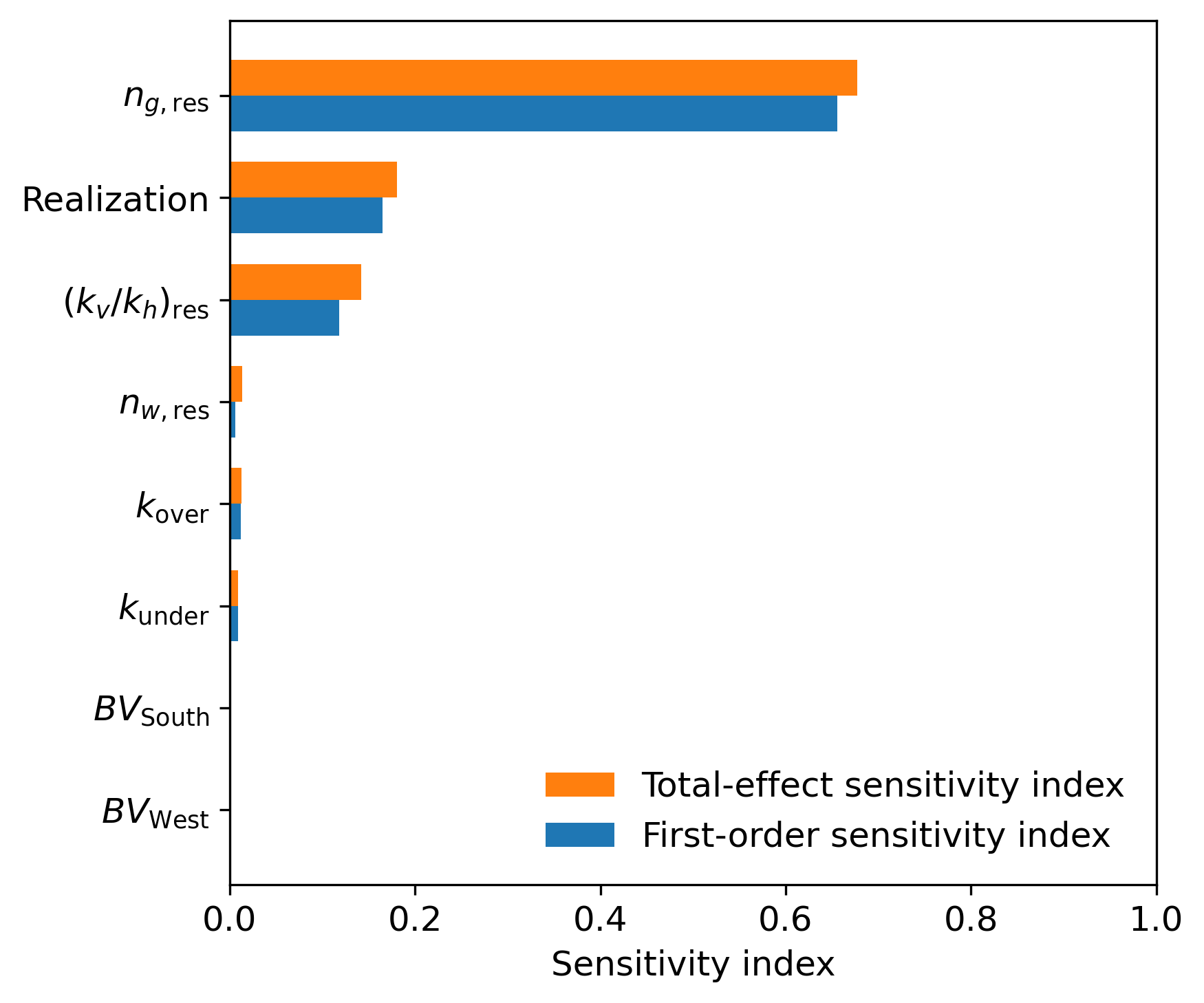}
        \caption{Total plume area}
    \end{subfigure}
    \caption{First-order and total-effect sensitivity indices for parameters and realizations on total CO$_2$ injected and total plume area at 50~years.}
    \label{fig:sobol index}
\end{figure}

The first-order sensitivity indices for total CO$_2$ injected are shown in blue in Figure~\ref{fig:sobol index}(a). Total CO$_2$ injected is most sensitive to the geological realization, which accounts for 58.5\% of the total variance. This high sensitivity is likely due to uncertainty in the facies distribution around the planned injection wells. Because high target injection rates are specified for these wells (see Table~\ref{tab:injection_schedule}), uncertainty in geological properties can have a large impact on injectivity and larger-scale pressure effects. 
Injection is seen to be somewhat sensitive to (in decreasing order) $(k_v/k_h)_{\mathrm{res}}$, $n_{g,\mathrm{res}}$, $n_{w,\mathrm{res}}$, $k_{\mathrm{over}}$, and $k_{\mathrm{under}}$. Sensitivity to the bulk volumes of the southern and western boundaries, $BV_{\mathrm{south}}$ and $BV_{\mathrm{west}}$, is negligible. This is likely because the minimum values of these parameters, which correspond to the total bulk volume of the Broom Creek Formation in North Dakota (used in the base case), are sufficiently large that further increases have almost no effect on the total CO$_2$ injected. The total-effect sensitivity indices (orange bars) are only slightly larger than the corresponding first-order indices. Moreover, the sum of the first-order sensitivity indices for all eight uncertain variables is 0.98, indicating that 98\% of the total variance can be explained by the additive effects of the individual input variables, while only 2\% is attributable to higher-order interactions. 

Sensitivity indices for the total plume area are shown in Figure~\ref{fig:sobol index}(b). The plume area is most sensitive to $n_{g,\mathrm{res}}$, which accounts for 65.6\% of the total output variance. This strong sensitivity is due to both mobility effects and the gas-spreading (shock--rarefaction) behavior of the gas-water system. Specifically, increasing $n_{g,\mathrm{res}}$ acts to reduce gas mobility and injectivity, thus potentially limiting the amount of CO$_2$ injected~\citep{YOSHIDA2016161}. In addition, for a given volume of injected CO$_2$, the value of $n_{g,\mathrm{res}}$ can have a strong impact on the saturation field, with larger $n_{g,\mathrm{res}}$ values resulting in less dispersed plumes. Thus, we see that $n_{g,\mathrm{res}}$, along with the realization and $(k_v/k_h)_{\mathrm{res}}$, account for 93.9\% of the total variance. As with the total CO$_2$ injected, the total-effect sensitivity indices are only slightly larger than the corresponding first-order indices, indicating that the interaction effects are small. Accordingly, the sum of the first-order sensitivity indices for all eight uncertain variables is 0.966.

This type of variance-based sensitivity analysis would be prohibitively expensive using numerical flow simulation, even with the coarse models constructed in this study. Specifically, to perform 655,360 numerical simulations with coarse models, each of which needs about 605~seconds of simulation time, would require about 12.6~years of serial computation (parallelization would of course reduce the elapsed time). By contrast, the surrogate model requires only 0.22~seconds to perform the same 655,360 evaluations using a batch size of 10,000 on an NVIDIA A100 GPU. The serial time required to simulate the 4400 training samples is about 739~hours ($\sim$31~days). These runs are performed in parallel, so the elapsed time is less than a week. Finally, as noted earlier, the training time for this network is less than 3~minutes.

\section{Quantification of Inter-Project Interference Effects}
\label{sec:single-project}
In previous sections, we developed and applied a unified model that captures the interactions between all active and planned projects in the Broom Creek Formation. In this section we assess the impact of these basin-scale interactions by conducting separate (standalone) simulations for each project, with all other projects eliminated. We first describe the setup of the standalone models and then compare the single-project results to those of the unified model.

Separate single-project models are constructed for each of the five projects. For each of these standalone studies, 300~cases are randomly sampled from the set of 4400~models employed in the global sensitivity analysis described in Section~\ref{sec:sensitivity analysis using coarse model}. The same nested LGRs and upscaling parameters as were used in all previous runs are used in these assessments. Therefore, the only difference in the setup between the single-project models and the unified model is that, in the single-project models, CO$_2$ injection from the other projects is set to zero. This simulation setup is similar to that previously used in North Dakota modeling, where numerical simulations submitted as part of the Storage Facility Permit (SFP) applications typically consider only the project for which the permit is sought~\citep{NDIC2022DGC, NDIC2023BF, NDIC2023WestTundra, NDIC2024SCS1}. In the DCCW SFP, the numerical simulation also included the DCCE project to account for pressure interference between the two projects~\citep{NDIC2021EastTundra, NDIC2023WestTundra}. However, as described in Section~\ref{sec:geol_setting}, DCCW and DCCE are both developed by the same entity. Therefore, although some pressure interference effects were considered, the simulations did not account for interference with projects run by other operators. Similarly, the SCS 1--3 SFP applications considered pressure interference only among projects developed by the same entity, not those of other operators~\citep{NDIC2024SCS1, NDIC2024SCS2, NDIC2024SCS3}.

We now compare the simulation results obtained from the single-project models and the unified model. Results for the total CO$_2$ injected at 50~years for each project, under both modeling approaches, are provided in Table~\ref{tab:SvsP_injected_co2}. 
Values are reported as mean $\pm$ standard deviation computed over the 300~cases considered. 
The total CO$_2$ injected is identical for the BF, DGC, and RTE projects in both the unified and single-project models. This is because their target injection rates are relatively low (0.18--0.36~MTPA, as listed in Table~\ref{tab:injection_schedule}), so the BHPs do not reach the maximum allowable pressure, even in the unified model. Substantial reduction in the total CO$_2$ injected is observed for the DCC and SCS projects, however, indicating significant inter-project interference. In particular, the mean injected CO$_2$ for the DCC project decreases from 270.3~MT in the single-project model to 195.0~MT in the unified model, corresponding to a reduction of 75.3~MT. This occurs because the DCC (and SCS) projects have high target rates, so maximum BHP constraints are more likely to be reached in the unified model, which acts to limit total injection. 

\begin{figure}[h!]
    \centering

    \begin{subfigure}{0.45\textwidth}
        \centering
        \includegraphics[width=\textwidth]{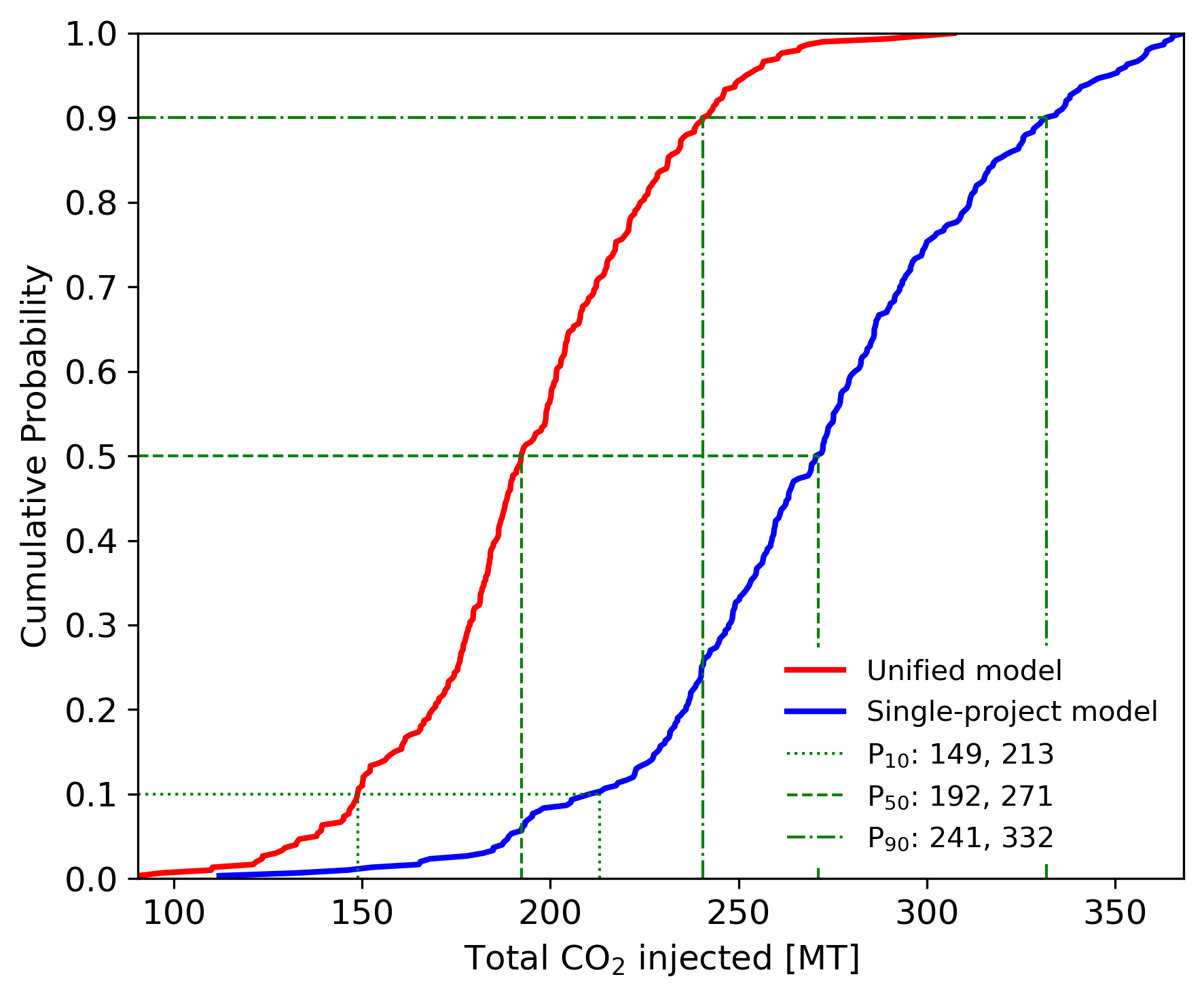}
        \caption{Total CO$_2$ injected in the DCC project}
    \end{subfigure}
    \hfill
    \begin{subfigure}{0.45\textwidth}
        \centering
        \includegraphics[width=\textwidth]{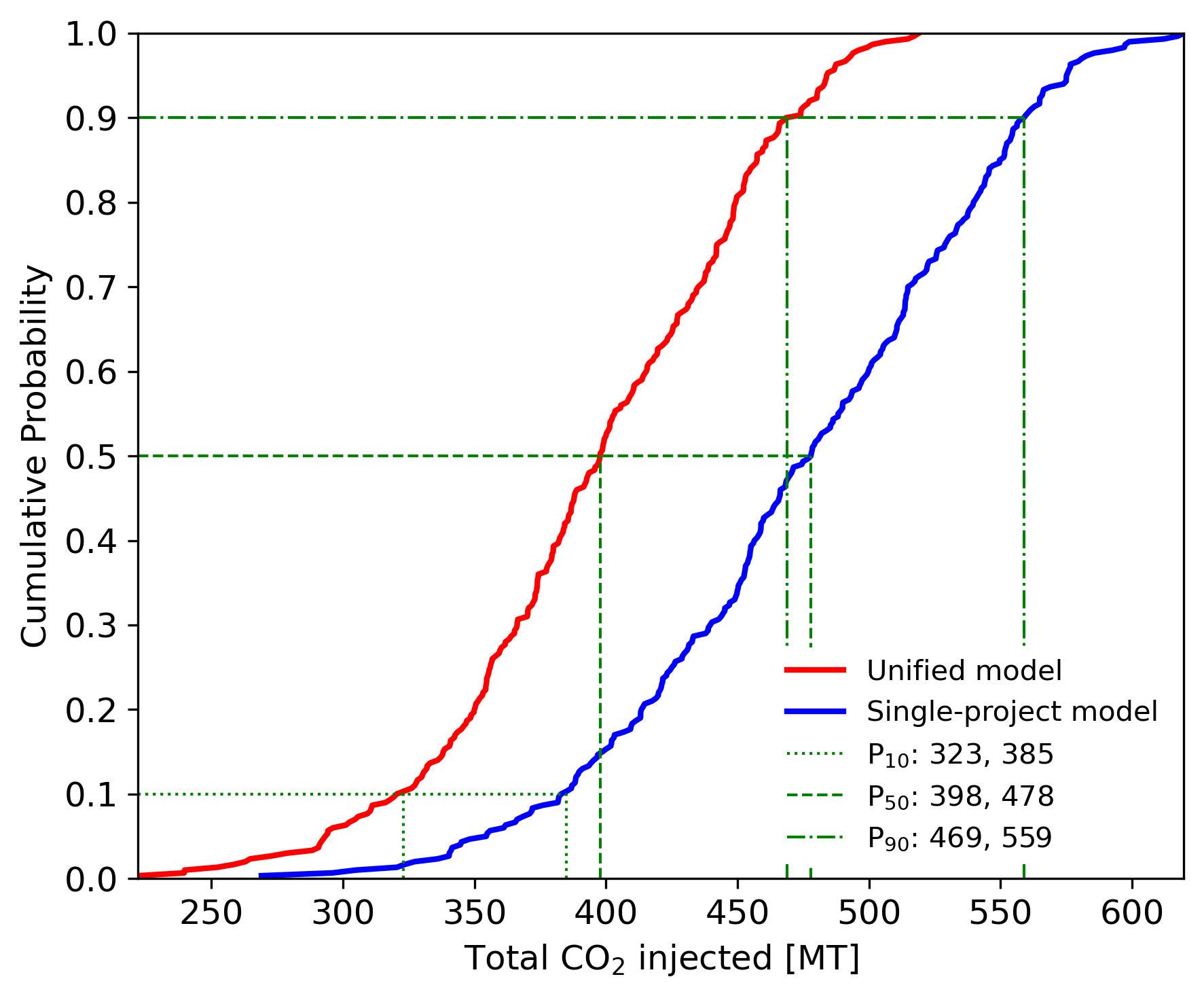}
        \caption{Total CO$_2$ injected in the SCS project}
    \end{subfigure}

    \begin{subfigure}{0.45\textwidth}
        \centering
        \includegraphics[width=\textwidth]{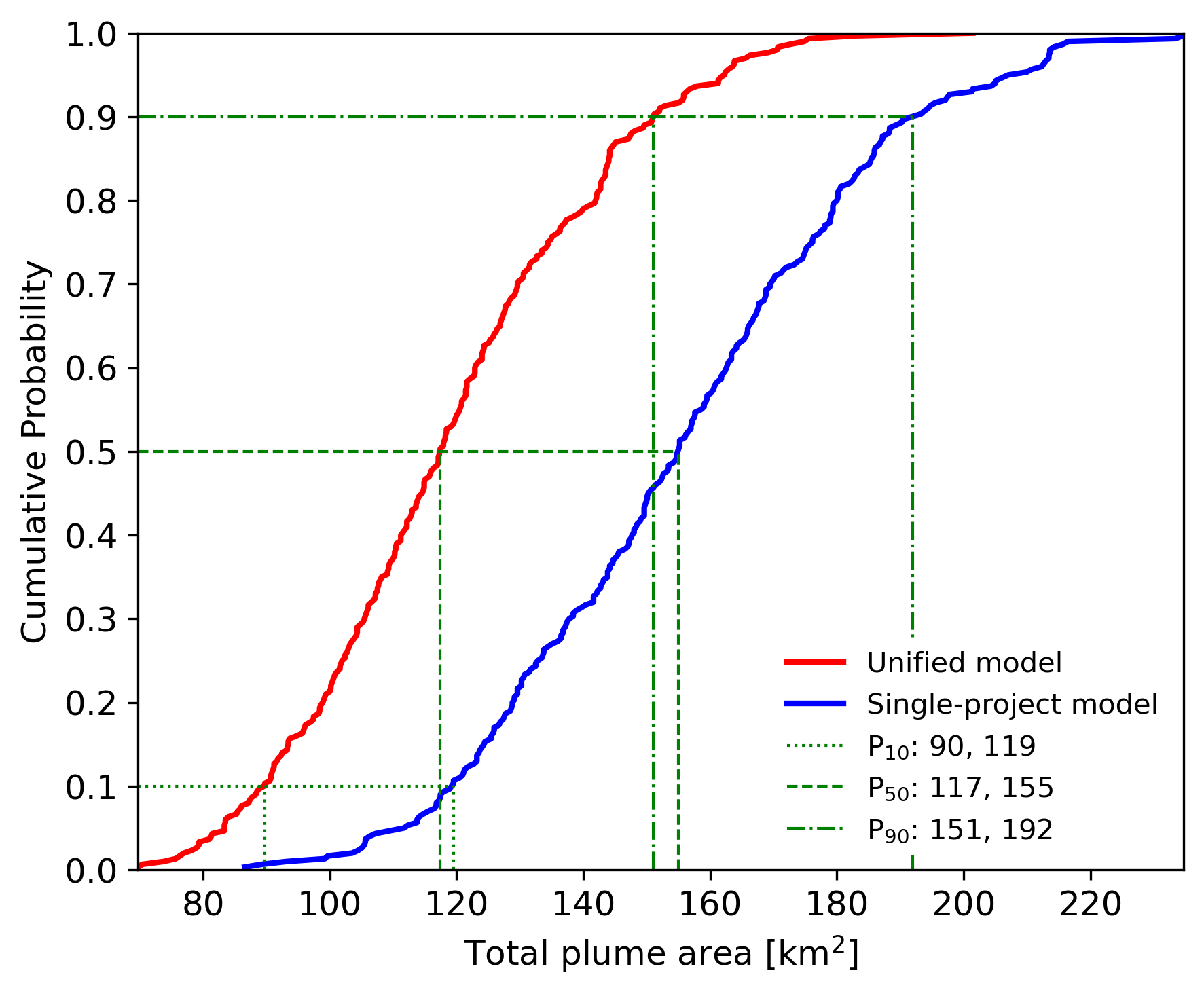}
        \caption{Total plume area in the DCC project}
    \end{subfigure}
    \hfill
    \begin{subfigure}{0.45\textwidth}
        \centering
        \includegraphics[width=\textwidth]{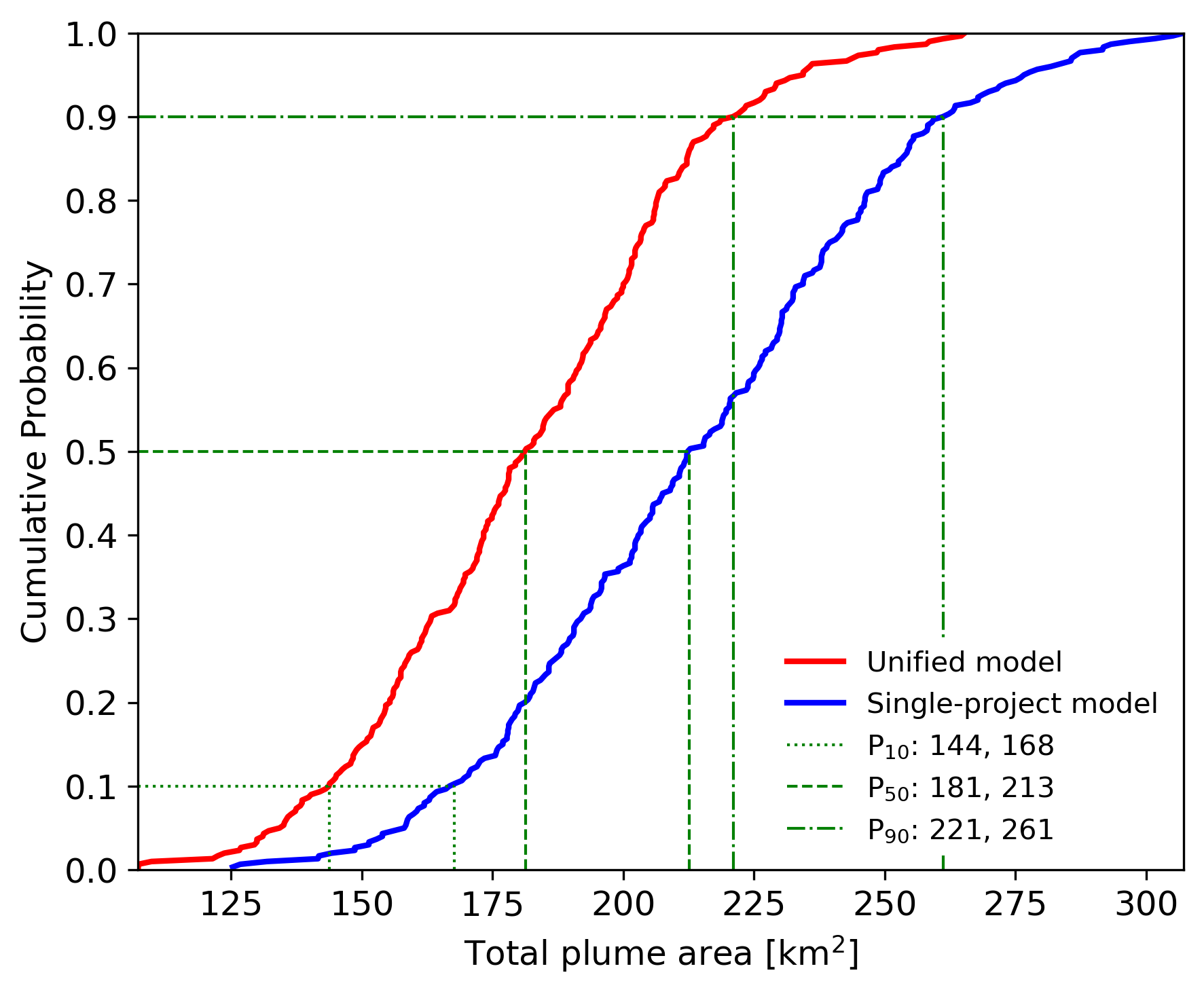}
        \caption{Total plume area in the SCS project}
    \end{subfigure}

    \caption{Cumulative distribution functions for total CO$_2$ injected (top row) and total plume area (bottom row) at 50~years for the DCC and SCS projects. Results are shown for unified and single-project models. The numerical values next to each percentile correspond to results for the two cases.}
    \label{fig:SvsP_CDF}
\end{figure}

Cumulative distribution functions (CDFs) of total CO$_2$ injected for the DCC and SCS projects are shown in Figure~\ref{fig:SvsP_CDF}(a) and (b). Results are presented for both the unified and single-project models. Clear shifts in the CDFs are evident, with differences in the P$_{50}$ responses of about 80~MT in both cases. Note that, for both projects, all 300~cases (i.e., the full CDF) exhibit a reduction in the total CO$_2$ injected in the unified model compared to the single-project model.

\begin{table}[t]
    \centering
    \caption{Comparison of total CO$_2$ injected at 50~years for single-project and unified models. Values reported as mean $\pm$ standard deviation.}
    \label{tab:SvsP_injected_co2}
    \begin{tabular}{lccc}
    \toprule
    Project & Single-project model [MT] & Unified model [MT] & Ratio of mean responses [-] \\
    \midrule
    DCC & $270.3 \pm 46.3$& $195.0 \pm 35.1$& 0.722 \\ 
    SCS & $474.7 \pm 68.3$& $396.9 \pm 58.0$& 0.836 \\ 
    BF  & $9.8 \pm 0.0$    & $9.8 \pm 0.0$    & 1.000 \\ 
    DGC & $102.5 \pm 0.0$  & $102.5 \pm 0.0$  & 1.000 \\ 
    RTE & $9.0 \pm 0.0$    & $9.0 \pm 0.0$    & 1.000 \\ 
    \bottomrule
    \end{tabular}
\end{table}

\begin{table}[t]
    \centering
    \caption{Comparison of total plume area at 50~years for single-project and unified models. Values reported as mean $\pm$ standard deviation.}
    \label{tab:SvsP_plume_area}
    \begin{tabular}{lccc}
    \toprule
    Project & Single-project model [km$^2$] & Unified model [km$^2$] & Ratio of mean responses [-] \\
    \midrule
    DCC & $155.6 \pm 28.8$& $119.0 \pm 23.7$& 0.766\\ 
    SCS & $214.6 \pm 36.6$& $182.4 \pm 30.6$ & 0.850\\ 
    BF  & $21.3 \pm 4.3$   & $19.6 \pm 4.0$   & 0.920\\ 
    DGC & $64.1 \pm 9.7$   & $62.2 \pm 9.6$& 0.969 \\ 
    RTE & $9.5 \pm 2.1$    & $9.3 \pm 2.0$    & 0.986\\ 
    \bottomrule
    \end{tabular}
\end{table}

Analogous results for total plume area at 50~years are provided in Table~\ref{tab:SvsP_plume_area} and Figure~\ref{fig:SvsP_CDF}(c) and (d). The smaller mean plume areas for the DCC and SCS projects in the unified model are largely due to the decreases in the total amount of CO$_2$ injected for these cases. We again see shifts of the full CDFs for both projects. Interestingly, slight reductions in the total plume area are also observed for the other three projects (BF, DGC, and RTE), even though their total CO$_2$ injected is unchanged (see Table~\ref{tab:SvsP_injected_co2}). This shows that pressure effects from larger projects can influence plume migration in smaller projects, even though total injection is not affected.

\begin{figure}[h!]
    \centering

    \begin{subfigure}{0.4\textwidth}
        \centering
        \includegraphics[width=\textwidth]{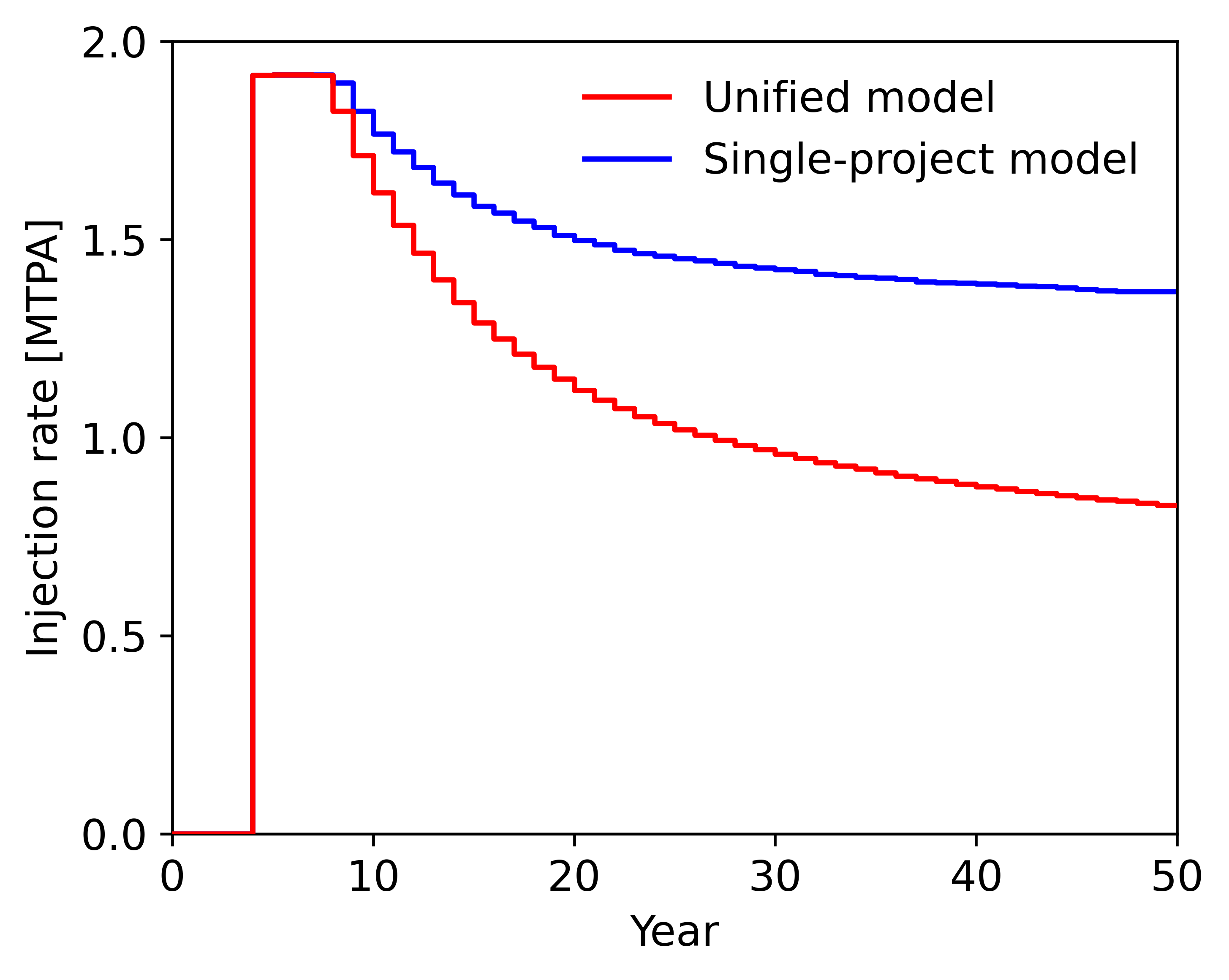}
        \caption{Well Liberty~1}
    \end{subfigure}
    \begin{subfigure}{0.4\textwidth}
        \centering
        \includegraphics[width=\textwidth]{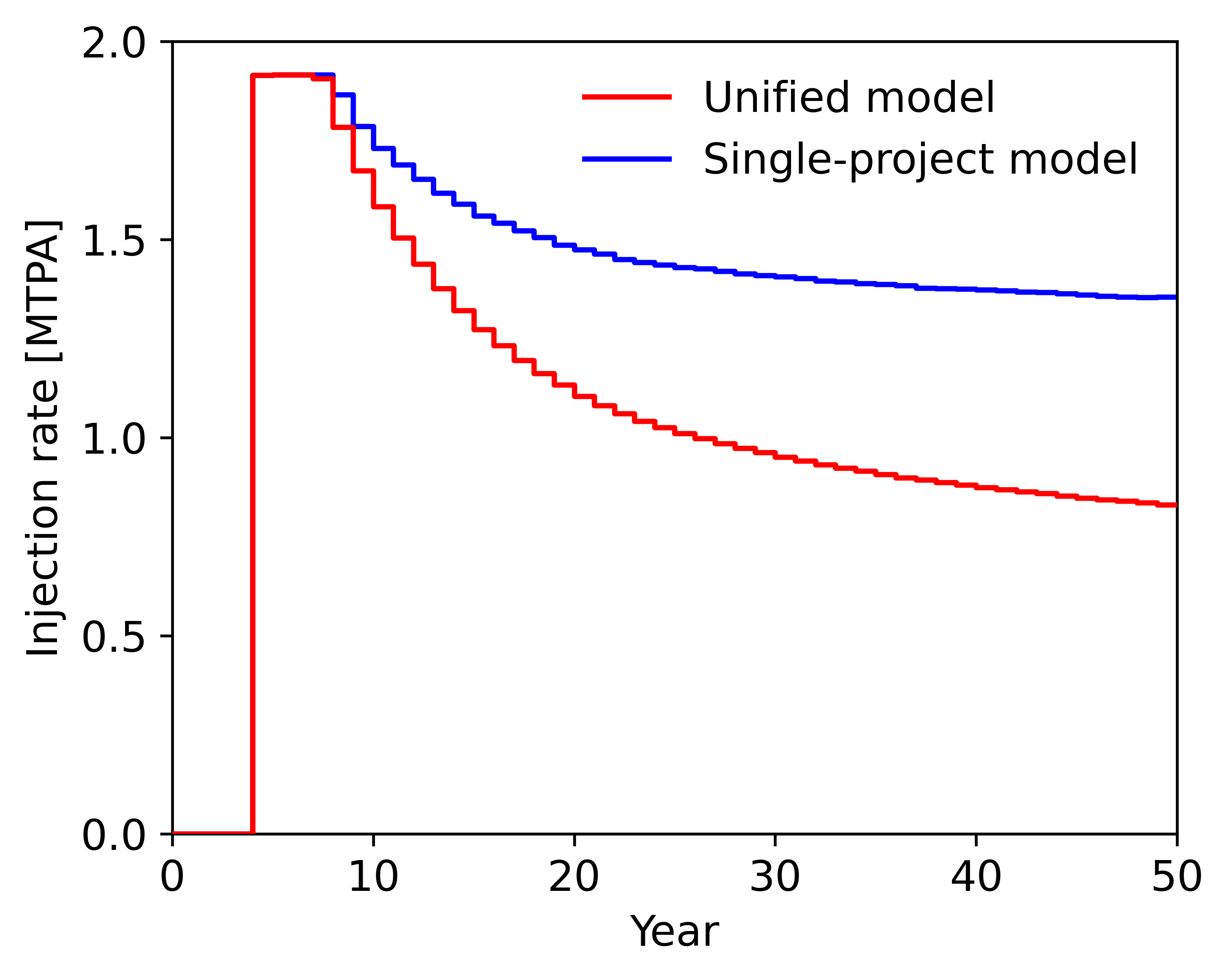}
        \caption{Well Unity~1}
    \end{subfigure}

    \vspace{0.5em}

    \begin{subfigure}{0.4\textwidth}
        \centering
        \includegraphics[width=\textwidth]{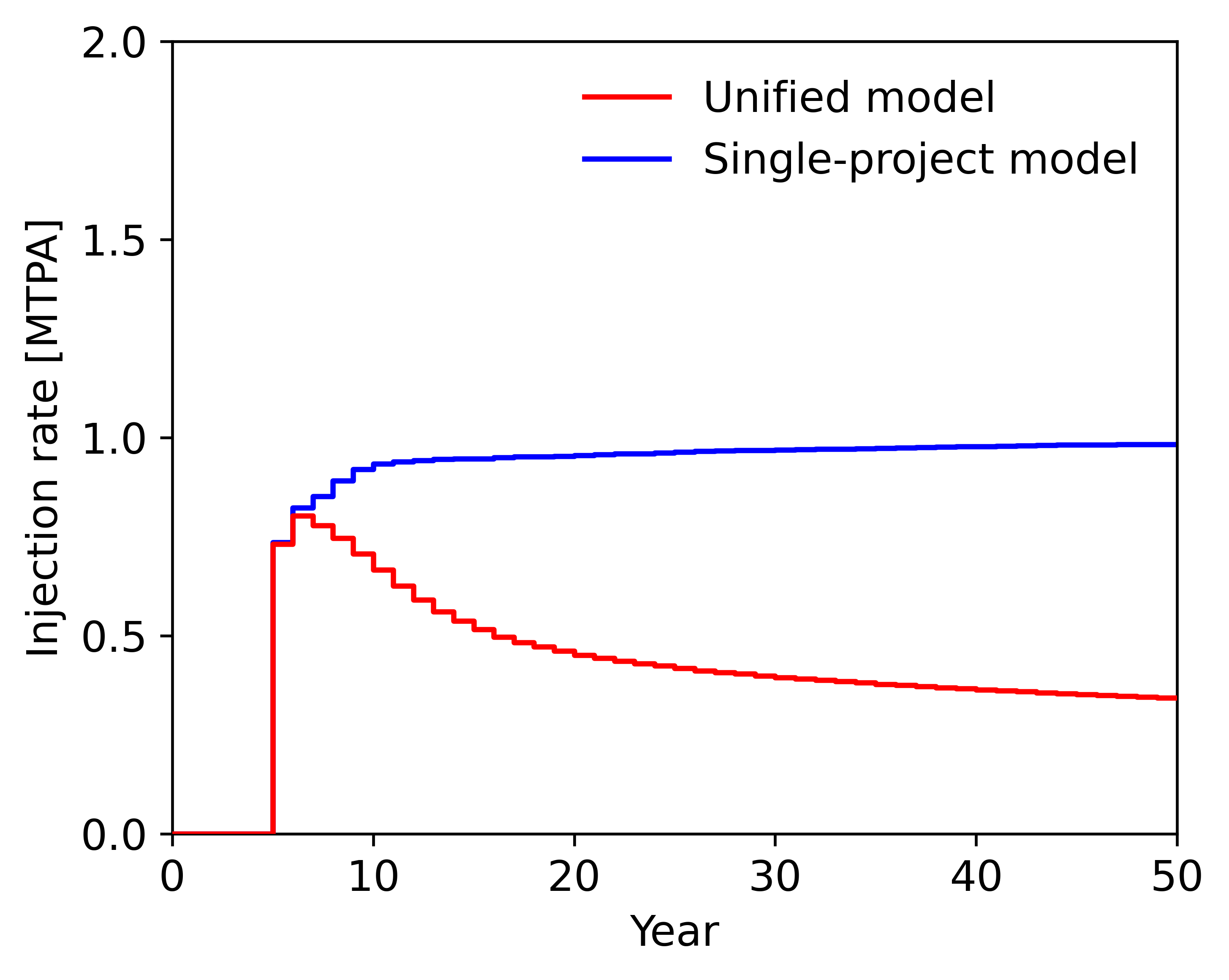}
        \caption{Well IIW-N}
    \end{subfigure}
    \begin{subfigure}{0.4\textwidth}
        \centering
        \includegraphics[width=\textwidth]{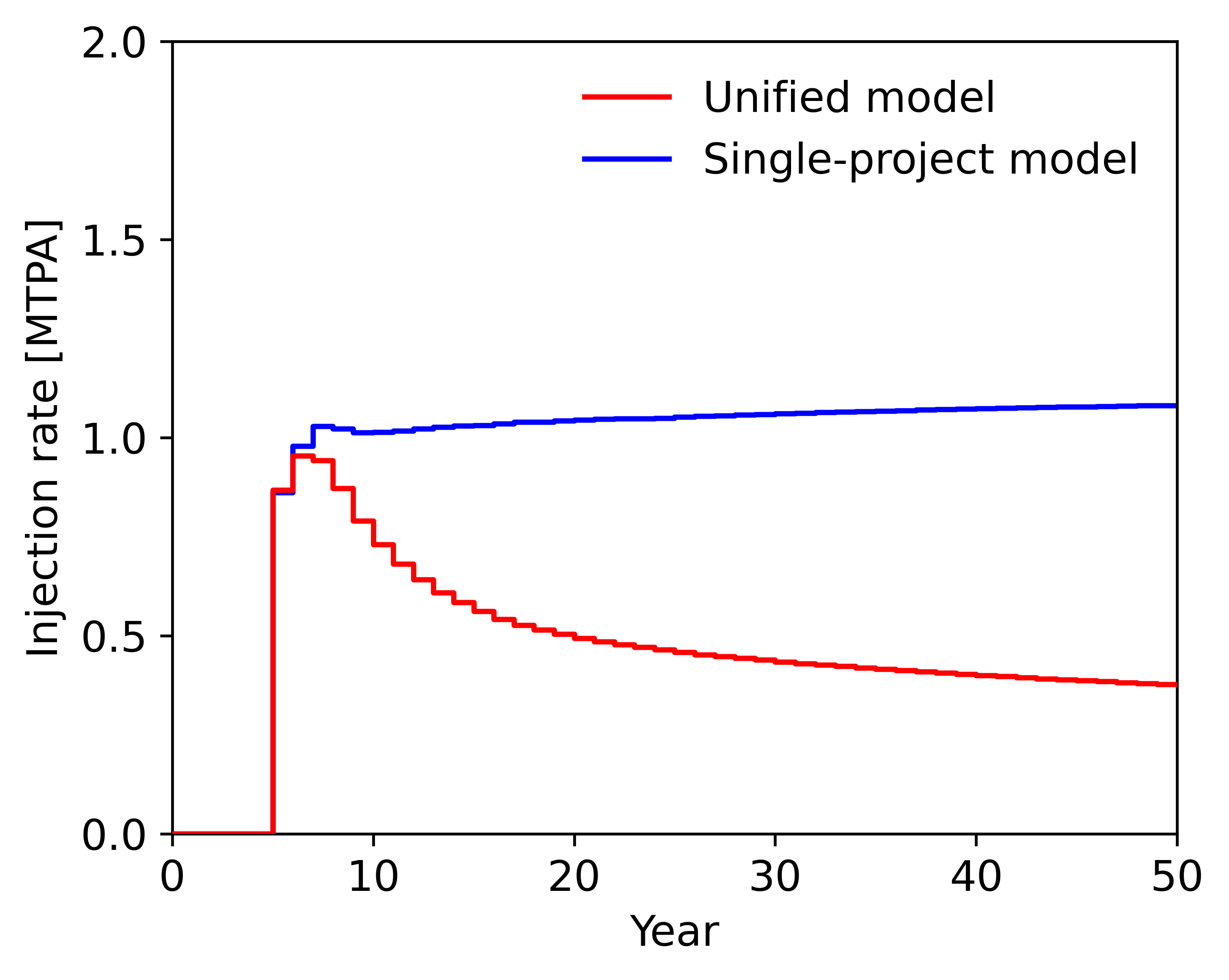}
        \caption{Well IIW-S}
    \end{subfigure}

    \caption{Comparison of injection rates between the unified model and the single-project model for the DCC project. Results for all four DCC wells are shown.}
    \label{fig:single_project_injection_rates}
\end{figure}

Finally, we compare the injection profiles and plume migration predicted by the single-project and unified models for the DCC project. One particular case (corresponding approximately to the P$_{10}$ response for total CO$_2$ injected in the unified model) is used for this comparison. Injection rate predictions for the four DCC wells are shown in Figure~\ref{fig:single_project_injection_rates}. All wells inject less CO$_2$ in the unified model than in the single-project model. This is due both to reaching the BHP limit earlier, and to lower injection rates when the well is operating under BHP control. Differences in the injection rates grow significantly in time. The increase in rate at early time observed for wells IIW-N and IIW-S in the single-project model (Figure~\ref{fig:single_project_injection_rates}(c) and (d)) is likely due to mobility effects, i.e., the presence of CO$_2$ near these injectors increases mobility enough to allow them to operate under rate control.

\begin{figure}[h!]
    \centering

    \begin{subfigure}{0.47\textwidth}
        \centering
        \includegraphics[height=4.5cm]{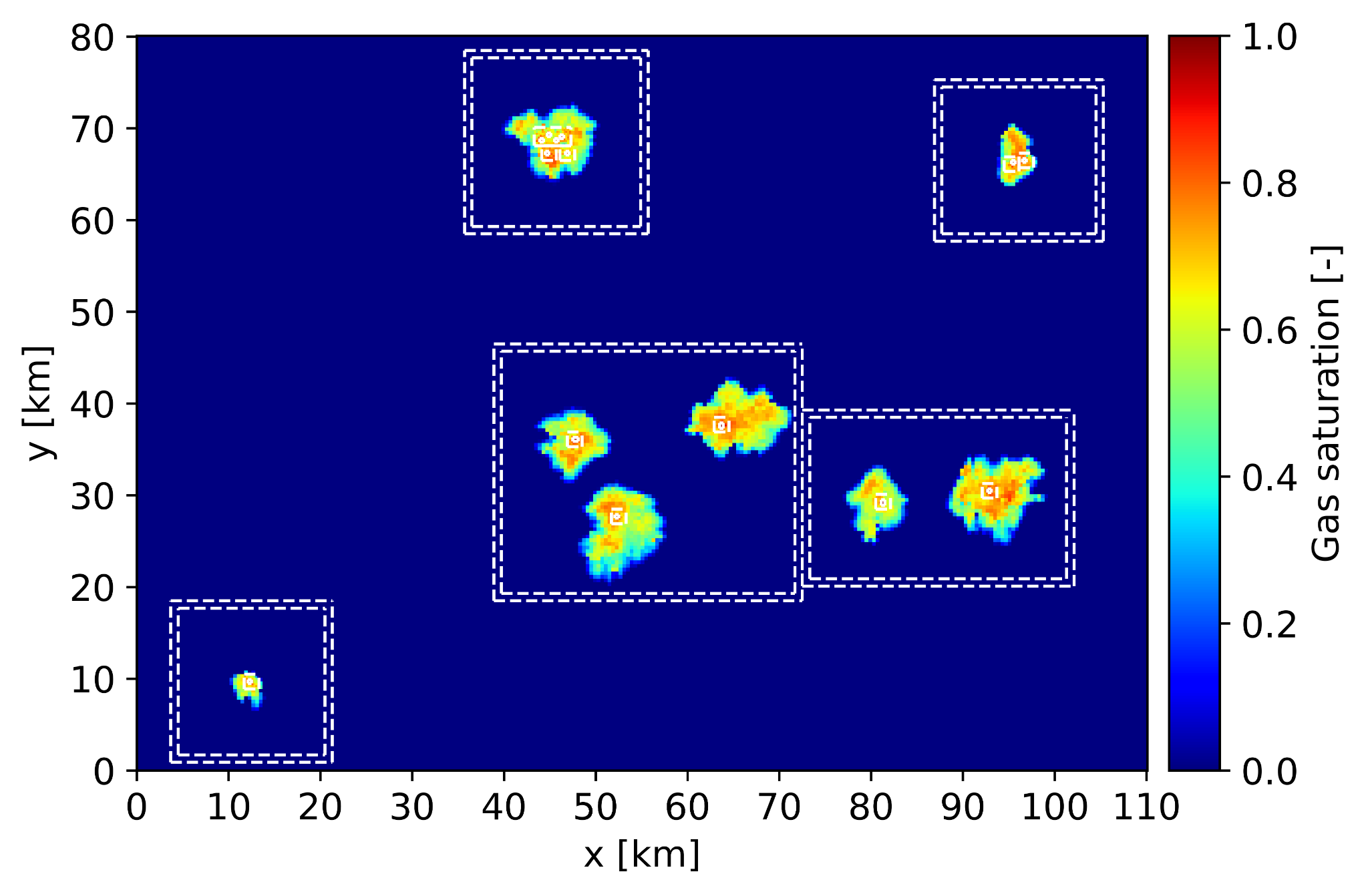}
        \caption{Overall domain of the unified model}
    \end{subfigure}
    \begin{subfigure}{0.47\textwidth}
        \centering
        \includegraphics[height=4.5cm]{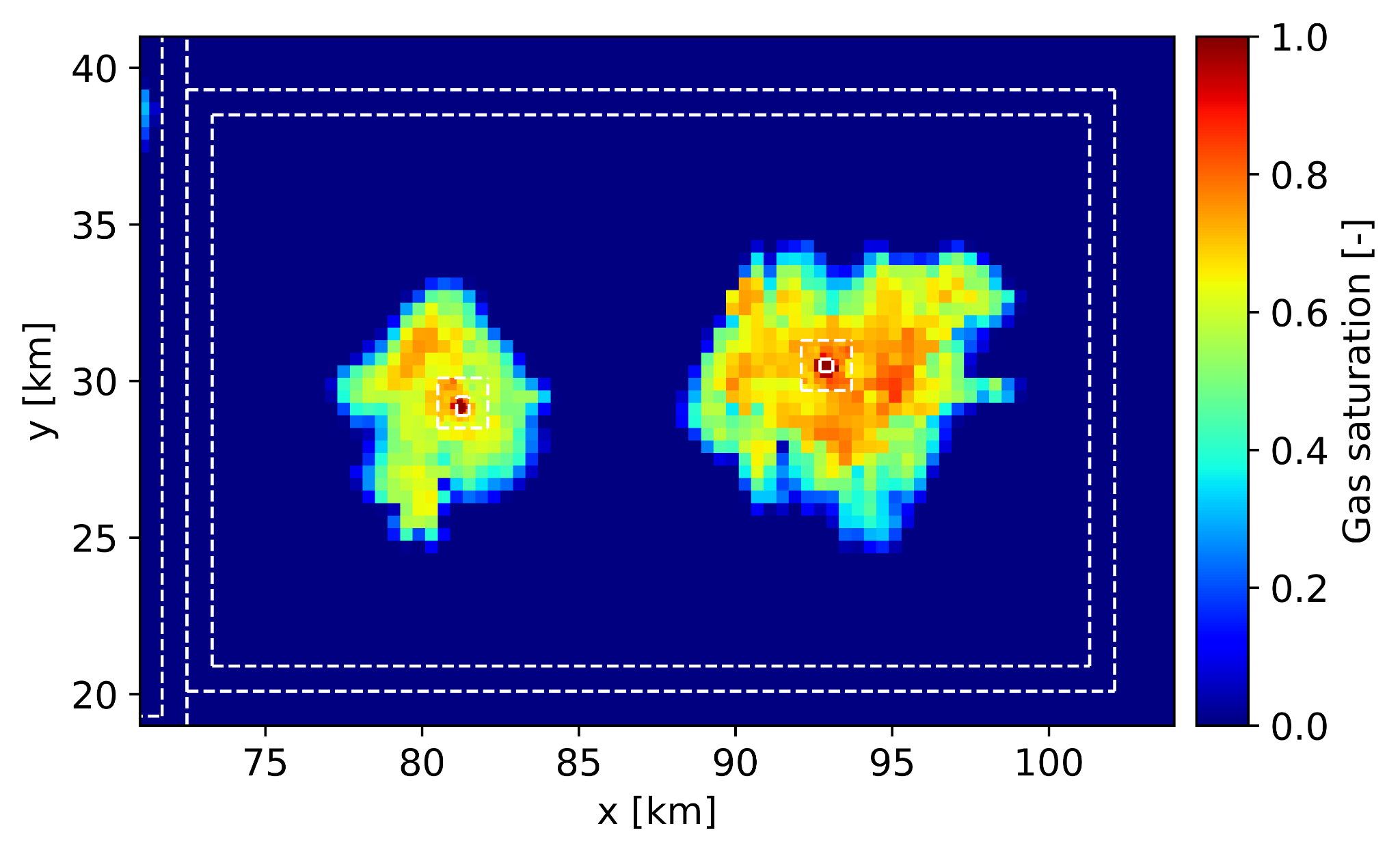}
        \caption{DCC project area of the unified model}
    \end{subfigure}
    \vspace{0.5em}
    \begin{subfigure}{0.47\textwidth}
        \centering
        \includegraphics[height=4.5cm]{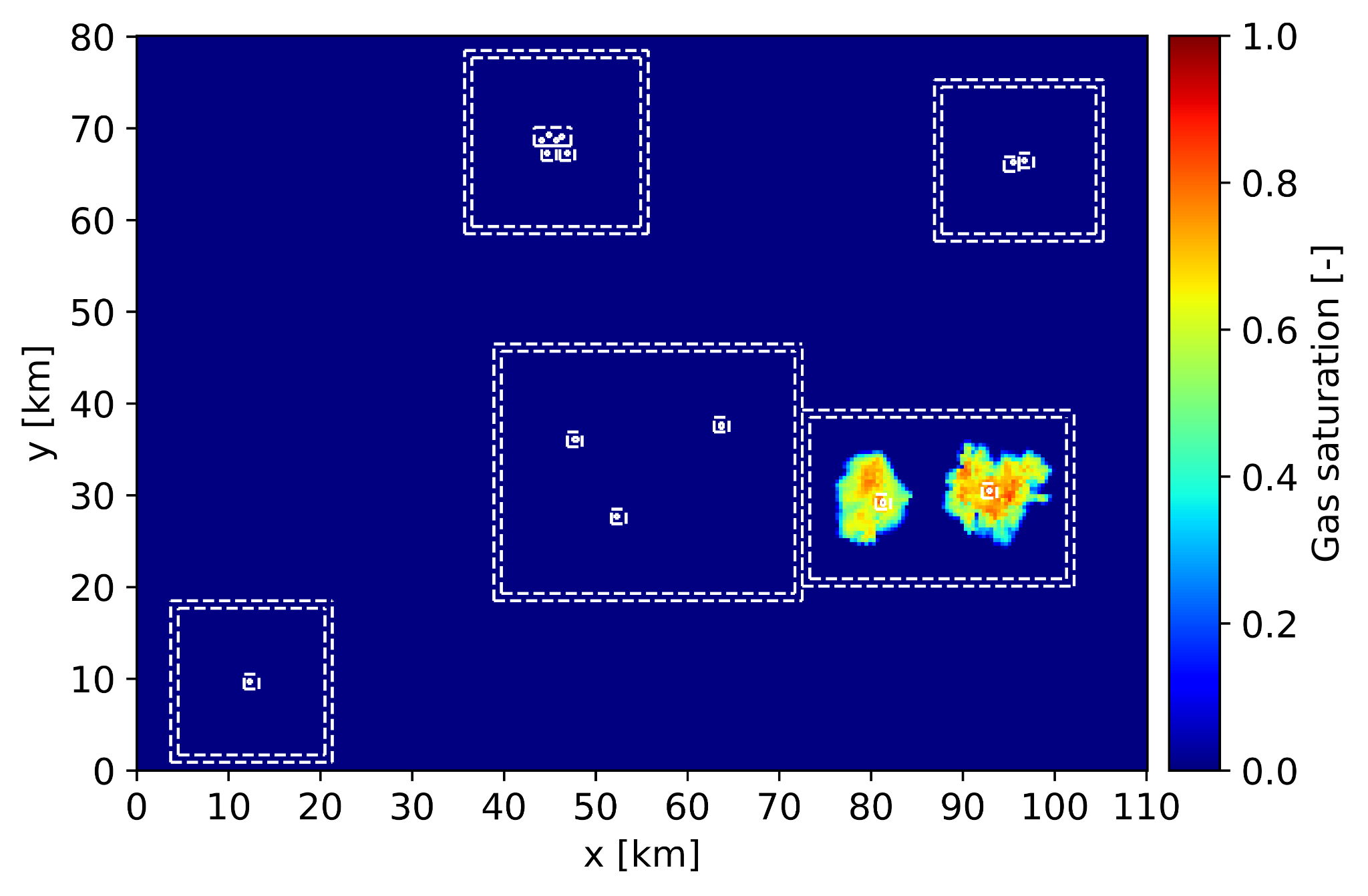}
        \caption{Overall domain of the single-project model}
    \end{subfigure}
    \begin{subfigure}{0.47\textwidth}
        \centering
        \includegraphics[height=4.5cm]{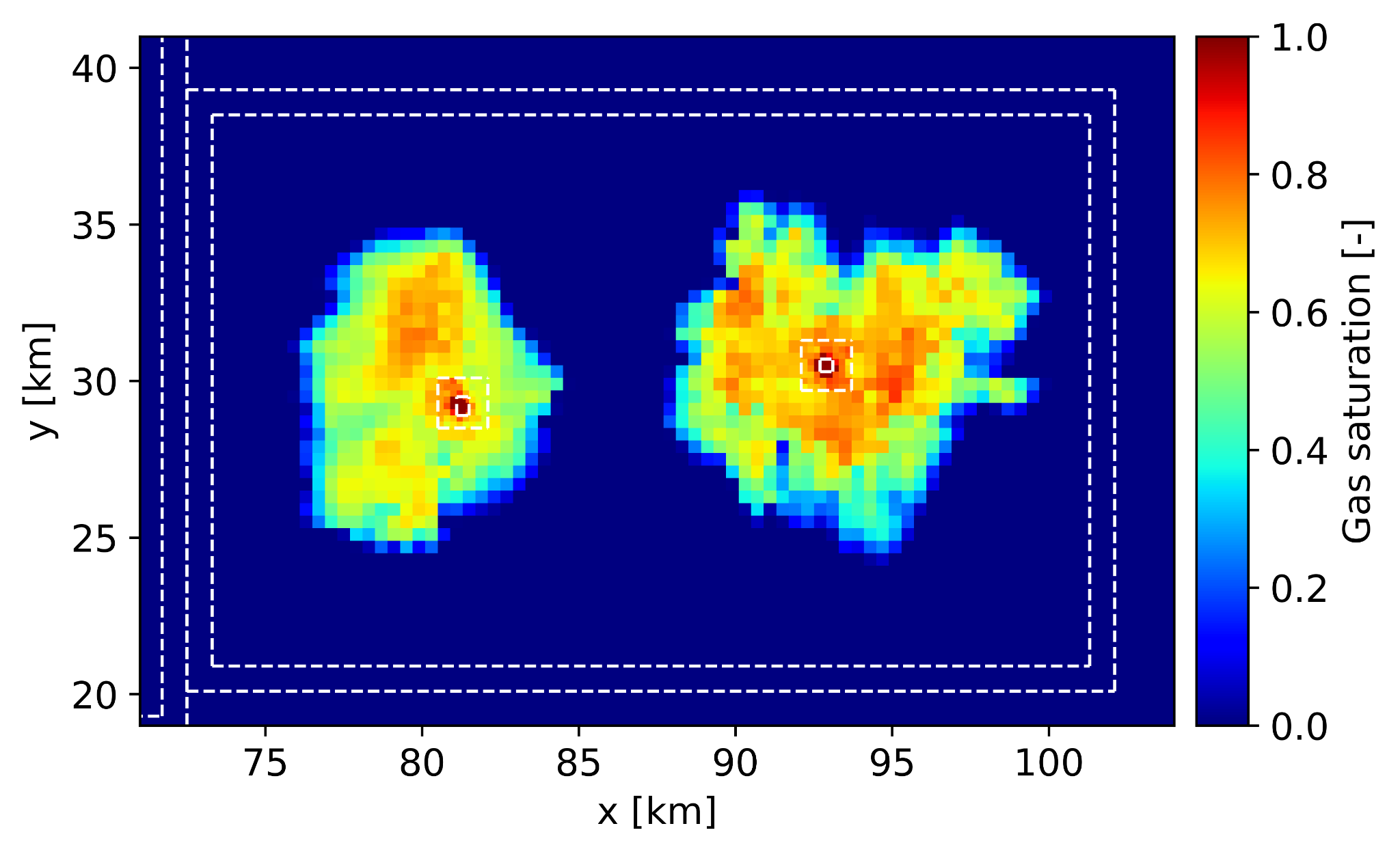}
        \caption{DCC project area of the single-project model}
    \end{subfigure}
    
    \caption{Comparison of gas saturation distributions at 50~years between the unified model (a, b) and the single-project model (c, d) for the DCC project. White dashed lines indicate the boundaries of the nested LGR regions. In (c), all projects other than DCC have zero injection.}
    \label{fig:Gas saturation: single vs unified}
\end{figure}

The CO$_2$ plume distributions at 50~years predicted by the unified and single-project models for the DCC project are shown in Figure~\ref{fig:Gas saturation: single vs unified}. Results for the unified model appear in the top row, and those for the DCC-only model in the bottom row. The overall plume geometries are qualitatively similar between the two models, though the plume extent in the unified model (114.7~km$^2$) is noticeably less than that in the single-project model (162.9~km$^2$). This is largely due to the decreased total CO$_2$ injected in the unified model (147.8~MT) relative to the single-project model (229.2~MT).

\FloatBarrier

\section{Concluding Remarks}
\label{sec:conclusion}
In this work, we developed a unified basin-scale model for the Broom Creek Formation in North Dakota. This model was constructed by integrating publicly available datasets from multiple CO$_2$ storage projects targeting this formation. Our model, which includes operating and approved projects, involves 18~wells with a maximum overall injection rate of 30~MTPA. 
The wells inject for 42--50~years, which is considerably longer than the actual (initial) permitted time window. This allows us to quantify inter-project interference for large-scale storage projects operating over extensive time frames. Wells operate under specified injection rates, though they are subject to maximum BHP constraints. The fine model contains $44\times10^6$~grid cells and takes over 2~days to run on 32~cores. We therefore introduced a multilevel upscaling procedure that combines nested LGRs, which retain higher resolution in the vicinity of wells, with power averaging for permeability in coarsened regions. A constraint-satisfaction problem is solved to construct the nested LGR regions. The power averaging parameters were determined by minimizing the mismatch between fine and coarse-scale model responses for key quantities of interest (QoIs). The resulting coarse model contains about 400,000~cells and runs in about 10~minutes, corresponding to a speedup of $\sim$300$\times$. For the base-case model, the errors in total CO$_2$ injected, total plume area, and individual well injection rates are about 0.9\%, 2.6\%, and 1.2\%, respectively. 

Uncertainty quantification was accomplished via a global sensitivity analysis. This required us to perform 4400~coarse-scale simulations. 
Seven model parameters, namely the bulk volumes of the boundary regions, overburden and underburden permeabilities, anisotropy ratio, and relative permeability variables, were treated as uncertain, along with the geological realization. Substantial uncertainty was observed for both the total CO$_2$ injected and plume area. The P$_{10}$--P$_{90}$ range for total CO$_2$ injected was 170~MT, corresponding to 24\% of the median response (714~MT), while that for total plume area was 146~km$^2$, corresponding to 37\% of the median response (393~km$^2$). The sensitivity analysis was conducted using the coarse model with the same nested LGR structure and power-averaging parameters as in the base case. To assess upscaling accuracy over an appropriate range of models and responses, 29~additional fine-scale simulation runs were performed. The models simulated at the fine scale were selected to span the cumulative probability distributions for the QoIs, as determined from the global sensitivity analysis. For the three QoIs (total CO$_2$ injected at 50~years, total plume area at 50~years, and injection rates at individual wells), both the mean and median upscaling errors ranged from 1.0\% to 1.6\%, demonstrating the robustness of our nested-LGR gridding and upscaling workflow. 

A deep learning-based surrogate model was developed to predict the total CO$_2$ injected and total plume area at 50~years for any sample of the seven model parameters and geological realizations. The surrogate model achieved mean relative errors of 0.1\% for total CO$_2$ injected and 0.3\% for the plume area. Variance-based global sensitivity analysis using the surrogate model was then performed. This assessment required over 650,000~function evaluations, so the surrogate model was essential. The analysis showed that the particular geological realization had the strongest impact on total CO$_2$ injected, accounting for 59\% of the total variance. For total plume area, a model parameter controlling the shape of the gas relative permeability curve ($n_{g,\mathrm{res}}$) had the largest impact, accounting for 66\% of the total variance. Finally, we quantified the impact of the other projects in the basin on the performance of each individual project. This assessment also considered uncertainty. Clear shifts in the CDFs for total CO$_2$ injected and plume area were observed between the single-project models and the unified model for projects with high target injection rates, indicating significant basin-scale interactions. For example, the P$_{50}$ total CO$_2$ injected for the DCC project decreased by 29\%, from 271~MT in the single-project model to 192~MT in the unified model, as a result of pressure interference.

There are several directions for future work in this area. Additional effects, such as variogram parameters and the relationship between permeability and porosity, could be included in the sensitivity analysis. In our ongoing work, we are using the unified model and the multilevel upscaling workflow developed here within a basin-scale optimization framework. This will allow us to determine optimal well locations and injection strategies for new storage projects in the Broom Creek Formation while accounting for pressure interference with existing projects. We are also using this framework to investigate economic metrics to promote the efficient and equitable use of storage resources. Within these optimizations, the constraint satisfaction problem is solved for each candidate well configuration, thus enabling us to use coarse-scale flow simulations. It will also be of interest to extend our workflow to handle horizontal and deviated wells. This will require some modification of the procedures for constructing the nested LGRs. Finally, the overall workflow developed here can be extended for use in other subsurface applications. Examples include combined enhanced oil production -- geologic carbon storage operations, geothermal reservoir simulation, and underground hydrogen storage.

\section*{CRediT authorship contribution statement}
\textbf{Keisuke Yamamura}: Conceptualization, Formal analysis, Methodology, Software, Visualization, Validation, Writing -- original draft. \textbf{Louis J. Durlofsky}: Supervision, Conceptualization, Resources, Writing -- review \& editing.

\section*{Declaration of competing interest}
\noindent The authors declare that they have no known competing financial interests or personal relationships that could have appeared to influence the work reported in this paper.

\section*{Acknowledgments}
\noindent We are grateful to INPEX Corporation and the Stanford Center for Carbon Storage for funding this work. We thank Bret Fossum at the Energy \& Environmental Research Center (EERC) for assistance with the geological models. We acknowledge the SDSS Center for Computation for providing the computational resources used in this work.

\section*{Code availability}
\noindent The basin-scale Broom Creek Formation simulation model and the CSP code developed in this study will be made available upon publication of this paper.


\bibliographystyle{elsarticle-harv}
\bibliography{reference}


\clearpage

\setcounter{section}{0}
\renewcommand{\thesection}{S\arabic{section}}
\renewcommand{\thefigure}{S\arabic{figure}}
\renewcommand{\thetable}{S\arabic{table}}
\renewcommand{\theequation}{S\arabic{equation}}

\setstretch{1.05}
\setlength{\parindent}{0pt}
\setlength{\parskip}{0.5em}

\pagestyle{fancy}
\fancyhf{}
\fancyfoot[C]{\thepage}
\renewcommand{\headrulewidth}{0.4pt}
\renewcommand{\footrulewidth}{0pt}

\begin{center}

{\Large\bfseries
Supplementary Material
}

\vspace{1.0em}

{\large\bfseries
Basin-Scale Modeling of Multiple Storage Projects in the
Broom Creek Formation, North Dakota, USA:
Unified Model and Uncertainty Quantification
}

\vspace{1.2em}

{\normalsize
Keisuke Yamamura$^{1}$ and Louis J. Durlofsky$^{1}$
}

\vspace{0.7em}

{\small
$^{1}$Department of Energy Science and Engineering,
Stanford University, Stanford, CA 94305, USA
}

\vspace{1.0em}

\end{center}

\hrule
\vspace{1.2em}
\setcounter{section}{0}
\setcounter{figure}{0}
\setcounter{table}{0}
\setcounter{equation}{0}
\section*{S1.~Geological and Simulation Model Details}
\subsection*{S1.1.~Structural Framework}

\begin{table}[H]
    \centering
    \caption{Wells with facies logs used as conditioning data for geostatistical facies simulation. N/G denotes the net-to-gross ratio. ``Manual (this study)'' indicates wells for which facies logs were manually constructed in this study rather than obtained from previous interpretations.}
    \begin{tabular}{llll}
    \toprule
    Well Name & NDIC File No. & Facies Interpretation Source & N/G \\
    \midrule
    BNI~1 & 34244 & \citet{osti_1606011} & 0.525 \\
    Flemmer-1 & 34243 & \citet{osti_1606011} &  0.736 \\
    MAG~1 & 37833 & BF~\citep{NDIC2023BF} &  0.755 \\
    MAG~2 & 39960 & Manual (this study)&  0.556 \\
    RTE-10 & 37229 & RTE~\citep{NDIC2021RTE} &  0.687 \\
    RTE-10.2 & 37858 & Manual (this study) &  0.610 \\
    Coteau~1 & 37379 & DGC~\citep{NDIC2022DGC} &  0.787 \\
    Coteau~2 & 38916 & Manual (this study) &  0.729 \\
    Coteau~3 & 38917 & Manual (this study) &  0.661 \\
    Coteau~4 & 38918 & Manual (this study) &  0.670 \\
    Coteau~5 & 39418 & Manual (this study) &  0.633 \\
    Coteau~6 & 40027 & Manual (this study) &  0.511 \\
    J-LOC~1 & 37380 & DCC~\citep{NDIC2023WestTundra} &  0.621 \\
    Liberty~1 & 37672 & DCC~\citep{NDIC2021EastTundra} &  0.566 \\
    Archie Erickson~2 & 38622 & SCS~\citep{NDIC2024SCS2} & 0.687 \\
    Milton Flemmer~1 & 38594 & SCS~\citep{NDIC2024SCS1} &  0.629 \\
    Slash Lazy H~5 & 38701 & SCS~\citep{NDIC2024SCS2} &  0.717 \\
    \bottomrule
    \end{tabular}
    \label{table:facieslog}
\end{table}

\begin{figure}[H]
    \centering
    \begin{subfigure}{0.48\textwidth}
        \centering
        \includegraphics[width=\textwidth]{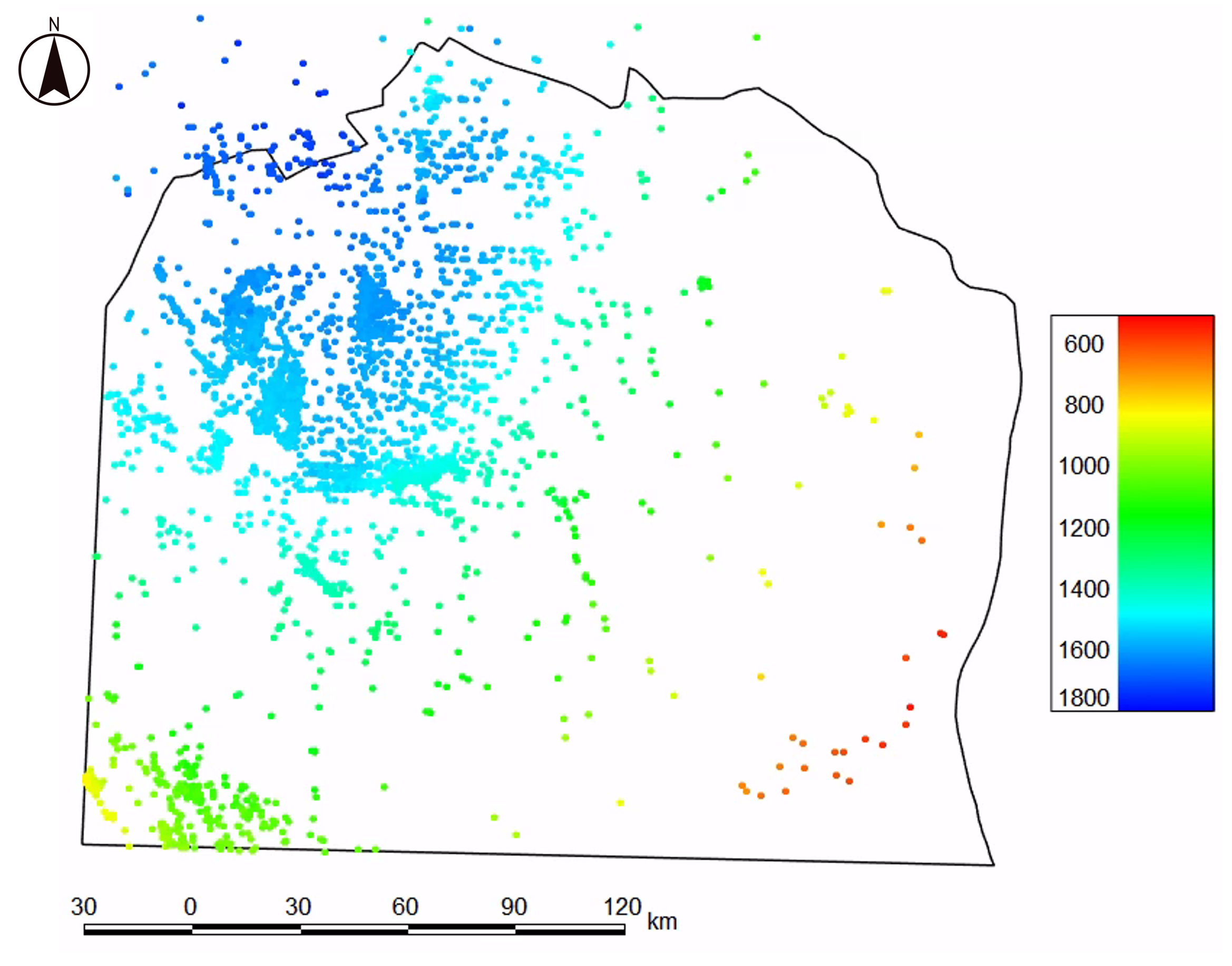}
        \caption{Broom Creek Formation}
    \end{subfigure}
    \hfill
    \begin{subfigure}{0.48\textwidth}
        \centering
        \includegraphics[width=\textwidth]{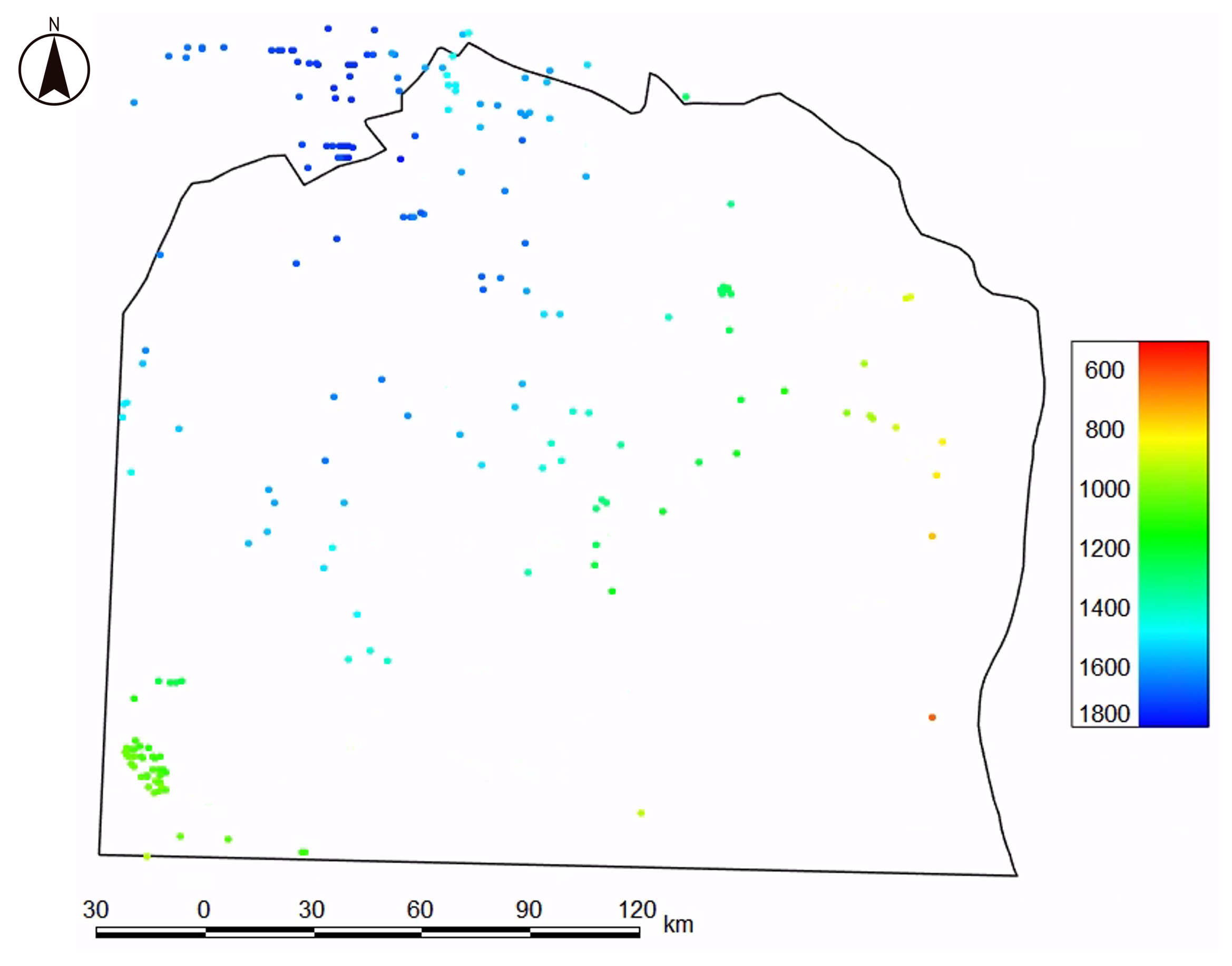}
        \caption{Amsden Formation}
    \end{subfigure}
    \caption{Locations of wells with formation-top data used to construct the Broom Creek and Amsden horizons. Colors indicate formation-top depth (mSSL), and black lines delineate the extent of the structural framework.}
    \label{fig:FormationTopData}
\end{figure}

\begin{figure}[H]
    \centering
    \includegraphics[width=0.5\textwidth]{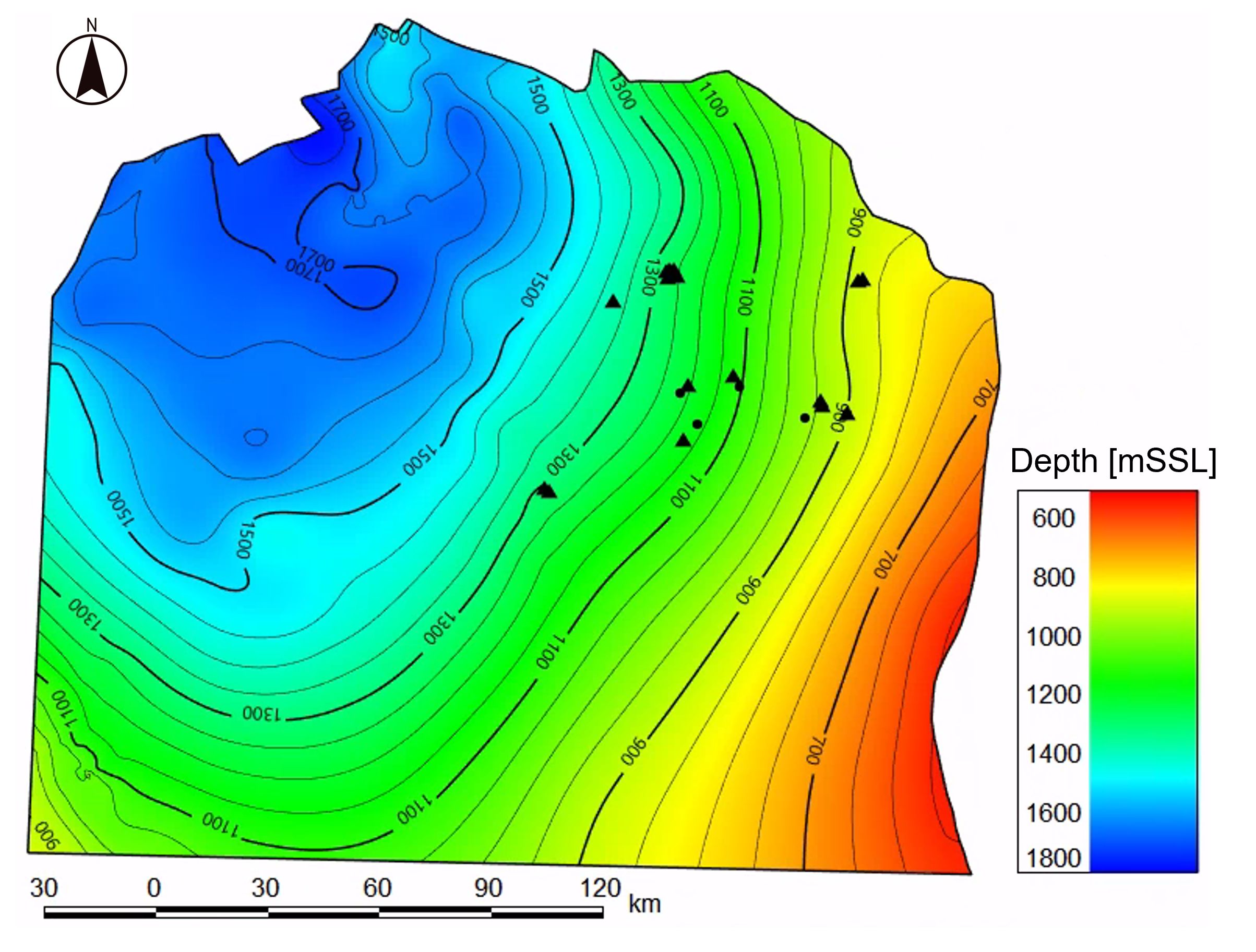}
    \caption{Top of the Amsden Formation derived from well formation-top data (mSSL). Black triangles and circles indicate drilled and planned CCS-related wells, respectively.}
    \label{fig:Amsden_top}
\end{figure}

\FloatBarrier

\subsection*{S1.2.~Lithofacies}

\begin{figure}[H]
    \centering
    \includegraphics[width=0.8\textwidth, height=0.75\textheight, keepaspectratio]{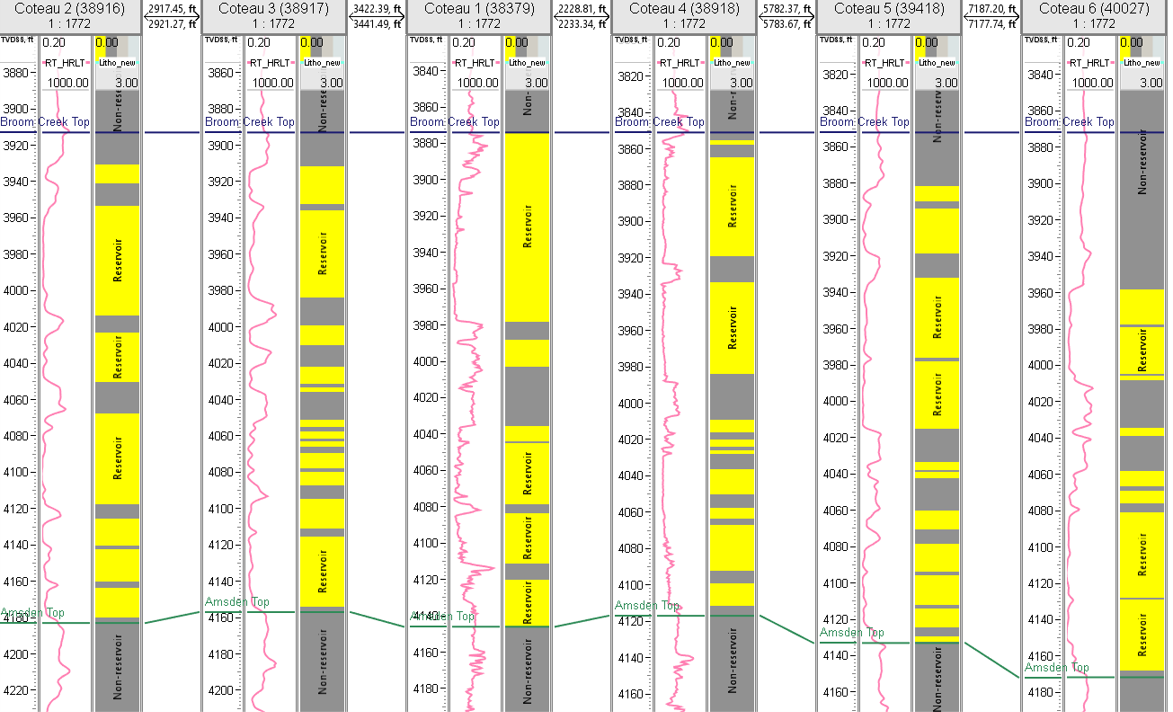}
    \caption{Facies logs for wells Coteau 1--6 in the DGC project. Yellow and gray indicate reservoir and nonreservoir rock, respectively. Pink lines show the resistivity logs, while blue and green lines indicate the tops of the Broom Creek and Amsden formations, respectively. Values above the facies logs indicate well-to-well distances.}
    \label{fig:FaciesCoteau1-6}
\end{figure}

\FloatBarrier
\subsection*{S1.3.~Porosity and Permeability}

\begin{figure}[H]
    \centering
    \includegraphics[width=0.5\textwidth]{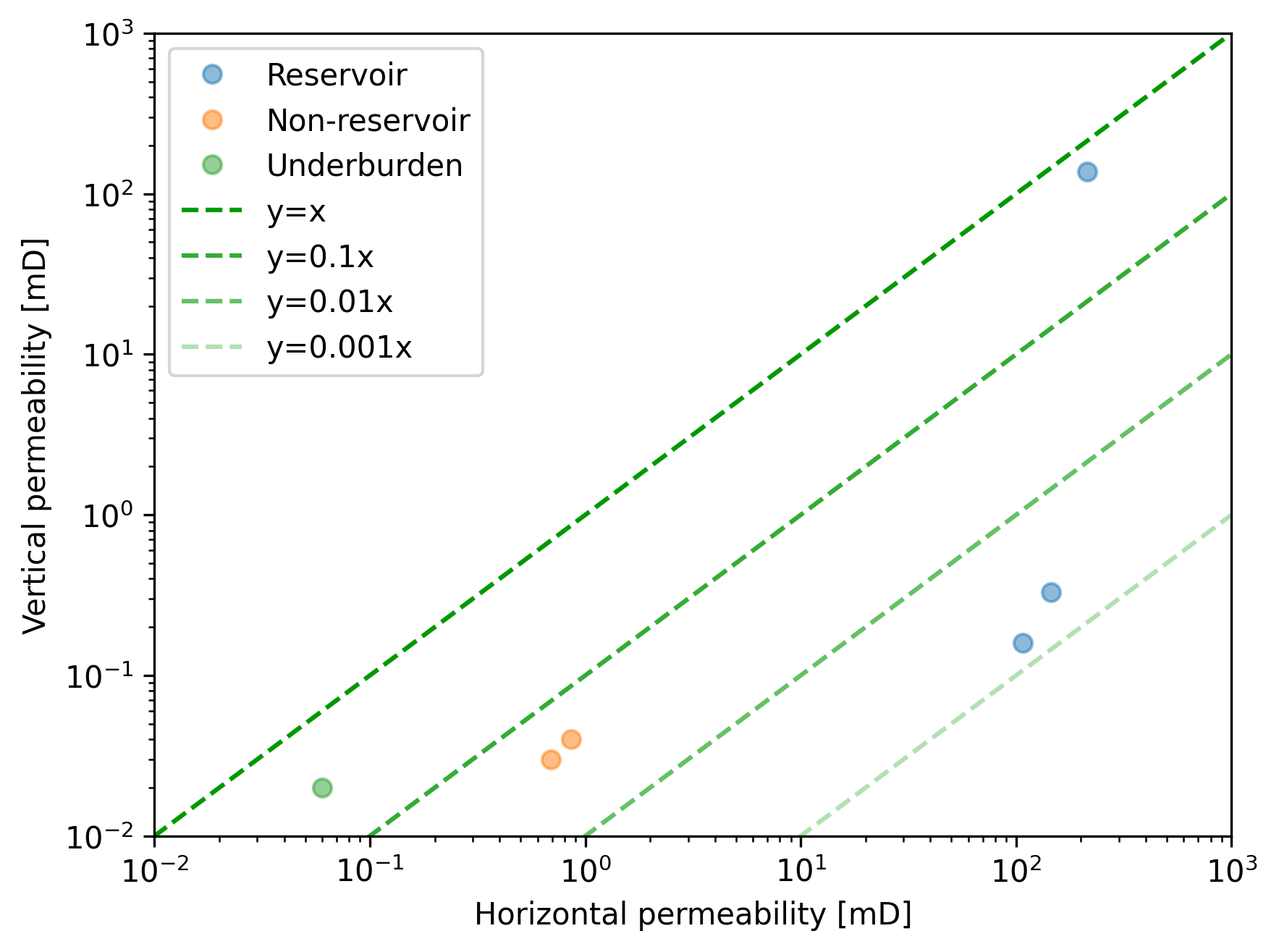}
    \caption{Horizontal and vertical air permeability measured at 5.52~MPa (800~psi).}
    \label{fig:KvKh-RCA}
\end{figure}

\begin{figure}[H]
    \centering
    \includegraphics[width=0.7\textwidth]{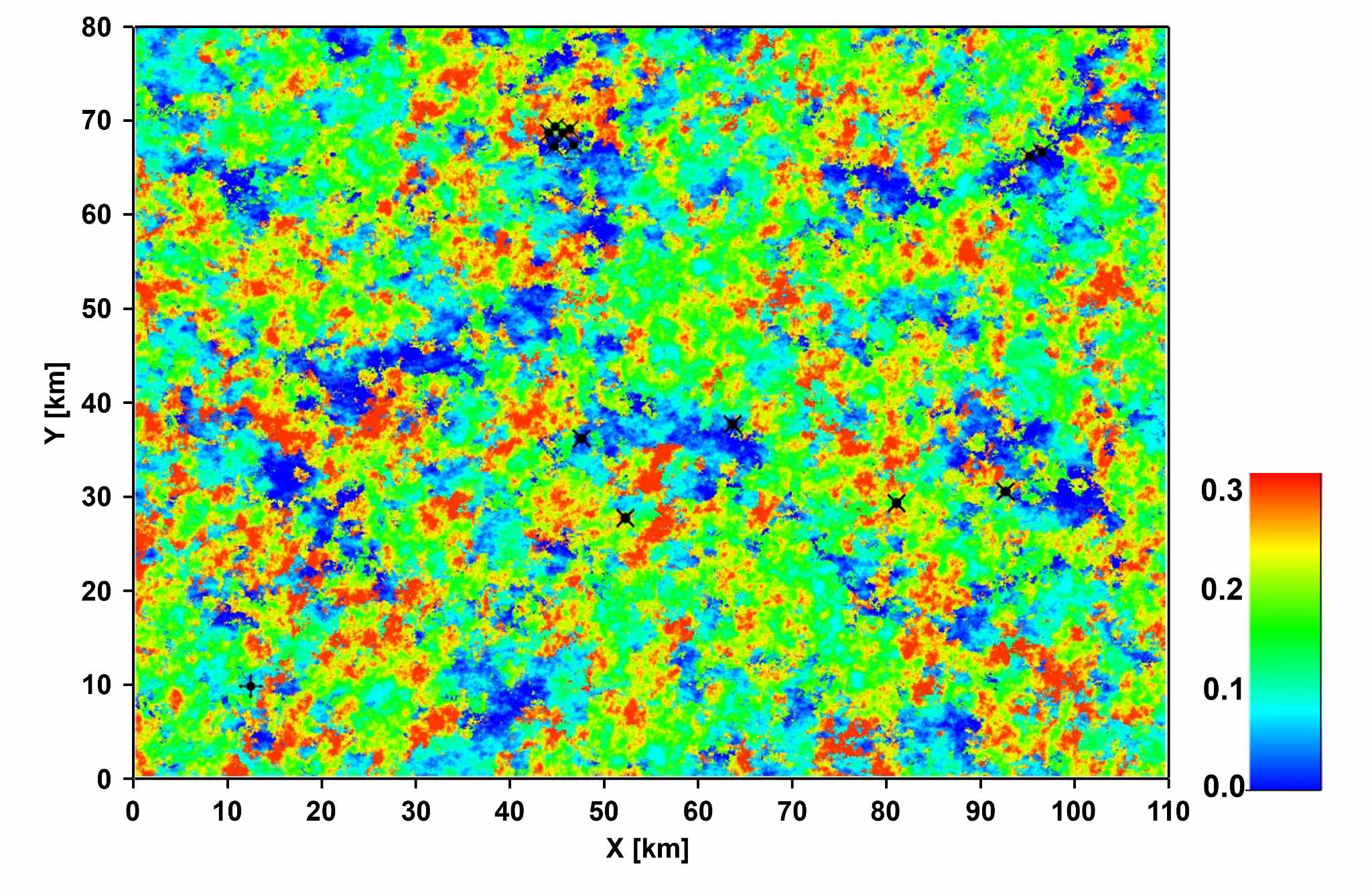}
    \caption{Porosity distribution of the top layer of the Broom Creek Formation in the base case.}
    \label{fig:Phi-Distribution}
\end{figure}

\FloatBarrier
\subsection*{S1.4.~Other Reservoir Properties}

\begin{figure}[H]
    \centering
    \begin{subfigure}[t]{0.48\textwidth}
        \centering
        \includegraphics[height=8cm]{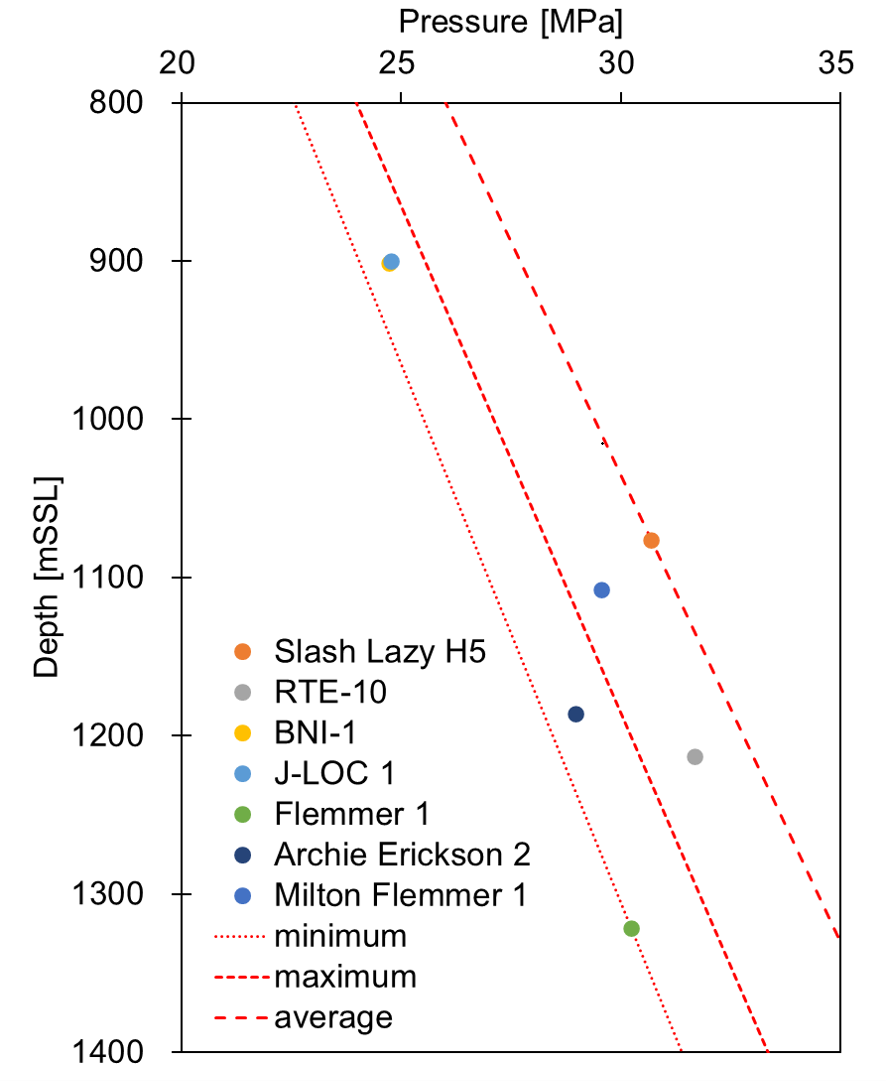}
        \caption{Fracture gradient in the Broom Creek Formation. The fracture gradients are determined from step rate tests. Labels next to circles indicate well names. The red dashed lines represent the minimum, average, and maximum fracture gradients of 0.0147, 0.0156, and 0.017~MPa/m (0.65, 0.69, and 0.75~psi/ft), respectively.}
    \end{subfigure}
    \hfill
    \begin{subfigure}[t]{0.48\textwidth}
        \centering
        \includegraphics[height=8cm]{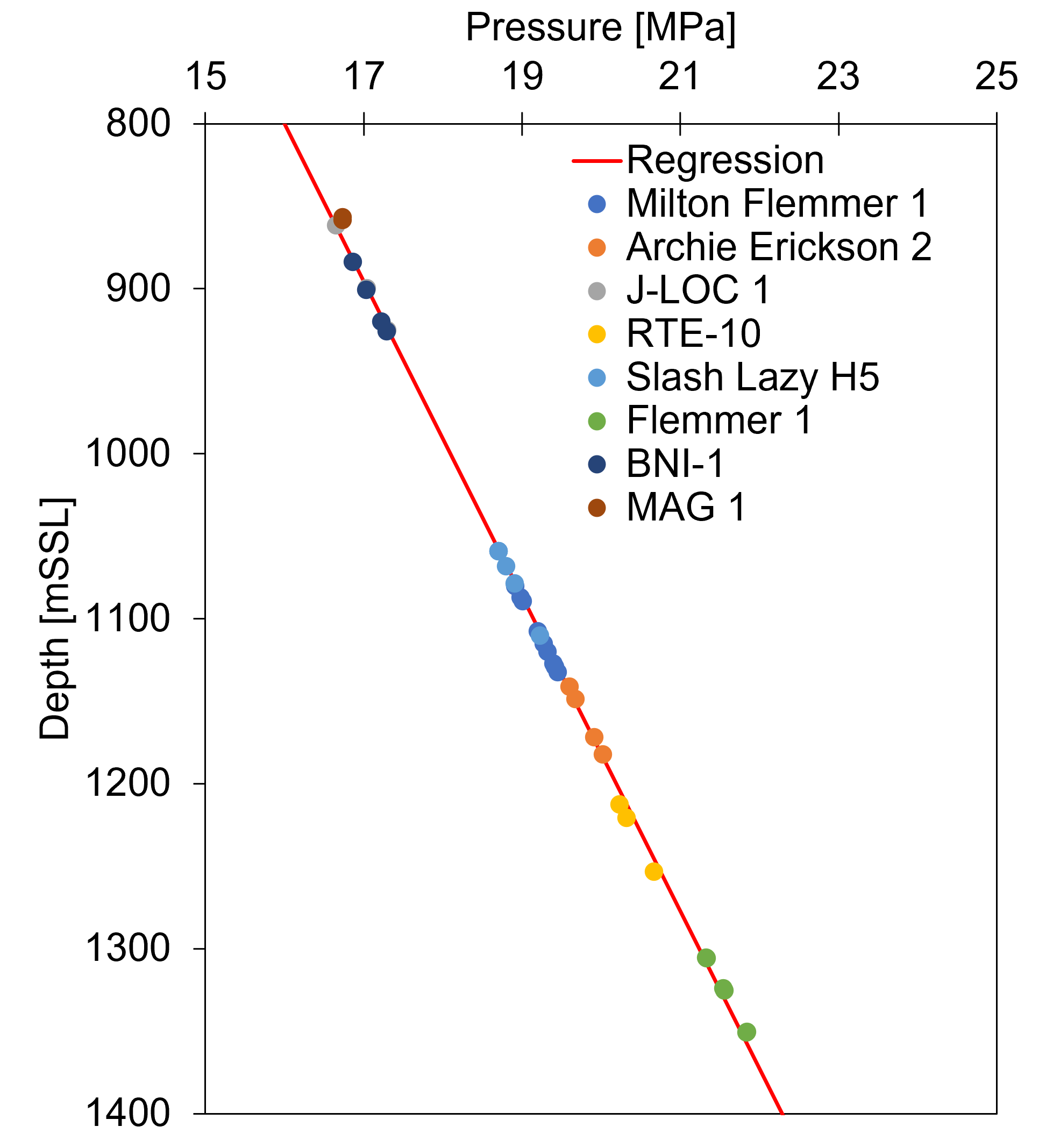}
        \caption{Measured reservoir pressure of the Broom Creek Formation. The data fall on a single regression line, indicating a pressure gradient of 0.0105~MPa/m.}
    \end{subfigure}
    \caption{Fracture gradient and reservoir pressure in the Broom Creek Formation.}
    \label{fig:FractureAndReservoirPressure}
\end{figure}

\begin{figure}[H]
    \centering
    \includegraphics[width=0.7\textwidth]{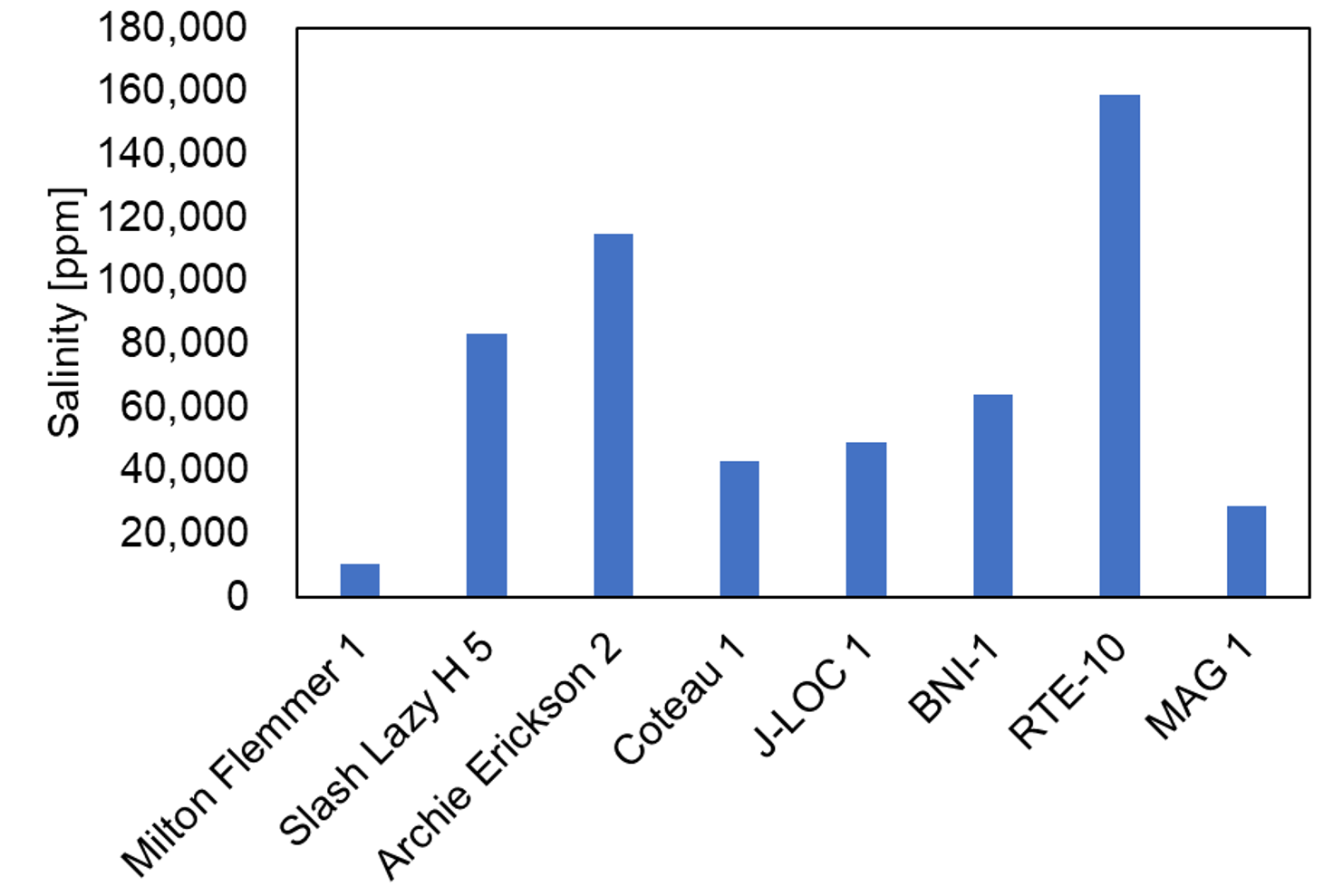}
    \caption{Salinity of the Broom Creek Formation based on fluid measurements. The average and median values are 69,025~ppm and 56,550~ppm, respectively.}
    \label{fig:MeasuredSalinity}
\end{figure}

\FloatBarrier
\section*{S2.~Details on Nested Local Grid Refinement Procedures}
\subsection*{S2.1.~Global Region Definition}

While the upscaling factors, $l_i$ and $l_j$, are applied uniformly within the LGR regions, two exceptions are allowed in the global region to maintain grid conformity. First, when the number of remaining fine cells is not divisible by the target upscaling factor, all remaining cells are combined into a single coarse block. For example, in the gray region between $x=53.3$ and $55.7~\mathrm{km}$ in Figure~6 in the main paper, the overlapping of the level-1 LGRs for the DCC and SCS projects along the $x$-axis leaves 24~fine cells between their boundaries. These 24~cells are therefore combined into a single coarse block, resulting in a local upscaling factor of~24 instead of the target value of~16. Second, to prevent combining cells with and without pore-volume multipliers, boundary cells are not combined normal to the boundaries, though they are combined along boundaries.

\subsection*{S2.2.~Constraints Included in the Constraint Satisfaction Problem}
Here we describe the five constraints handled in the Constraint Satisfaction Problem (Eq.~9 in the main text). The notation follows that used in the paper.

\begin{enumerate}
\item Each LGR region must contain at least one well. Specifically, at least one well must satisfy

\begin{equation}
    i_{\min} \le i_{\mathrm{well}} < i_{\max}, \qquad j_{\min} \le j_{\mathrm{well}} < j_{\max}.
\end{equation}

\item The coordinates of each child LGR region must conform to those of its parent LGR region. Specifically,
\begin{equation}
    \{i_{\min}^{c}, i_{\max}^{c}\} \subseteq \mathbf{i}^{p},
    \qquad
    \{j_{\min}^{c}, j_{\max}^{c}\} \subseteq \mathbf{j}^{p},
\end{equation}
and
\begin{equation}
    \mathbf{i}^{p} \cap [i_{\min}^{c},i_{\max}^{c}]
    \subseteq \mathbf{i}^{c},
    \qquad
    \mathbf{j}^{p} \cap [j_{\min}^{c},j_{\max}^{c}]
    \subseteq \mathbf{j}^{c}.
\end{equation}
Here $\mathbf{i}^{p}$ and $\mathbf{j}^{p}$ denote the sets of parent-LGR $(i,j)$ grid-line indices in the $x$ and $y$-directions, and $\mathbf{i}^{c}$ and $\mathbf{j}^{c}$ denote the corresponding sets for the child LGR. In addition, $\mathbf{i}^{p} \cap [i_{\min}^{c},i_{\max}^{c}]$ and $\mathbf{j}^{p} \cap [j_{\min}^{c},j_{\max}^{c}]$ denote the subsets of parent-LGR coordinates within the extent of the child LGR region in the $x$ and $y$-directions.

\item Each child LGR region must be completely enclosed within its parent LGR region. This constraint requires

\begin{equation}
    i_{\min}^{p} \le i_{\min}^{c} < i_{\max}^{c} \le i_{\max}^{p},
    \qquad
    j_{\min}^{p} \le j_{\min}^{c} < j_{\max}^{c} \le j_{\max}^{p}.
\end{equation}

\item For every pair of sibling LGRs, the coordinates within their overlapping extent must be identical. Consider an arbitrary pair of sibling LGRs, $s_k$ and $s_l$, with coordinate sets $(\mathbf{i}^{s_k},\mathbf{j}^{s_k})$ and $(\mathbf{i}^{s_l},\mathbf{j}^{s_l})$. Their overlapping extents in the $i$ and $j$-directions are defined as
\begin{equation}
    i_{\mathrm{o},\min}
    = \max\left(i_{\min}^{s_k},i_{\min}^{s_l}\right),
    \qquad
    i_{\mathrm{o},\max}
    = \min\left(i_{\max}^{s_k},i_{\max}^{s_l}\right),
\end{equation}
and
\begin{equation}
    j_{\mathrm{o},\min}
    = \max\left(j_{\min}^{s_k},j_{\min}^{s_l}\right),
    \qquad
    j_{\mathrm{o},\max}
    = \min\left(j_{\max}^{s_k},j_{\max}^{s_l}\right).
\end{equation}
For $i_{\mathrm{o},\min} < i_{\mathrm{o},\max}$, the $i$-coordinates within the overlapping extent must satisfy
\begin{equation}
    \mathbf{i}^{s_k}
    \cap [i_{\mathrm{o},\min},i_{\mathrm{o},\max}]
    =
    \mathbf{i}^{s_l}
    \cap [i_{\mathrm{o},\min},i_{\mathrm{o},\max}].
\end{equation}
Similarly, for $j_{\mathrm{o},\min} < j_{\mathrm{o},\max}$, the $j$-coordinates must satisfy
\begin{equation}
    \mathbf{j}^{s_k}
    \cap [j_{\mathrm{o},\min},j_{\mathrm{o},\max}]
    =
    \mathbf{j}^{s_l}
    \cap [j_{\mathrm{o},\min},j_{\mathrm{o},\max}].
\end{equation}

\item Each LGR region must exclude the boundary cells. Assuming that the grid lines are indexed starting from 1 in both the $x$ and $y$-directions, the grid-line indices range from 1 to $N_x+1$ and from 1 to $N_y+1$ (for a grid containing $N_x \times N_y$ fine-scale cells). Accordingly, this constraint requires
\begin{equation}
    1 < i_{\min} < i_{\max} < N_x+1,
    \qquad
    1 < j_{\min} < j_{\max} < N_y+1.
\end{equation}
This constraint prevents cells with and without pore-volume multipliers from being combined. It can be omitted if no special treatment is applied to boundary cells.

\end{enumerate}

\FloatBarrier

\section*{S3.~Surrogate Models}

\begin{figure}[h!]
    \centering
    \includegraphics[width=0.5\textwidth]{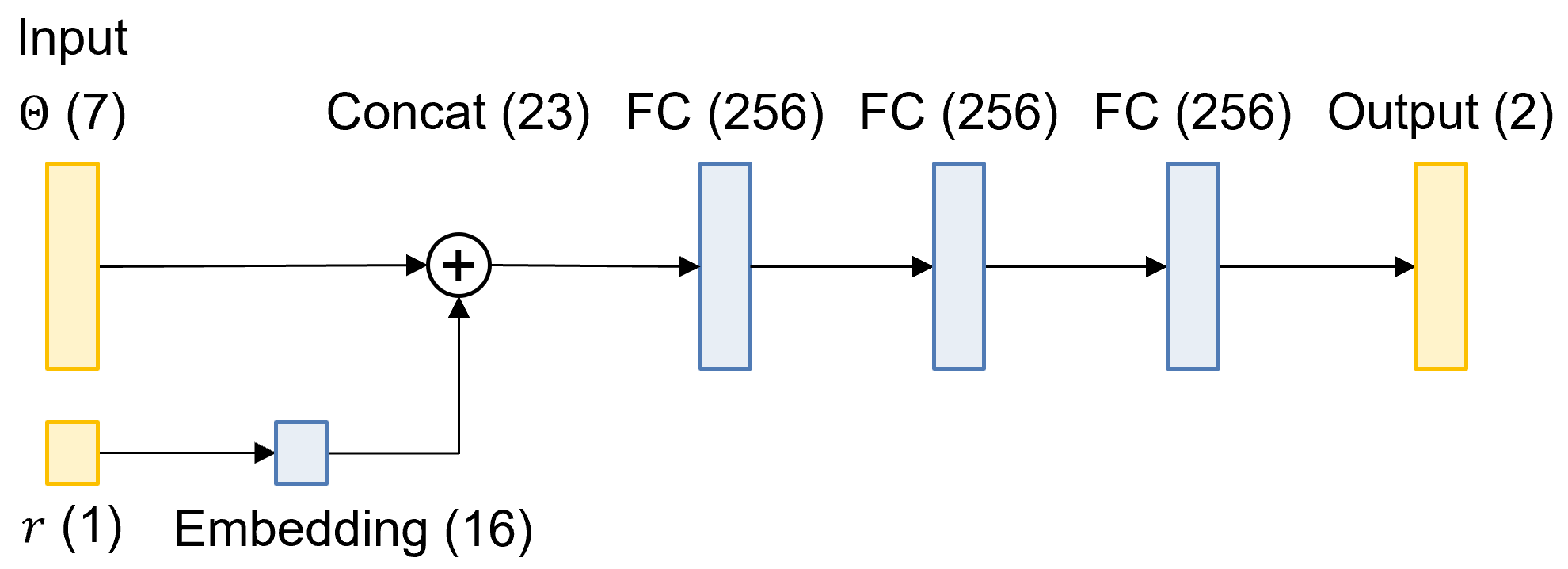}
    \caption{Schematic of the surrogate model architecture. Here $\theta$ denotes the set of input parameters with continuous values, while $r$ indicates a categorical input parameter. FC denotes a fully connected layer. Further details are provided in Table~\ref{tab:surrogate_architecture}.}
    \label{fig:surrogate_architecture}
\end{figure}

\begin{table}[t]
    \centering
    \caption{Architecture of the surrogate model ($p$ denotes the dropout probability).}
    \label{tab:surrogate_architecture}
    \begin{tabular}{lll}
    \toprule
    Component & Description & Dimension \\
    \midrule
    Continuous input & Seven continuous parameters & 7 \\
    Categorical input & Realization index & 1 \\
    Embedding & Learnable embedding vector & 16 \\
    Feature vector & Concatenated input & 23 \\
    FC layer 1 & Linear + SiLU + Dropout ($p=0.005$) & 256 \\
    FC layer 2 & Linear + SiLU + Dropout ($p=0.005$) & 256 \\
    FC layer 3 & Linear + SiLU + Dropout ($p=0.005$) & 256 \\
    Output layer & Linear & 2 \\
    \bottomrule
    \end{tabular}
\end{table}

\begin{figure}[h!]
    \centering

    \begin{subfigure}{0.35\textwidth}
        \centering
        \includegraphics[width=\textwidth]{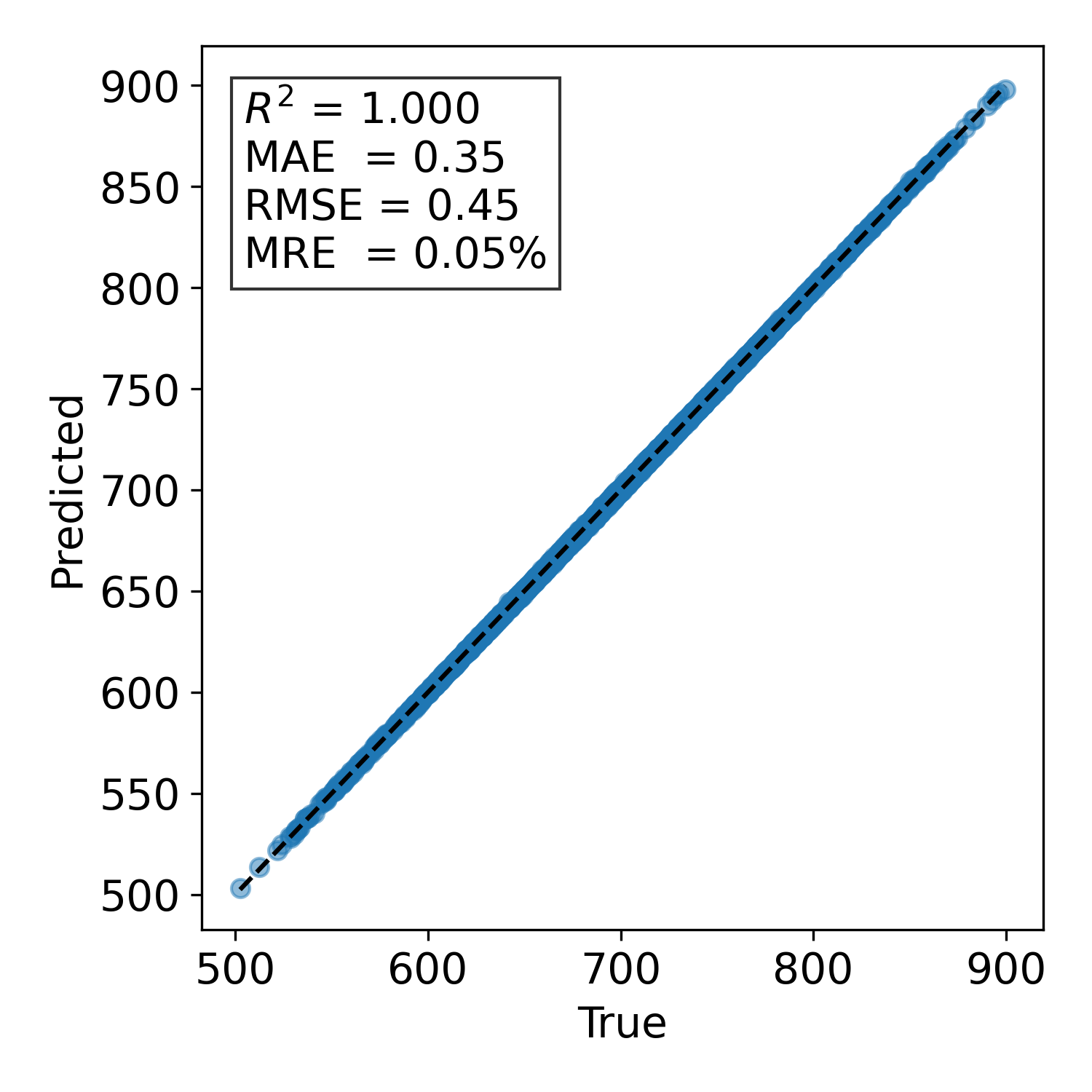}
        \caption{Total CO$_2$ injected (training)}
    \end{subfigure}
    \begin{subfigure}{0.35\textwidth}
        \centering
        \includegraphics[width=\textwidth]{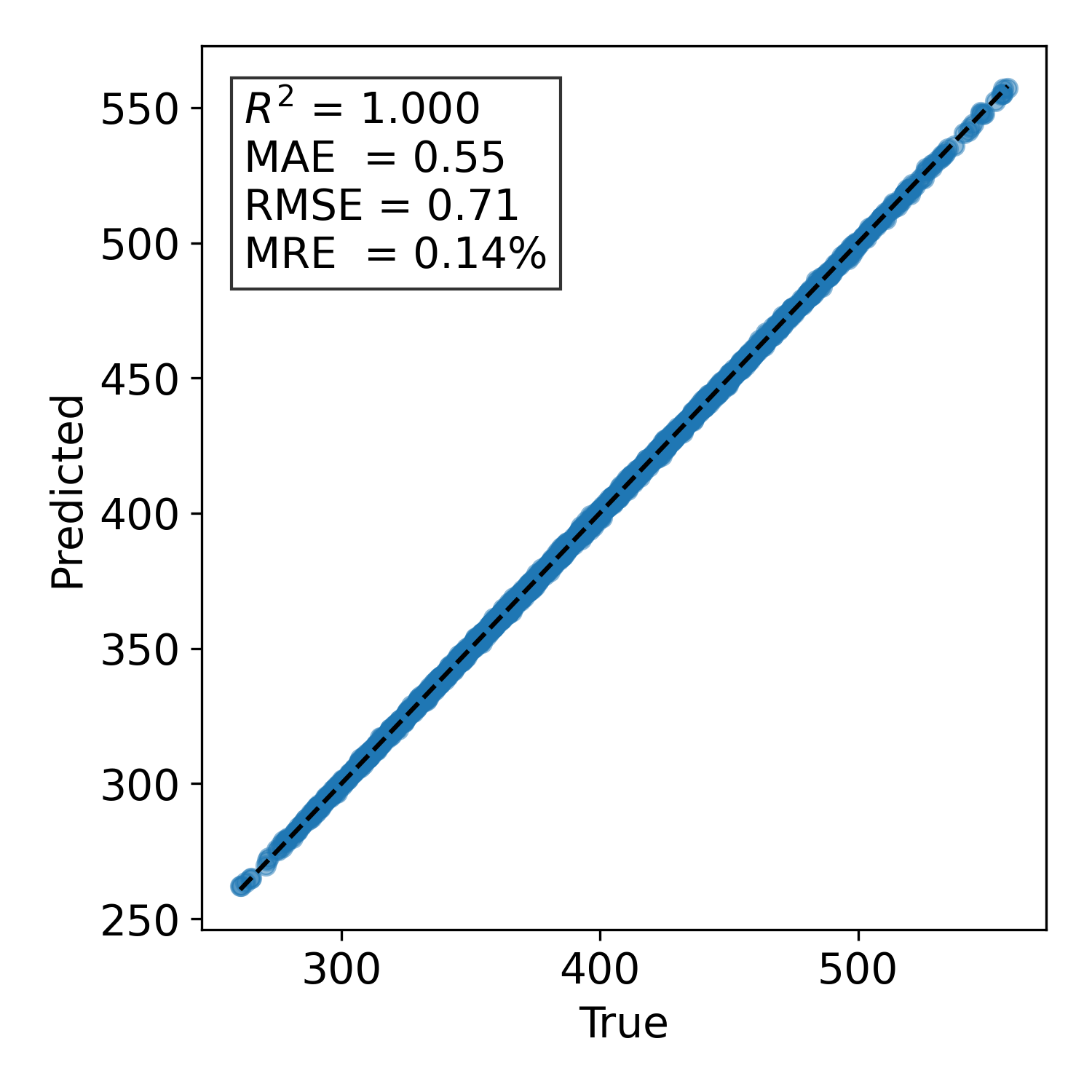}
        \caption{Total plume area (training)}
    \end{subfigure}

    \vspace{0.5em}

    \begin{subfigure}{0.35\textwidth}
        \centering
        \includegraphics[width=\textwidth]{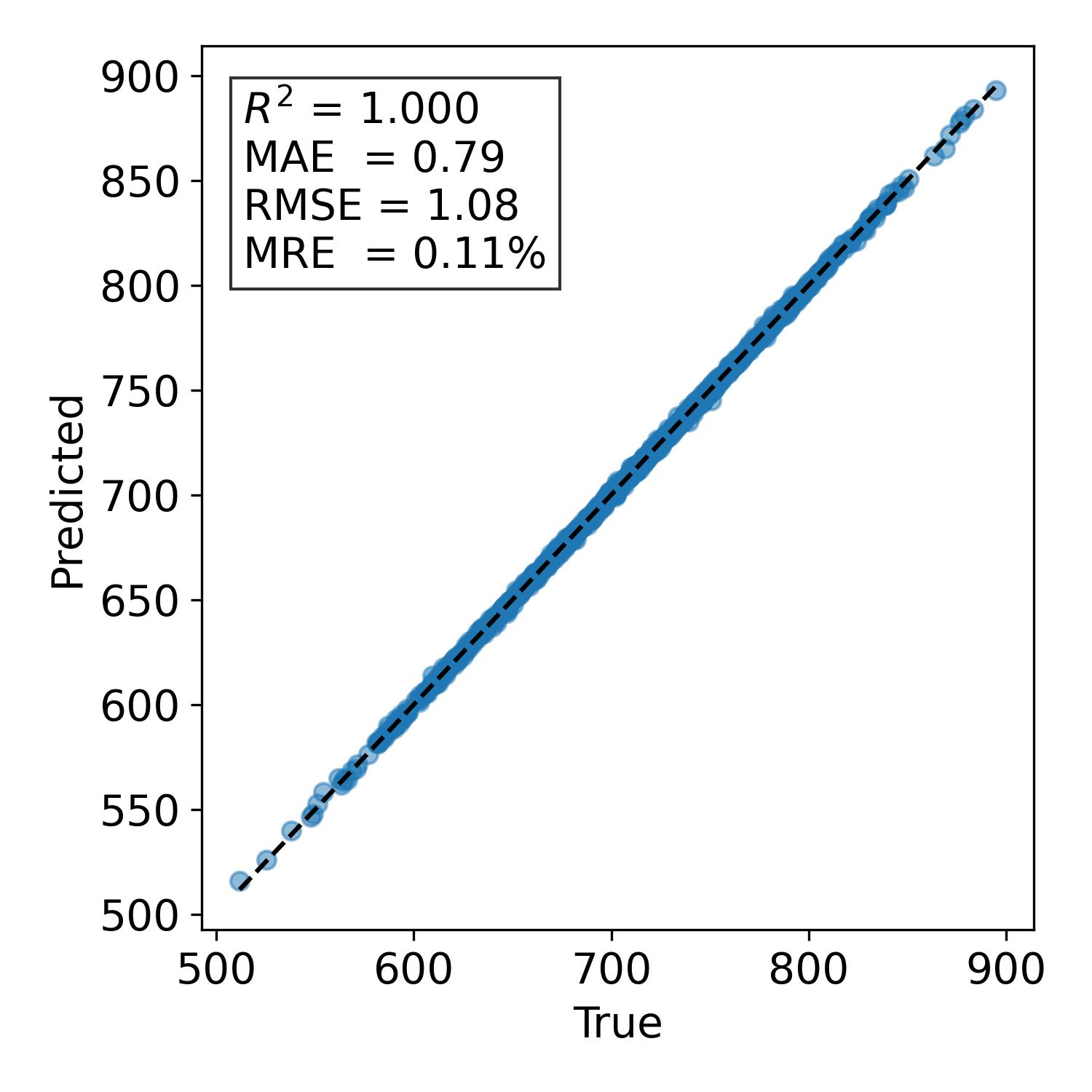}
        \caption{Total CO$_2$ injected (validation)}
    \end{subfigure}
    \begin{subfigure}{0.35\textwidth}
        \centering
        \includegraphics[width=\textwidth]{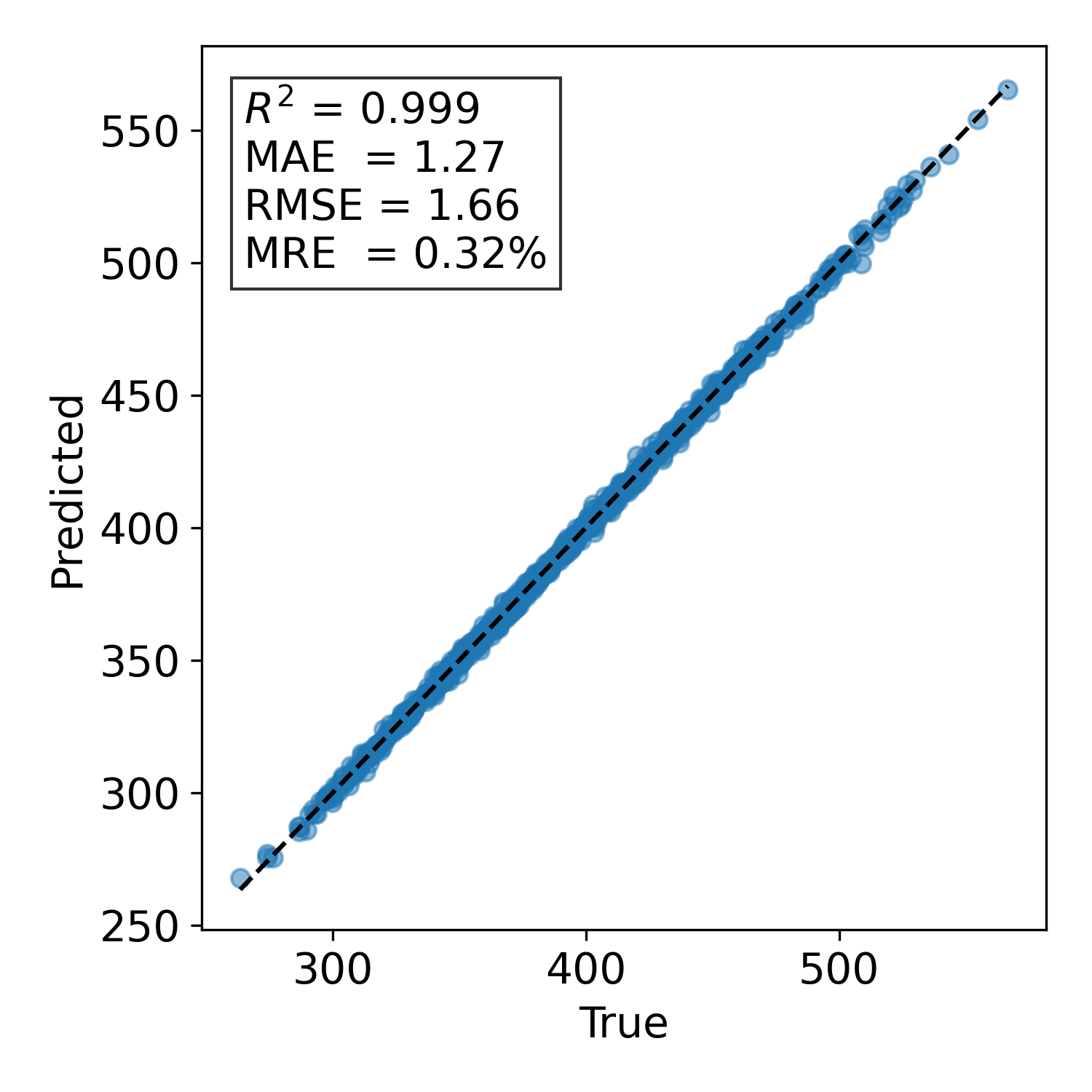}
        \caption{Total plume area (validation)}
    \end{subfigure}

    \caption{Prediction accuracy of the surrogate model. Results are at 50~years. Total CO$_2$ injected in units of MT, total plume area in units of km$^2$. MAE, RMSE, and MRE denote the mean absolute error, root mean square error, and mean relative error. Black dashed lines are of unit slope.}
    \label{fig:surrogate_model_performance}
\end{figure}

\FloatBarrier

\end{document}